\documentclass[aps,prb, superscriptaddress,twocolumn,longbibliography]{revtex4-1}
\usepackage{amsmath}
\usepackage{amssymb}
\usepackage{graphicx}
\usepackage[usenames,dvipsnames]{xcolor}
\usepackage{tikz}
\usepackage{pgffor}
\usepackage{verbatim}
\usepackage{bbm}
\usepackage{float}
\usepackage{mathrsfs}
\usepackage[colorlinks, breaklinks=true,linkcolor=red, citecolor=blue, linktocpage=true]{hyperref}

\renewcommand{\>}{\rangle}

\begin{document}

\title{Bulk-boundary correspondence for three-dimensional symmetry-protected topological phases}
\author{Chenjie Wang}
\affiliation{James Franck Institute and Department of Physics, University of Chicago, Chicago, Illinois 60637, USA}
\affiliation{Perimeter Institute for Theoretical Physics, Waterloo, Ontario N2L 2Y5, Canada}

\author{Chien-Hung Lin}
\affiliation{James Franck Institute and Department of Physics, University of Chicago, Chicago, Illinois 60637, USA}
\affiliation{Department of Physics, University of Alberta, Edmonton, Alberta T6G 2E1, Canada}

\author{Michael Levin}
\affiliation{James Franck Institute and Department of Physics, University of Chicago, Chicago, Illinois 60637, USA}

\date{\today}

\begin{abstract}
We derive a bulk-boundary correspondence for three-dimensional (3D) symmetry-protected topological (SPT) phases with unitary symmetries. The correspondence consists of three
equations that relate bulk properties of these phases to properties of their gapped, symmetry-preserving surfaces. 
Both the bulk and surface data appearing in our correspondence are defined via a procedure in which we gauge the symmetries of the system of 
interest and then study the braiding statistics of excitations of the resulting gauge theory. The bulk data is defined 
in terms of the statistics of bulk excitations, while the surface data is defined in terms of the statistics of surface 
excitations. An appealing property of this data is that it is plausibly complete in the sense that the bulk data uniquely distinguishes each 3D SPT phase, 
while the surface data uniquely distinguishes each gapped, symmetric surface.
Our correspondence applies to any 3D bosonic SPT phase with finite Abelian unitary symmetry group. It applies to any surface that 
(1) supports only Abelian anyons and (2) has the property that the anyons are not permuted by the symmetries.
\end{abstract}
\maketitle


\section{Introduction}

A gapped quantum many-body system is said to belong to a non-trivial symmetry-protected topological (SPT) phase if it 
satisfies three conditions: First, the Hamiltonian is invariant under some set of internal (on-site) symmetries, none of which are broken 
spontaneously. Second, the ground state is short-range entangled --- that is, the ground state can be transformed into a 
product state or atomic insulator using a local unitary transformation. Third, it is impossible to continuously 
connect the ground state to a product state or atomic insulator, by varying some parameter in the Hamiltonian, without breaking one of 
the symmetries or closing the energy gap.\cite{gu09, pollmann10, fidkowski11,chen11a,chen11b, schuch11,chen13} 
Famous examples of nontrivial SPT phases include the one dimensional Haldane spin chain\cite{haldane83}, which is protected by time reversal
symmetry, and the two-dimensional (2D) and three-dimensional (3D) topological insulators\cite{hasan10, qi11} which are protected 
by time reversal and charge conservation symmetry.

Perhaps the most interesting property of nontrivial SPT phases is that they have protected boundary 
modes. Here, the precise meaning of ``protected'' depends on dimensionality. For example, in 
the two dimensional case, the edges of SPT phases are believed to be protected in the sense that they cannot be 
both gapped and symmetric\cite{kane05,xu06,wu06, chen11c, levin12, else14}. On the other hand, 
in the three dimensional case, the surfaces of SPT phases are believed to be protected in the sense that 
any surface that is both gapped and symmetric must also support anyon excitations\cite{vishwanath13, wangc13, burnell14,
chen14,  bonderson13, wangc13b, chen14a,metlitski15, fidkowski13, metlitski14, wangc14}. 

For some SPT phases, we can not only establish the existence of a protected boundary, but we can derive a ``bulk-boundary correspondence.'' 
Let us clarify what we mean by this term since there are at least two different types of bulk-boundary correspondences discussed in the 
literature. One type of bulk-boundary correspondence is a construction that provides a particular (i.e. non-unique) field theory description of
the boundary for each bulk phase\cite{wen_advances_in_physics,levin12a,lu12,vishwanath13,xu13}. Another type of bulk-boundary correspondence is a 
\emph{universal} relation between \emph{measurable} properties of the bulk and boundary. In this paper, we will be interested in bulk-boundary correspondences
of the second kind. 

The classic example of such a bulk-boundary correspondence appears in the context of 2D non-interacting 
fermion systems with charge conservation symmetry. For these systems, one can relate the bulk electric Hall conductivity $\sigma_{xy}$, 
measured in units of $e^2/h$, to the number $N_R, N_L$ of right-moving and left-moving edge modes\cite{halperin82}:
\begin{equation}
\sigma_{xy} = N_R - N_L
\label{chernbulkbound}
\end{equation}
Similar relations, which connect bulk topological band invariants to the properties of boundary modes, are known for other non-interacting 
fermion systems.\cite{hasan10, qi11}  

Less is known about such bulk-boundary correspondences for \emph{interacting} SPT phases. 
One place where it would be particularly useful to have such a correspondence is in the context of
3D SPT phases with gapped symmetric surfaces. This case is interesting because surfaces of this kind are relatively easy 
to characterize due to the energy gap, but at the same time they exhibit nontrivial 
structure associated with surface anyon excitations. It is natural to ask: what are the general 
constraints that relate the bulk and surface properties of these systems? 

There are several cases where this question has been answered --- at least partially. In particular, 
in the case of 3D topological insulators, Refs.~\onlinecite{metlitski13,wangc13} derived constraints connecting the properties 
of the surface to properties of monopoles in the bulk. Similarly, it is possible to derive constraints
for other 3D SPT phases with at least one $U(1)$ symmetry and one anti-unitary symmetry\cite{senthil15}. 
Unfortunately, however, these constraints rely on a special combination of symmetries and therefore do not
give insight into the more general structure of the bulk-boundary correspondence.

In this paper, we take a step towards a more general theory by deriving a bulk-boundary correspondence for a large 
class of 3D SPT phases. More specifically, we consider general 3D \emph{bosonic} SPT phases with \emph{unitary} \emph{Abelian} symmetries. 
To simplify the discussion, we focus on gapped symmetric surfaces with the property that (1) the surface anyons are Abelian, and (2) these anyons are
not permuted\cite{bombin10,fidkowski15, barkeshli14} by the symmetries. We denote the symmetry group by $G=\prod_{i=1}^K \mathbb Z_{N_i}$, and the group of 
surface anyons by $A = \prod_{\mu=1}^{M} \mathbb Z_{N_\mu}$ --- with the group law in $A$ corresponding to fusion of anyons.
For this class of systems, we derive a bulk-boundary correspondence analogous to Eq. (\ref{chernbulkbound}).

Before we can explain our correspondence, we need to describe the bulk and surface data that we use.
The bulk data was originally introduced by Ref. \onlinecite{wangcj15} and consists of three tensors
\begin{equation}
\text{ Bulk data} = \{\Theta_{i,l}, \Theta_{ij,l}, \Theta_{ijk,l}\}, \label{bdata}
\end{equation}
where the indices $i,j,k,l$ range over $1,\dots, K$. These quantities are defined via a simple recipe.
Suppose we are given a lattice boson model belonging to an SPT phase with symmetry group $G$. To find the corresponding
bulk data, the first step is to minimally couple the model to a dynamical 
lattice gauge field with gauge group $G$.\cite{kogut79} 
After gauging the model 
in this way, the second step is to study the braiding statistics of the ``vortex loop'' excitations of the resulting gauge theory\cite{threeloop, ran14}. 
The tensors $\{\Theta_{i,l}, \Theta_{ij,l}, \Theta_{ijk,l}\}$ are then defined in terms of the braiding statistics of these vortex loops,
as reviewed in more detail in section \ref{sec:bulk_data}.

The surface data has a similar character and consists of five tensors
\begin{equation}
\text{Surface data} = \{\Phi_{\mu}, \Phi_{\mu\nu}, \Omega_{i\mu}, \Omega_{ij\mu}, x_{il}^\mu\} \label{sdata}
\end{equation}
where the indices $i,j,l$ range over $1,\dots, K$ and the indices $\mu,\nu$ range over $1, \dots, M$. Like the bulk data, the surface
data is defined by gauging the lattice boson model and studying the braiding statistics of the excitations of the resulting gauge theory.
The only difference is that we consider the braiding statistics of \emph{surface} excitations instead of bulk excitations. In particular,
the tensors $\{\Phi_{\mu}, \Phi_{\mu\nu}, \Omega_{i\mu}, \Omega_{ij\mu}, x_{il}^\mu\}$ are defined in terms of the braiding statistics
of surface anyons and vortex lines ending at the surface.

The reason we use the above bulk and surface data is that this data has a number of appealing properties. First,
the quantities in (\ref{bdata}) and (\ref{sdata}) are \emph{measurable} in the sense that they can be extracted from a 
microscopic model by following a well-defined procedure. Second, these quantities are 
\emph{topological invariants}: that is, they remain fixed under continuous, symmetry-preserving deformations of the (ungauged) 
Hamiltonian that do not close the bulk or surface gap, respectively.\footnote{To see this, recall that our gauging prescription maps gapped 
lattice boson models onto gapped gauge theories.\cite{wangcj15,levin12} 
Therefore, if two lattice boson models can be connected without closing the gap, the corresponding gauge theories can also be 
connected without closing the gap and hence must have the 
same braiding statistics data.} Finally, there is reason to think that the bulk data and surface data are \emph{complete} 
in the sense that the bulk data uniquely distinguishes every 3D SPT phase, while the surface data uniquely distinguishes every gapped 
symmetric surface (we discuss the evidence for this claim in sections \ref{sec:bulkinv_def} and \ref{sec:surfaceinv_def}). 

The main result of this paper is a set of three equations (\ref{formula1}-\ref{formula3}) that connect the bulk data (\ref{bdata}) to the 
surface data (\ref{sdata}). We derive these equations by relating the bulk braiding processes that define (\ref{bdata}) to the surface 
braiding processes that define (\ref{sdata}) using topological invariance and other properties of braiding statistics. 

Our results are closely related to a conjecture of Chen, Burnell, Vishwanath, and Fidkowski\cite{chen14}. To describe this conjecture, we need to
recall two facts. The first fact is that many (possibly all) 3D SPT phases with finite unitary symmetry group $G$ can be 
realized by exactly soluble lattice boson models known as \emph{group cohomology models}\cite{chen13}. These models are parameterized by 
elements of the cohomology group $H^4(G,U(1))$. The second fact is that each 2D anyon system with unitary symmetry group $G$
is associated with an anomaly which takes values in $H^4(G,U(1))$\cite{etingof10,chen14}. (See Refs.~\onlinecite{vishwanath13, kapustin14,cho14,wangj15} for other related discussions of anomalies.) If this anomaly is nonzero then the corresponding anyon 
system cannot be realized in a strictly 2D lattice model. Given these two facts, 
Chen et al. conjectured that gapped symmetric surfaces of group cohomology model always have an anomaly $\nu$ that matches the
$\nu \in H^4(G,U(1))$ defining the bulk cohomology model. The authors checked that this conjecture gives correct predictions 
for a particular lattice model.

What is the relationship between our bulk-boundary correspondence and this conjecture? To make a connection, we use our 
bulk-boundary formulas (\ref{formula1}-\ref{formula3}) to obtain constraints on the surfaces of group 
cohomology models. We then compare these constraints to those predicted by the conjecture and we show that the 
two sets of constraints are mathematically \emph{equivalent}. Thus, our bulk-boundary correspondence gives a \emph{proof} of 
the conjecture for the case where $G$ is Abelian. Conversely, the conjecture implies our bulk-boundary correspondence, 
if we assume that the group cohomology models realize every possible 3D SPT phase. 

The rest of the paper is organized as follows. In Sec.~\ref{sec:bulk_data} and Sec.~\ref{sec:surface_data}, we define the bulk data 
and surface data, respectively. In Sec.~\ref{sec:correspondence}, we present the bulk-boundary correspondence (\ref{formula1}-\ref{formula3}) 
that relates the two sets of data to one another. We derive the correspondence in 
Sec.~\ref{sec:derivation}. We discuss the implications of the correspondence for purely 2D systems in 
Sec.~\ref{sec:pure2d}. In Sec.~\ref{sec:connection_set}, we explain the connection between our bulk-boundary correspondence and the conjecture of Chen et al\cite{chen14}.
Technical arguments and calculations are given in the Appendices.




\section{Bulk data: Review}
\label{sec:bulk_data}

In this section we review the definition of the bulk data $\{\Theta_{i,l}, \Theta_{ij,l}, \Theta_{ijk,l}\}$:\cite{wangcj15} that is, we explain how to compute these quantities given an arbitrary
3D lattice boson model belonging to a SPT phase with unitary Abelian symmetry group $G = \prod_{i=1}^K \mathbb Z_{N_i}$. 

As discussed in the introduction, the computation/definition proceeds in two steps. The first step is to minimally couple the lattice 
boson model of interest to a dynamical lattice gauge field with gauge group $G$.\cite{kogut79} The details of this gauging procedure 
are not important for our purposes; the only requirement is that the gauge coupling constant is small so that the resulting gauge theory 
is gapped and deconfined. (See Refs.~\onlinecite{wangcj15,levin12} for a precise gauging prescription in which the coupling constant is 
chosen to be exactly zero). The second step is to study the excitations of the gauged model. The bulk data is defined in terms of the braiding statistics of these excitations. 

In what follows, we focus the second step of this procedure. First, we discuss the excitations of the gauged models and review their braiding statistics. After this preparation, we give the precise 
definition of $\{\Theta_{i,l}, \Theta_{ij,l}, \Theta_{ijk,l}\}$. 

\subsection{Bulk excitations}
We begin by reviewing the excitation spectrum of the gauged models.
The gauged models support two types of excitations in the bulk: particle-like \emph{charges} and loop-like \emph{vortices}. Charge excitations are characterized by their gauge charge
\begin{equation}
q=(q_1,\dots, q_K)
\end{equation}
where each component $q_i$ is an integer defined modulo $N_i$. Similarly, vortex loop excitations are characterized by their gauge flux
\begin{equation}
\phi = (\phi_{1}, \dots, \phi_{K})
\end{equation}
where each component $\phi_{i}$ is a multiple of $\frac{2\pi}{N_i}$ and is defined modulo $2\pi$. An important point is that while charge 
excitations are uniquely characterized by the amount of gauge charge that they carry, there are typically \emph{many} topologically 
distinct types of vortex loops that carry the same gauge flux $\phi$. These different loops can be obtained from one another by attaching charge excitations.

Some comments on notation: we will use Greek letters $\alpha, \beta,\dots$ to denote vortex excitations, and we will use $\phi_\alpha$ to denote the amount of gauge flux carried by $\alpha$. We will use $q,q',\dots$ to denote charge excitations, and we will use the same symbols $q,q',\dots$ to denote the amount of gauge charge that they carry.


Let us now discuss the braiding statistics of these excitations. There are several types of braiding processes one can consider:
(i) braiding of two charges, (ii) braiding of a charge around a vortex loop, and (iii) braiding involving multiple vortex loops. The
first kind of braiding process is not very interesting because the charges are all bosons: one way to see this is to
note that the charges can be viewed as local excitations of the original \emph{ungauged} model, and the ungauged model has only bosonic
excitations since it is short-range entangled. As for the second kind of process, it is easy to see that if we braid a charge $q$ around a vortex
loop $\alpha$, the resulting statistical phase is given by the Aharonov-Bohm law
\begin{equation}
\theta = q\cdot \phi_{\alpha} \label{eq_ab}
\end{equation}
where ``$\cdot$'' denotes the vector inner product.

\begin{figure}
\centering
\includegraphics{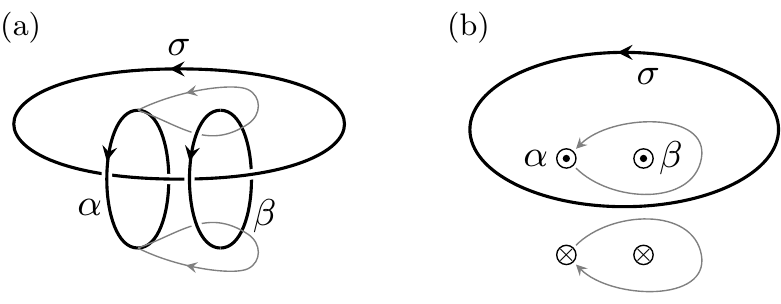}
\caption{(a) Three-loop braiding process. (b) Cross-section of the braiding process in the plane containing the base loop $\sigma$.}\label{fig_threeloop}
\end{figure}

All that remains is the last type of braiding process, involving multiple loops.
There are several ways to braid vortex loops, but in this paper we will primarily be interested in a braiding process in which a
loop $\alpha$ is braided around a loop $\beta$ while both loops are linked to a ``base loop'' $\sigma$ (Fig.~\ref{fig_threeloop})\cite{threeloop,ran14}.
We will also consider more general braiding processes involving multiple loops $\alpha_1,\alpha_2,\dots, \alpha_N$, all of which are linked to a single base loop $\sigma$.
The reason we focus on these kinds of braiding processes is that the associated Berry phases are \emph{not} fixed by the Aharonov-Bohm law, but instead probe more interesting properties of the bulk.

One technical complication is that in some models the vortex loop excitations have \emph{non-Abelian} braiding statistics even though the gauge group $G$ is Abelian.
Therefore, even if we specialize to the case where $G$ is Abelian as we do here, it is still important to have a more complete theory of loop braiding statistics that includes the concepts of fusion rules, quantum dimensions, and so on.

Fortunately this formalism can be developed rather easily. The key point is that there is a direct analogy between 3D loop braiding and 2D
particle braiding. This analogy can be seen by considering a 2D cross-section of the loop braiding
process [Fig.~\ref{fig_threeloop}(b)]. We can see that a braiding process involving two loops linked with a single base loop
can be mapped onto a process involving two particles in two dimensions. More generally, a process involving $N$ loops linked
to a single base loop can be mapped into a process involving $N$ particles in two dimensions.
In addition to braiding, the analogy also carries over to fusion processes. Just as two particles can be fused together to form another particle,
two loops $\alpha, \beta$ that are linked to the same loop $\sigma$ can be fused to form a new loop that is also linked to $\sigma$.

With this analogy, we can immediately generalize the notation and results of 2D anyon theory\cite{kitaev06} to
3D loop braiding. In particular, we can define $F$ symbols, $R$ symbols, and quantum dimensions  of loops
in the same way as for particles. We will denote these quantities by $F_{\alpha\beta\gamma,c}^\delta$, $R_{\alpha\beta,c}^\delta$,
and $d_{\alpha,c}$. Here $\alpha,\beta,\gamma$ are loops linked with a base loop $\sigma$, while $c$ is an integer
vector that parameterizes the gauge flux carried by $\sigma$:
\begin{equation}
\phi_\sigma = (\frac{2\pi}{N_1}c_1, \dots, \frac{2\pi}{N_K} c_K)
\end{equation}
(The reader may wonder why we use `$c$' instead of `$\sigma$' in our notation for $F_{\alpha\beta\gamma,c}^\delta$,
$R_{\alpha\beta,c}^\delta$, etc. The reason is that these quantities only depend on the \emph{gauge flux}
carried by $\sigma$ and this gauge flux is conveniently parameterized by $c$).

An additional quantity that we will need below is the \emph{topological spin} of a loop $\alpha$ linked to a base loop $\sigma$. We will denote this quantity by $s_{\alpha,c}$ where
$0\le s_{\alpha,c}<1$. Here, $s_{\alpha,c}$ is defined in the same way as for 2D anyon theories:
\begin{equation}
e^{i2\pi s_{\alpha,c}} = \frac{1}{d_{\alpha,c}}\sum_{\delta} d_{\delta, c}{\rm Tr}\left( R_{\alpha\alpha,c}^\delta \right)
\end{equation}
where the summation runs over all fusion channels $\delta$ of two $\alpha$ loops, both of which are linked to $\sigma$.

\subsection{Definition of bulk data}
\label{sec:bulkinv_def}
With this preparation, we are now ready to define the bulk data. 
This data consists of three tensors $\{\Theta_{i,l}, \Theta_{ij,l}, \Theta_{ijk,l}\}$,
where the indices $i,j,k,l$ range over $1,\dots, K$. To define these tensors, let $\alpha, \beta, \gamma$ 
be vortex loops linked to a base loop $\sigma$. Suppose that $\alpha$ carries unit type-$i$ flux, that is $\phi_\alpha = \frac{2\pi}{N_i} \mathbf e_i$ where $\mathbf e_i$ is the vector $(0,\dots,1,\dots,0)$ with the $i$th entry being $1$ and all other entries being $0$. Similarly, suppose that
$\beta, \gamma,\sigma$ carry unit flux $\frac{2\pi}{N_j} \mathbf e_j,\frac{2\pi}{N_k} \mathbf e_k,\frac{2\pi}{N_l}\mathbf e_l$, respectively. Then, $\Theta_{i,l}, \Theta_{ij,l}, \Theta_{ijk,l}$ are defined as follows:
\begin{enumerate}
\item $\Theta_{i,l} = 2\pi N_i s_{\alpha,\mathbf e_l}$, where $s_{\alpha,\mathbf e_l}$ is the topological spin of $\alpha$ when it is linked to $\sigma$;
\item $\Theta_{ij,l}$ is the Berry phase associated with braiding the loop $\alpha$ around $\beta$ for $N^{ij}$ times, while both are linked to $\sigma$;
\item $\Theta_{ijk,l}$ is the Berry phase associated with the following braiding process: $\alpha$ is first braided around $\beta$, then around $\gamma$, then around $\beta$ in the opposite direction, and finally around $\gamma$ in the opposite direction. Here $\alpha,\beta,\gamma$ are all linked to $\sigma$.
\end{enumerate}
Above, we have used $N^{ij}$ to denote the least common multiple of $N_i$ and $N_j$. (Throughout this paper, we will use $N^{ij\dots k}$ and $N_{ij\dots k}$ to denote the least common multiple and greatest common divisor of integers $N_i,N_j,\dots, N_k$ respectively.)

These definitions deserve a few comments. First, we would like to point out that one needs to do some work to show that the above quantities are well-defined. In particular, one needs to establish
two results: (1) $\Theta_{ij,l},\Theta_{ijk,l}$ are Abelian phases, and (2) $\Theta_{i,l}, \Theta_{ij,l}, \Theta_{ijk,l}$ depend only on $i,j,k,l$ and not on the choice of the loops $\alpha, \beta, \gamma, \sigma$. The first fact is not obvious since vortex loops can have non-Abelian braiding statistics in some cases so that the Berry phase associated with general braiding processes is actually \emph{non-Abelian}. The second fact is not obvious either since there are multiple topologically distinct loop excitations that carry the same gauge flux.\footnote{These loops differ from one another by the attachment of charge excitations.} The proof of these two properties is given in Ref.~\onlinecite{wangcj15}.

We should also explain our motivation for using this data to characterize bulk SPT phases. 
Much of our motivation comes from a result of Ref.~\onlinecite{wangcj15}, which showed that 
$\{\Theta_{i,l}, \Theta_{ij,l}, \Theta_{ijk,l}\}$ take different values in every group cohomology 
model with symmetry group $G =\prod_{i} \mathbb Z_{N_i}$. What makes this result especially 
interesting is that it has been conjectured that the group cohomology models realize every 3D SPT phase\cite{chen13}. 
If this conjecture is true, then we can conclude that the above data is \emph{complete} in the sense that it 
can distinguish all 3D SPT phases with finite Abelian unitary symmetry.



Finally, we would like to make a comment about the definition of $\Theta_{i,l}$. Unlike the other two quantities 
which are concretely defined through braiding processes, $\Theta_{i,l}$ is a rather abstract quantity since $s_{\alpha,c}$ 
does not have a simple interpretation in terms of physical processes. Fortunately there is an alternative definition of $\Theta_{i,l}$ 
which is more concrete\cite{wangcj15}:
\begin{enumerate}
\item $\Theta_{i,l}$ is equal to $(-1)$ times the phase associated with braiding $\alpha$ around its anti-vortex $\bar\alpha$ for $\frac{N_i}{2}$ times, while both $\alpha$ and $\bar\alpha$ are linked to $\sigma$.
\end{enumerate}
Obviously, this definition fails when $N_i$ is odd since $N_i/2$ is not an integer. However, when $N_i$ is odd, there is no need to find an alternative definition for $\Theta_{i,l}$: the reason is that $\Theta_{i,l}$ is known to satisfy the constraints (see Appendix \ref{sec:app_bulkin_con})
\begin{equation}
N_i\Theta_{i,l} =0 \pmod{2\pi}, \quad 2\Theta_{i,l} = \Theta_{ii,l} \pmod{2\pi} \label{spinconstraint}
\end{equation}
Together with another constraint $N_i\Theta_{ii,l}=0 \pmod{2\pi}$, it is easy to see that these constraints completely determine $\Theta_{i,l}$ in terms of $\Theta_{ii,l}$ when $N_i$ is odd. 
More specifically, one can derive the following relation:
\begin{equation}
\Theta_{i,l} = \frac{N_i+1}{2} \Theta_{ii,l} \pmod{2\pi}
\label{thetailodd}
\end{equation}


\section{Surface data}

Next we define the surface data $\{\Phi_{\mu}, \Phi_{\mu\nu}, \Omega_{i\mu}, \Omega_{ij\mu}, x_{il}^\mu\}$. Our task is as follows: suppose we are given a 3D lattice boson model belonging to an SPT phase with symmetry group $G = \prod_{i=1}^K \mathbb Z_{N_i}$. Suppose that this model is defined in a geometry with a surface and that this surface has three properties: (1) it is gapped and symmetric, (2) it supports only Abelian anyons, and (3) it has the property that the surface anyons are not permuted by the symmetries. Given a microscopic model of this kind, we need to explain how to compute the corresponding surface data.

The computation/definition proceeds in the same way as for the bulk data. First, we couple the lattice boson model to a dynamical lattice gauge field with gauge group $G$. Then, after gauging the model, we study the surface excitations of the resulting gauge theory. The surface data is defined in terms of the braiding statistics of these surface excitations.

In what follows, we build up to the definition of 
$\{\Phi_{\mu}, \Phi_{\mu\nu}, \Omega_{i\mu}, \Omega_{ij\mu}, x_{il}^\mu\}$ in several steps. First, we discuss the surface excitations of the original \emph{ungauged} boson models. Then we discuss the excitations of the \emph{gauged} models. Finally, after this preparation, we explain the definition of the surface data.

\label{sec:surface_data}


\subsection{Surface excitations of ungauged models}
\label{sec:surex_ungauged}

By assumption the surfaces of the ungauged boson models only support Abelian anyons. The purpose of this section is to introduce some notation for labeling these anyons and describing their braiding statistics.

Our notation for labeling the anyons is based on the observation that the anyons form an Abelian group under fusion. Denoting this Abelian group by $A = \prod_{\mu=1}^M\mathbb Z_{N_\mu}$, we label each anyon by a group element $x\in A$, or equivalently, an $M$ component integer vector
\begin{equation}
x=(x^1, \dots, x^M)
\label{anyonlabel}
\end{equation}
where each component $x^\mu$ is defined modulo $N_\mu$. The vacuum anyon $\mathbbm 1$ corresponds to the zero vector $(0,\dots,0)$, while the fusion product of two anyons $y,y'$ is given by
$y+y'$.

To describe the braiding statistics of these particles, we focus on the \emph{unit} anyons --- that is, the anyons labeled by vectors
$\boldsymbol\epsilon_\mu = (0,\dots,1,\dots,0)$ with the $\mu$th entry being $1$ and all other entries being $0$. We denote the exchange 
statistics of the unit anyon $\boldsymbol\epsilon_\mu$ by $\Phi_\mu$ and we denote the mutual statistics of the unit anyons $\boldsymbol\epsilon_\mu$ and $\boldsymbol\epsilon_\nu$ by $\Phi_{\mu\nu}$. From $\Phi_\mu$ and $\Phi_{\mu\nu}$, we can reconstruct the exchange statistics $\theta_x$ and mutual statistics $\theta_{xy}$ of any anyons $x,y \in A$:
\begin{align}
\theta_{x} & = \sum_\mu (x^\mu)^2 \Phi_\mu  + \sum_{\mu<\nu} x^\mu x^\nu \Phi_{\mu\nu}   \label{exchange} \\
\theta_{xy} & = \sum_{\mu\nu} x^\mu y^\nu \Phi_{\mu\nu}  \label{mutual}
\end{align}
Here, the above formulas (\ref{exchange}, \ref{mutual}) follow immediately from the linearity relations for Abelian statistics:
\begin{align*}
\theta_{x+y} &= \theta_x+\theta_y+\theta_{xy}\nonumber\\
\theta_{x(y+y')}& = \theta_{xy}+\theta_{xy'} 
\end{align*}

\subsection{Surface excitations of gauged models}
\label{sec:surex_gauged}

\begin{figure}
\centering
\includegraphics{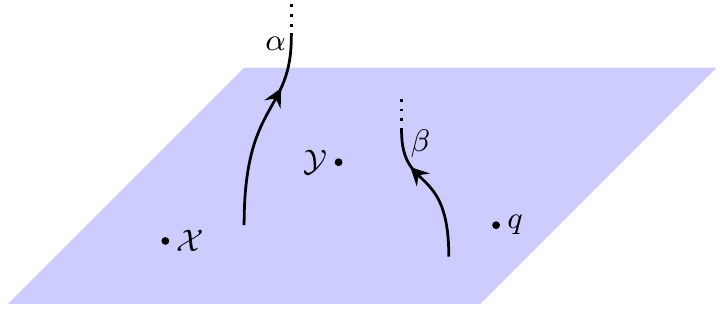}
\caption{ Sketch of excitations on and near the surface. Above the surface is the gauged SPT model and below is the vacuum. 
}\label{fig_surfexcitation}
\end{figure}

We now move on to discuss the surface excitations in the \emph{gauged} models.
The gauged models support three types of excitations on or near the surface: (1) point-like charge excitations that can exist in either
the bulk or the surface; (2) point-like surface anyons that are confined to the surface; (3) vortex line excitations that end on the surface. See Fig.~\ref{fig_surfexcitation} for a sketch of these excitations.

Let us discuss each of these excitations in turn. We begin with the charges and vortex lines. Like their bulk counterparts, the charges can be labeled by their
gauge charge $q = (q_1,...,q_K)$ while the vortex lines can be labeled by their gauge flux $\phi = (\phi_1,...,\phi_K)$. Just as in 
the bulk, the gauge charge uniquely characterizes the charge excitations, while the gauge flux does \emph{not} uniquely characterize the vortex lines: there are multiple topologically distinct vortex lines that carry the same gauge flux $\phi$, which differ from one another by the attachment of surface anyons and charges.

There are three types of braiding processes we can perform with charges and vortex lines. First, we
can braid or exchange charges with one another. These processes are not very interesting since the
charges are all bosons. Second, we can braid charges around vortex lines. As in the bulk, the Berry
phase for such processes is given by the Aharonov-Bohm formula (\ref{eq_ab}). Finally, we can braid
vortex lines around one another.\footnote{To make sense of braiding processes involving multiple vortex lines, we have to
keep track of the \emph{other} ends of the vortex lines.} The latter processes are highly non-trivial and can
even give non-Abelian Berry phases. We will see some examples of these processes later when we derive the bulk-boundary correspondence
in Sec.~\ref{sec:derivation}.

Let us now turn to the surface anyons. To understand the structure of these excitations, it is helpful to think about their relationship to
the surface anyons in the \emph{ungauged} models. In particular, we will argue below that
each surface anyon $\mathcal{X}$ in the gauged model is naturally associated with a corresponding surface anyon in the ungauged model,
which we denote by $\xi_{\mathcal X}$. 
We will think of the mapping $\mathcal X \rightarrow \xi_{\mathcal X}$ as being analogous to the mapping between vortices
$\alpha$ and their gauge flux $\phi_\alpha$; therefore we will use the terminology that $\xi_{\mathcal X}$ is the ``anyonic flux'' carried by 
$\mathcal X$. Like the gauge flux, $\xi_{\mathcal X}$ can be concretely represented as a vector --- or more precisely, an $M$ component 
integer vector, as in Eq. \ref{anyonlabel}.

At an intuitive level, $\xi_{\mathcal{X}}$ is obtained by ``ungauging'' the excitation $\mathcal{X}$;
to define $\xi_{\mathcal{X}}$ more precisely, let $|\tilde{\psi}\>$ be an excited state of the gauged model which contains an anyon
$\mathcal{X}$ localized near some point $r$. Let us suppose that $|\tilde{\psi}\>$ has vanishing gauge flux through every plaquette in the
lattice. (We can make this assumption without any loss of generality since $\mathcal{X}$ is not a vortex excitation).
Then we can find a gauge in which the state $|\tilde{\psi}\>$ has a vanishing lattice gauge field on every link of the lattice.
In this gauge, the state $|\tilde{\psi}\>$ can be written as a tensor product $|\tilde{\psi}\> = |\psi\> \otimes |a=0\>$ where $|a=0\>$ is a ket
describing the configuration of the lattice gauge fields, and $|\psi\>$ is a ket describing the configuration of the matter fields. Since
$|\psi\>$ only involves matter fields, we can think of it as an excited state of the \emph{ungauged} model. By construction this state
contains a localized excitation near the point $r$; we define $\xi_{\mathcal{X}}$ to be the
anyon type of this localized excitation. \footnote{The reader may worry that the localized excitation in {$|\psi\rangle$} might not have a
definite anyon type --- that is, it might be a linear superposition of different anyons; however, one can argue that this situation
does not happen if the symmetry does not permute the anyons, as we assume here.}

The mapping between anyons $\mathcal{X}$ and their anyonic flux $\xi_{\mathcal{X}}$ has several important properties.
The first property is that the mapping is not one-to-one: distinct anyons
$\mathcal X \neq \mathcal{X}'$ can carry the same anyonic flux, $\xi_\mathcal{X} = \xi_{\mathcal{X}'}$. The simplest
example of this is given by the charge excitations: all the charges share the same (trivial) flux, $\xi_q = (0,\dots,0)$. More
generally, two anyons $\mathcal{X}, \mathcal{X'}$ have the same anyonic flux if and only if $\mathcal{X}'$ can be obtained from
$\mathcal{X}$ by attaching a charge excitation, that is $\mathcal{X}' = \mathcal{X} \times q$ for some $q$.

Another important property of the mapping $\mathcal{X} \rightarrow \xi_{\mathcal X}$ is that the braiding statistics of anyons
$\mathcal{X}, \mathcal{Y}$, etc. is identical to the braiding statistics of $\xi_{\mathcal X}, \xi_{\mathcal Y}$, etc.
More specifically, it can be shown that the mutual statistics between any two anyons $\mathcal X$ and $\mathcal Y$ is Abelian and is
given by
\begin{equation}
\theta_{\mathcal X\mathcal Y} = \theta_{\xi_\mathcal X \xi_\mathcal Y} \label{eq_AB_stat}
\end{equation}
Similarly, the exchange statistics of $\mathcal X$, or equivalently its topological spin $2\pi s_\mathcal X$, is given by
\begin{equation}
2\pi s_{\mathcal X} = \theta_{\mathcal X} = \theta_{\xi_\mathcal X}\label{eq_spin}
\end{equation}

Given these results, one might be tempted to conclude that the surface anyons in the gauged model have Abelian statistics.
However, this conclusion is incorrect: the surface anyons in the gauged model can be \emph{non-Abelian} in general. This
non-Abelian character is not manifest when we focus on braiding processes that only involve surface anyons but it becomes apparent
when we consider processes in which a surface anyon is braided around a vortex line. Such processes will play an important role in
the definition of the surface data in the next section.

The last property of the mapping $\mathcal{X} \rightarrow \xi_\mathcal X$ involves fusion rules. Because the surface anyons can be
non-Abelian, they can have complicated fusion rules of the general form
\begin{equation}
\mathcal X \times \mathcal Y = \sum_\mathcal Z N_{\mathcal X \mathcal Y}^\mathcal Z \mathcal Z\label{eq_fusion}
\end{equation}
However, these rules have a special structure: all the fusion outcomes $\mathcal Z$ carry the same anyonic flux $\xi_\mathcal Z$ which is given by
\begin{equation}
\xi_\mathcal Z = \xi_\mathcal X + \xi_\mathcal Y \label{eq_fusion2}
\end{equation}

\subsection{Definition of surface data}
\label{sec:surfaceinv_def}

\begin{figure}
\centering
\includegraphics{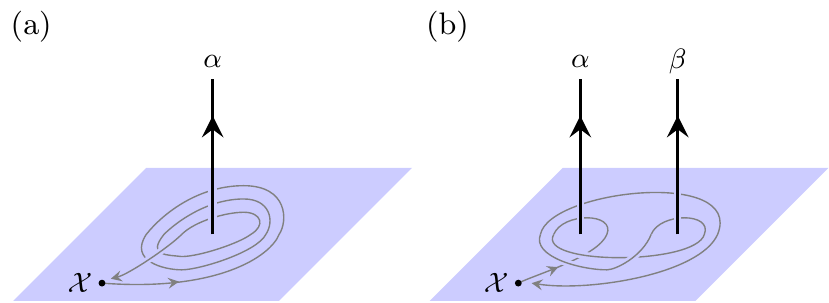}
\caption{Braiding processes that define the surface data $\Omega_{i\mu}$ [panel (a)] and $\Omega_{ij\mu}$ [panel (b)].}\label{fig_trajectories}
\end{figure}

Now that we have discussed the surface excitations and their statistics, we are ready to define the surface data. This data consists of five
tensors $\{\Phi_{\mu\nu}, \Phi_\mu, \Omega_{i\mu}, \Omega_{ij\mu}, x_{il}^\mu \}$,
where the indices $i,j,l$ run over $1,...,K$ while $\mu,\nu$ run over $1,...,M$. We have already defined $\Phi_{\mu\nu}, \Phi_\mu$
in Sec.~\ref{sec:surex_ungauged}; below we define the other two quantities
$\Omega_{i\mu}$ and $\Omega_{ij\mu}$. We discuss $x_{il}^\mu$ in the next section.

The two quantities $\Omega_{i\mu}$ and $\Omega_{ij\mu}$ are defined in terms of braiding statistics of surface anyons and vortex lines.
Let $\mathcal X$ be a surface anyon with unit type-$\mu$ anyonic flux, i.e. $\xi_\mathcal X = \boldsymbol \epsilon_\mu$, where
$\boldsymbol\epsilon_\mu = (0,\dots,1,\dots,0)$ with the $\mu$th entry being $1$.
Let $\alpha,\beta$ be vortex lines carrying unit type-$i$ and type-$j$ flux, i.e.
$\phi_\alpha = \frac{2\pi}{N_i}\mathbf  e_i$ and $\phi_\beta = \frac{2\pi}{N_j} \mathbf  e_j$. Then $\Omega_{i\mu}$ and $\Omega_{ij\mu}$ are defined
by:
\begin{enumerate}
\item $\Omega_{i\mu}$ is the Berry phase associated with braiding $\mathcal X$ around $\alpha$ for
$N^{i\mu}$ times;

\item $\Omega_{ij\mu}$ is the Berry phase associated with the following process: $\mathcal X$ is first braided around
$\alpha$, then around $\beta$, then around $\alpha$ in the opposite direction, and finally
around $\beta$ in the opposite direction.
\end{enumerate}
The above braiding processes are shown in Fig.~\ref{fig_trajectories}. Similarly to Sec. \ref{sec:bulkinv_def}, we use the notation
$N^{i\mu}$ to denote the least common multiple of $N_i$ and $N_\mu$, where the $N_i$ are the cyclic factors in the symmetry group 
$G = \prod_{i=1}^K \mathbb{Z}_{N_i}$, and the $N_\mu$ are the cyclic factors in the anyon group $A = \prod_{\mu=1}^M \mathbb{Z}_{N_\mu}$.

\begin{figure*}
\centering
\includegraphics{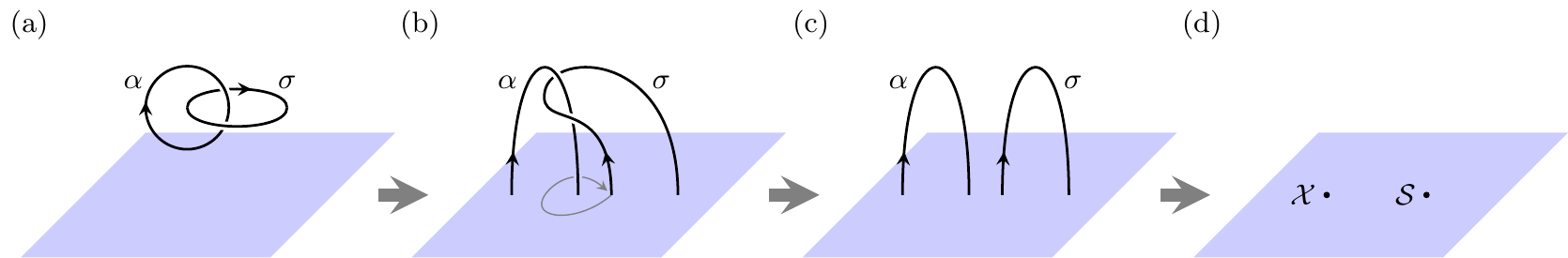}
\caption{ Interpreting $x_{il}^\mu$ through a 3D thought experiment. See the main text for details. }\label{fig_xdef}
\end{figure*}

To show that the above quantities are well-defined, we need to prove that (i) $\Omega_{i\mu}, \Omega_{ij\mu}$ are
always Abelian phases and (ii) they only depend on the indices $i,j,\mu$, i.e, on the gauge flux of the vortex lines and the anyonic flux
of $\mathcal{X}$. In other words, if we choose another anyon $\mathcal X'$ with $\xi_{\mathcal X'} =\boldsymbol \epsilon_\mu$
and another pair of vortex lines $\alpha',\beta'$ with $\phi_{\alpha'} = \frac{2\pi}{N_i}\mathbf  e_i$,
$\phi_{\beta'} = \frac{2\pi}{N_j} \mathbf  e_j$, we will obtain the same phases
$\Omega_{i\mu}$ and $\Omega_{ij\mu}$. The proof of these two points is technical, and hence is given separately in
Appendix \ref{sec:app_def}.

Our motivation for using the above surface data is twofold. First, there
is reason to think that the surface data is \emph{complete} in the sense that it can distinguish every
gapped symmetric surface satisfying our two assumptions, namely that (1) the surface supports only Abelian anyons and
(2) these anyons are not permuted by the symmetry. The main evidence for this
is that our data $\{\Phi_{\mu\nu}, \Phi_\mu,\Omega_{i\mu}, \Omega_{ij\mu},x_{il}^\mu\}$ is mathematically
equivalent to another set of data\cite{kitaev06, essin13,fidkowski15, barkeshli14} $\{F_{xyz},R_{xy},\omega \in H^2(G,A)\}$ that characterizes
2D anyon systems with symmetry group $G$. We discuss this equivalence in Sec.~\ref{sec:connection_set}. 
The latter data is plausibly complete in the context of the above class of surfaces\footnote{Here it is important to distinguish between strictly 2D systems and surfaces of 3D systems. 
In the context of strictly 2D systems, the data {$\{F_{xyz},R_{xy},\omega \in H^2(G,A)\}$} needs to be supplemented by an additional piece of information, namely an element of the
cohomology group {$\nu \in H^3(G,U(1))$}.\cite{kitaev06,fidkowski15,barkeshli14} However, we believe that this additional piece of data is unnecessary for characterizing surfaces of 3D systems, at least if one defines the notion of ``topologically equivalent'' surfaces in an appropriate manner.}
and if this is the case, then our data must be complete as well. Our second source of motivation for using the above surface data is that it can be naturally related to the bulk data via a
bulk-boundary correspondence, as we discuss below.


\subsection{An auxiliary surface quantity}
\label{sec:defx}

The last piece of surface data is a three index tensor $x_{il}^\mu$ where the indices $i,l$ run from $1,...,K$ while $\mu$ runs from $1,...,M$. We define this quantity \emph{implicitly} via the equation
\begin{equation}
\sum_\nu x_{il}^\nu \Phi_{\mu\nu} = \Omega_{il\mu} \pmod{2\pi} \label{eq_x7}
\end{equation}
More precisely, we define $x_{il}^\nu$ to be the unique integer tensor that satisfies the above equation and has components $x_{il}^\nu \in \{0,1,...,N_\nu-1\}$.

To understand the physical meaning of Eq. (\ref{eq_x7}), let us imagine fixing the indices $i,l$. Then $x_{il}^\nu$ reduces to an 
$M$ component integer vector, which we can think of as describing a \emph{surface anyon} $x_{il}$ in the ungauged model. Likewise, 
the left hand side of Eq. (\ref{eq_x7}) can be interpreted as the statistical phase associated with braiding a type-$\mu$ unit anyon 
around the anyon $x_{il}$. In this interpretation, $x_{il}$ represents the unique surface anyon with the property that its mutual 
statistics with type-$\mu$ unit anyons is $\Omega_{il \mu}$. From this point of view, it is not hard to see why $x_{il}^\mu$ exists 
and is unique. Indeed, the uniqueness of $x_{il}^\mu$ follows from the general principle of ``braiding non-degeneracy'' which guarantees
that no two surface anyons have the same mutual statistics with respect to all other surface anyons. \cite{kitaev06} 
As for the existence of $x_{il}^\mu$, this is less obvious but can also be deduced, with some algebra, from braiding non-degeneracy 
together with the fact that $\Omega_{il\mu}$ is a multiple of $2\pi/N_\mu$. (See Eq. \ref{surf_constraint2} below).

From Eq. \ref{eq_x7} we can see that $x_{il}^\mu$ carries the same information as $\Omega_{ij\mu}$ (assuming we know $\Phi_{\mu \nu}$). Therefore, the reader may wonder why we bother to define this additional quantity. One reason is that $x_{il}^\mu$ is often more convenient to work with 
than $\Omega_{ij\mu}$. Another reason is that the quantity $x_{il}^\mu$ plays an important role in the bulk-boundary correspondence. This role can be traced to the following thought experiment. Imagine a configuration of two linked vortex loops, $\alpha$ and $\sigma$ in the bulk (Fig.\ref{fig_xdef}) where $\alpha$ and $\sigma$ carry unit type-$i$ and type-$l$ flux, respectively, that is $\phi_\alpha = \frac{2\pi}{N_i}\mathbf e_i$ and $\phi_\sigma = \frac{2\pi}{N_l}\mathbf e_l$. Now imagine that we pull the linked loops down to the surface, absorbing a part of each loop into the surface, leaving a pair of tangled vortex ``arches'' [Fig.~\ref{fig_xdef}(b)]. We then disentangle the two arches by unwinding one of the two ends of $\sigma$ [the braiding path is shown in Fig.~\ref{fig_xdef}(b)]. The result is two separate arches [Fig.~\ref{fig_xdef}(c)]. We then shrink the two separated arches to the surface, leaving behind two localized surface excitations. We denote the resulting state by $|\psi\>$. In general, we know that any surface excitation can be written as a linear superposition of different states, each of which has a definite anyon type. Thus, we can write
\begin{equation}
|\psi\> = c |\mathcal X\>\otimes|\mathcal S\> + c' |\mathcal X' \> \otimes |\mathcal S'\> + \dots \label{wavefunction2}
\end{equation}
where $|\mathcal X\>\otimes|\mathcal S\>$ denotes a state with surface anyons $\mathcal X$ and $\mathcal S$, while
$|\mathcal X'\>\otimes|\mathcal S'\>$ denotes a state with surface anyons $\mathcal X'$ and $\mathcal S'$, and the coefficients $c,c'$, etc. are the corresponding complex amplitudes [Fig.~\ref{fig_xdef}(d)].
What does this thought experiment have to do with $x_{il}^\nu$? We show in Appendix \ref{sec:app_xil} that all the anyons $\mathcal{X}, \mathcal{X'},...$ carry the same anyonic flux $x_{il}$, while all the anyons $\mathcal{S}, \mathcal{S'},...$ carry anyonic flux $-x_{il}$:
\begin{align}
\xi_\mathcal X &= \xi_{\mathcal X'} = ... = x_{il} \nonumber \\
\xi_{\mathcal S} &= \xi_{\mathcal S'} = ... = -x_{il} \label{xileqs}
\end{align}
Thus the quantity $x_{il}$ naturally appears when we think about absorbing linked vortex loops into the surface.

\subsection{General constraints on surface data}

In this section, we discuss some of the constraints on $\{\Phi_{\mu\nu}, \Phi_\mu, \Omega_{i\mu}, \Omega_{ij\mu}, x_{il}^\mu\}$. Two important constraints are
\begin{align}
N_{\mu \nu} \Phi_{\mu\nu}  & = 0 \pmod{2\pi} \label{phi_constraint1}
\end{align}
and
\begin{equation}
N_\mu \Phi_\mu   = \begin{cases}
		0 \pmod{2\pi}  & \text{if $N_\mu$ is odd}\\
		0 \text{ or } \pi \pmod{2\pi}  & \text{if $N_\mu$ is even}
\end{cases}
\label{phi_constraint}
\end{equation}
To derive these constraints, consider the unit type-$\mu$ anyon $\boldsymbol \epsilon_\mu$. Using the fact that the fusion product of $N_\mu$ of these anyons gives the vacuum excitation, it is not hard to show that $N_\mu \Phi_{\mu\nu}  = 0 \pmod{2\pi}$,  and $N_\mu^2 \Phi_{\mu} = 0 \pmod{2\pi}$. With these relations in hand, the above constraints follow immediately.

Another set of constraints are
\begin{align}
N_{i\mu} \Omega_{i\mu}  & = 0 \pmod{2\pi} \label{surf_constraint1}\\
N_{ij\mu}  \Omega_{ij\mu} & = 0 \pmod{2\pi} \label{surf_constraint2}\\
\Omega_{ij\mu} + \Omega_{ji\mu} & = 0 \pmod{2\pi} \label{surf_constraint3}\\
\Omega_{ii\mu} &= 0 \pmod{2\pi} \label{surf_constraint4}
\end{align}
We derive these constraints in Appendix \ref{sec:app_surin_con}. A final set of constraints involve $x_{il}^\mu$:
\begin{align}
N_{il\mu} x_{il}^\mu& =  0  \pmod{N_\mu} \label{x_constraint1} \\
x_{il}^\mu + x_{li}^\mu & =0 \pmod{N_\mu} \label{x_constraint2} \\
x_{ii}^\mu & =0 \pmod{N_\mu} \label{x_constraint3}
\end{align}
These constraints follow from Eqs. (\ref{surf_constraint2}-\ref{surf_constraint4}) together with the definition of $x_{il}^\mu$ (\ref{eq_x7}).

Note there are additional constraints on the surface data beyond those listed above. In particular, the requirement that the surface anyons obey ``braiding non-degeneracy''\cite{kitaev06} gives extra constraints on $\Phi_{\mu \nu}$. We will not write out these constraints explicitly since they are not necessary for our purposes.


\section{Bulk-boundary correspondence}
\label{sec:correspondence}

\subsection{The correspondence}

Having defined the bulk data $\{\Theta_{i,l}, \Theta_{ij,l}, \Theta_{ijk,l}\}$ and the surface data $\{\Phi_{\mu\nu}, \Phi_{\mu},\Omega_{i\mu}, \Omega_{ij\mu}, x_{il}^\mu\}$, we are now ready to discuss the connection between the two. This connection is encapsulated by three equations, defined modulo $2\pi$:
\begin{widetext}
\begin{align}
\Theta_{i,l} &=  \sum_\mu \frac{N_{i}}{N^{i\mu}} x_{il}^\mu \Omega_{i \mu}  + \sum_\mu N_i \left(x_{il}^\mu\right)^2 \Phi_\mu  \label{formula1}  \\
\Theta_{ij,l} &= \sum_\mu \left(\frac{N^{ij}}{N^{j\mu}} x_{il}^\mu \Omega_{j \mu}  + \frac{N^{ij}}{N^{i\mu}} x_{jl}^\mu \Omega_{i \mu} \right)
+ \frac{N^{ij}(N^{ij}-1)}{2} \sum_\mu (x_{jl}^\mu - x_{il}^\mu) \Omega_{ij \mu}   \label{formula2}\\
\Theta_{ijk,l} &= \sum_\mu \left(x_{k l}^\mu \Omega_{ij \mu} + x_{i l}^\mu \Omega_{jk \mu} + x_{j l}^\mu \Omega_{ki \mu} \right) \label{formula3}
\end{align}
\end{widetext}
These equations are the main results of this paper. We will present their derivation in section \ref{sec:derivation}, but before doing that, we make some comments about these formulas and their implications:

\begin{enumerate}

\item{Note that the left hand side of equations (\ref{formula1}-\ref{formula3}) consists of the bulk data $\{\Theta_{i,l}, \Theta_{ij,l}, \Theta_{ijk,l}\}$, while the right hand side is built entirely out of the surface data $\{\Phi_{\mu\nu}, \Phi_{\mu},\Omega_{i\mu}, \Omega_{ij\mu}, x_{il}^\mu\}$. Thus, these equations allow us to completely determine the bulk data from the surface data. They also provide some \emph{constraints} on the surface data given the bulk data. This asymmetry between bulk and surface, which is also manifest in Eq. (\ref{chernbulkbound}), is not surprising since we expect that a given bulk phase can support many different types of surfaces.}

\item{Equations (\ref{formula1}-\ref{formula3}) have an important corollary: any 3D short-range entangled bosonic model that has nonzero values for $\{\Theta_{i,l}, \Theta_{ij,l}, \Theta_{ijk,l}\}$, has a \emph{protected} surface, i.e. its surface cannot be both gapped and symmetric unless it supports anyon excitations. To derive this corollary, we note that if we could find gapped symmetric surface without anyon excitations, then the right hand sides of equations (\ref{formula1}-\ref{formula3}) would vanish for this surface since the sum over $\mu$ would run over the empty set. Clearly this vanishing is inconsistent with nonzero values of $\{\Theta_{i,l}, \Theta_{ij,l}, \Theta_{ijk,l}\}$, so we conclude that such a surface is not possible.}

\item{Another important corollary is that the right hand sides of equations (\ref{formula1}-\ref{formula3}) vanish for any strictly 2D system. Indeed, we can derive this by thinking of the 2D system as living on the boundary of a 3D vacuum where $\{\Theta_{i,l}, \Theta_{ij,l}, \Theta_{ijk,l}\}$ are all zero. We discuss this point in more detail in section \ref{sec:pure2d}.}

\item{It is natural to ask whether there could be additional constraints relating bulk and surface data beyond equations (\ref{formula1}-\ref{formula3}). The analysis in this paper is not capable of answering this question definitively. That being said, if there are additional constraints, we can always replace the bulk data appearing in these constraints with surface data, using (\ref{formula1}-\ref{formula3}). Hence, any additional constraints can be written entirely in terms of surface data.}


\item{The coefficient of $\Omega_{i \mu}$ in Eq. (\ref{formula1}) is an \emph{integer}. This is not obvious, but can be proven using one of the general constraints on $x_{il}^\mu$, namely Eq. (\ref{x_constraint1}). This integrality property is important because $\Omega_{i \mu}$ is only defined modulo $2\pi$: therefore it is only because its coefficient is an integer that Eq. (\ref{formula1}) gives a well-defined phase $\Theta_{i,l}$. In a similar fashion one can check that all the coefficients of the phase factors $\Omega_{ij \mu}$, $\Phi_\mu$ in Eqs. (\ref{formula1}-\ref{formula3}) are integers, so all of these equations are well-defined. Finally, one can check that the coefficients of $x_{il}^\mu$ and $(x_{il}^\mu)^2$ in Eqs.~(\ref{formula1}-\ref{formula3}) satisfy appropriate conditions so that all three equations are well-defined even if $x_{il}^\mu$ is defined modulo $N_\mu$.}

\item{Several of the terms in these equations can only take two values: $0$ or $\pi$. In particular, this is the case for the second term on the right hand side of (\ref{formula1}) as well as the second term on the right hand of (\ref{formula2}). This property is interesting because it means that the above equations are simpler than they appear. These results can be established using the general constraints (\ref{x_constraint3}), (\ref{phi_constraint}), (\ref{surf_constraint2}).}

\end{enumerate}

How can we make use of the above bulk-boundary correspondence? We envision two types of applications. First, if we are given a surface theory and are able to extract all the surface
data, we can use the bulk-boundary correspondence to constrain the bulk SPT phase. Conversely, if we are given a bulk SPT phase, we can use the bulk-boundary correspondence to constrain the possible surfaces. Below we give two examples to illustrate these two ways of applying the bulk-boundary correspondence.

\subsection{Example 1}

\label{sec:example1}

In this section, we demonstrate the bulk-boundary correspondence by computing the bulk data corresponding to four different types of surfaces with 
$\mathbb Z_2\times \mathbb Z_2$ symmetry. These surface theories were originally introduced and analyzed by Ref.~\onlinecite{chen14} as we explain below.

To set up our example, imagine that we have a 3D lattice spin model that realizes an SPT phase with a symmetry group 
$G = \mathbb Z_2\times \mathbb Z_2$. Imagine that we study the model in a geometry with a boundary and we find that 
the surface is gapped and symmetric and that it supports two distinct types of Abelian anyons: a semion $s$ with exchange 
statistics $\theta_s = \pi/2$ and the vacuum excitation $\mathbbm 1$ with trivial statistics. Translating this information 
into our notation, this means that the surface anyons form a group $A = \mathbb Z_2$ while their statistics can be summarized by two quantities:
\begin{equation}
\Phi_{1} = \pi/2, \quad \Phi_{11} = \pi
\label{example_stat}
\end{equation}
Here the index $\mu$ can only take one value --- namely $\mu = 1$ --- since the group $A= \mathbb Z_2$ has only one generator.

Next, suppose we couple the system to a $\mathbb Z_2 \times \mathbb Z_2$ gauge field in order to probe its symmetry properties. After 
performing this gauging procedure, we take a surface anyon with anyonic flux $s$ and we braid it twice around a vortex line that carries 
gauge flux $(\pi,0)$. We find that the Berry phase associated with this process is $\pi$. We also braid the surface anyon twice around the vortex line $(0,\pi)$ and we find a Berry phase of $\pi$. Finally, we braid the surface anyon around one vortex line carrying flux $(\pi, 0)$ and one vortex line carrying flux $(0,\pi)$, then we braid in the opposite direction around both vortex lines as described in Sec. \ref{sec:surfaceinv_def} and we find that the associated Berry phase is again $\pi$. In our notation, this information is summarized by the surface data
\begin{equation}
\Omega_{11} = \pi, \ \ \ \Omega_{21} = \pi, \ \ \ \Omega_{121} = \pi
\label{CSL_surf}
\end{equation}
Here the indices $i,j$ can take two values $i,j=1,2$ since the symmetry group $G=\mathbb Z_2 \times \mathbb Z_2$ has two generators.

To complete the surface data, we still need several more quantities, namely $\{\Omega_{111}, \Omega_{221},\Omega_{211}\}$ and 
$\{x_{12}^1, x_{21}^1, x_{11}^1, x_{22}^1\}$. The first set of quantities can be completely fixed using general properties of $\Omega_{ij\mu}$. In particular, we know that $\Omega_{111} = \Omega_{221} = 0$ by  Eq. \ref{surf_constraint4}, while $\Omega_{211} = -\Omega_{121}$ by Eq. \ref{surf_constraint3}. The remaining quantities $\{x_{12}^1, x_{21}^1, x_{11}^1, x_{22}^1\}$ are then completely determined by the definition of $x_{il}^\mu$ (\ref{eq_x7}):
\begin{equation}
x_{12}^1 = x_{21}^1 = 1, \quad x_{11}^1 = x_{22}^1 = 0
\end{equation}

With the above surface data in hand, we can now illustrate the bulk-boundary correspondence (\ref{formula1}-\ref{formula3}). Let us focus on computing the bulk data $\Theta_{1,2}$ and $\Theta_{2,1}$.
Substituting the surface data into Eq. (\ref{formula1}) we obtain
\begin{align*}
\Theta_{1,2} &= x_{12}^1 \Omega_{11} + 2 (x_{12}^1)^2 \Phi_1 \\
&= \pi + \pi \\
&= 0
\end{align*}
and
\begin{align*}
\Theta_{2,1} &= x_{21}^1 \Omega_{21} + 2 (x_{21}^1)^2 \Phi_1 \\
&= \pi + \pi \\
&= 0
\end{align*}
Similarly, we can go ahead and compute the remaining bulk data, e.g. $\Theta_{1,1},\Theta_{2,2},\Theta_{11,2}$, etc., with the result being that they all vanish as well. Alternatively, we can obtain this result using the general constraints (\ref{eq_binv1})-(\ref{eq_binv10}) which completely determine these quantities in terms of $\Theta_{1,2}$ and $\Theta_{2,1}$. In other words, $\Theta_{1,2}$ and $\Theta_{2,1}$ are the only \emph{independent} bulk quantities.

For comparison, we now consider three other possibilities for the surface data, which we will refer to as `APS-X', `APS-Y' and `APS-Z' (we explain this terminology below):
\begin{align}
\text{APS-X}: \ \ \ \Omega_{11} = \pi, \ \ \ \Omega_{21} = 0, \ \ \ \Omega_{121} = \pi \nonumber \\
\text{APS-Y}: \ \ \ \Omega_{11} = 0, \ \ \ \Omega_{21} = \pi, \ \ \ \Omega_{121} = \pi \nonumber \\
\text{APS-Z}: \ \ \ \Omega_{11} = 0, \ \ \ \Omega_{21} = 0, \ \ \ \Omega_{121} = \pi
\end{align}
Here, as before, we assume that the surface anyons form a group $A = \mathbb Z_2$ with statistics (\ref{example_stat}). Applying the bulk-boundary formulas, we can compute $\Theta_{1,2}$ and $\Theta_{2,1}$ in the same way as above. The results are shown in Table \ref{tab1}, along with those corresponding to the surface data (\ref{CSL_surf}), which we will refer to as `CSL.'

What conclusions can we draw from these calculations? First, we can see from Table \ref{tab1} that $\Theta_{1,2}$ and $\Theta_{2,1}$ take different values in each of the four cases. Therefore, we can conclude that the corresponding bulk spin models all belong to distinct SPT phases. Also we see that at least one of $\Theta_{1,2}$ and $\Theta_{2,1}$ is nonzero for each of the APS-X, APS-Y, APS-Z cases listed above, which implies that the corresponding bulk spin models belong to \emph{non-trivial} SPT phases. Finally, since the bulk data vanishes for the `CSL' case, we can conclude that the corresponding bulk spin model belongs to a \emph{trivial} SPT phase --- if we make the additional assumption that the bulk data is complete.

\begin{table}
\caption{Surface and bulk data for four surfaces with $ G = \mathbb Z_2\times \mathbb Z_2$ and $ A = \mathbb Z_2$.
}\label{tab1}
\begin{tabular}{cccccc}
\hline\hline
& \multicolumn{3}{c}{Surface Data} & \multicolumn{2}{c}{Bulk Data} \\
Model$\ $& $\ \  \Omega_{1 1} \ \ $ & $\ \ \Omega_{2 1} \ \ $ & $ \ \ \Omega_{1 2 1} \ \ $ & $\ \ \Theta_{1,2} \ \ $ & $\ \  \Theta_{2, 1} \ \
$\\
\hline
CSL $\ $  & $\pi$ & $\pi$ & $\pi$ & 0 & 0 \\
APS-X $\ $ &$\pi$ & $0$ & $\pi$ & 0 & $\pi$ \\
APS-Y $\ $ &$0$ & $\pi$ & $\pi$ & $\pi$ & 0 \\
APS-Z $\ $ &$0$ & $0$ &   $\pi$ & $\pi$ & $\pi$ \\
\hline
\end{tabular}
\end{table}

As we mentioned above, these four types of surfaces were originally discussed by Ref.~\onlinecite{chen14}. Our terminology for these surfaces follows that of Ref.~\onlinecite{chen14}: the reason we refer to the first type of data as `CSL' is that a variant of the 2D Kalmeyer-Laughlin chiral spin liquid (CSL) state is described by this data; likewise, the reason we refer to the other types of data as `APS-X', `APS-Y' and `APS-Z' is because they correspond to the ``anomalous projective semion'' (APS) states of Ref.~\onlinecite{chen14}. Here the word ``anomalous'' signifies the fact that the last three types of surface data are incompatible with a pure 2D lattice model and can only exist on the boundary of a nontrivial 3D SPT phase.

It is worth pointing out that Ref.~\onlinecite{chen14} used a different language to describe the surface data than what we use here. In this alternate description, the symmetry properties of the surface are described by an element $\omega$ of the cohomology group $H^2(G,A)$ instead of the quantities $\{\Omega_{i\mu}, \Omega_{ij\mu}, x_{il}^\mu\}$. We explain this alternative language and its relationship with $\{\Omega_{i\mu}, \Omega_{ij\mu}, x_{il}^\mu\}$ in section \ref{sec:connection_set}.

\subsection{Example 2}
\label{sec:example2}

We can also use the bulk-boundary correspondence in the opposite direction: that is, we can use it to constrain the types of \emph{surfaces} 
that are compatible with a given \emph{bulk} Hamiltonian. For an example of this, imagine that we have a lattice boson model 
that realizes an SPT phase with symmetry group $G = \mathbb Z_2 \times\mathbb Z_2$. As we mentioned in the previous section, there are only 
two independent pieces of bulk data for this symmetry group: $\Theta_{1,2}$ and $\Theta_{2,1}$. Let us suppose that one or both of these 
quantities takes a nonzero value for our lattice spin model. Using this information we can constrain the possible surfaces 
of this system. In particular, assuming that the surface anyons are all Abelian and are not permuted by the symmetries, we 
can show that the group of surface anyons $A = \prod_{\mu=1}^M \mathbb{Z}_{N_\mu}$ has the property that $N_\mu$ is \emph{even} for at least one value of $\mu$.

One way to see this is to examine the general constraints (\ref{x_constraint1}-\ref{x_constraint3}) on $x_{il}^\mu$. In particular, from the constraint (\ref{x_constraint3}), we can see
that $x_{11}^\mu = x_{22}^\mu = 0$. Also, if $N_\mu$ is odd, then the constraint (\ref{x_constraint1}) implies that $x_{12}^\mu = x_{21}^\mu = 0$. Hence if $N_\mu$ were odd for all $\mu$ then $x_{il}^\mu$ would necessarily vanish completely. But then $\Theta_{1,2}$ and $\Theta_{2,1}$ would also have to vanish according to the bulk-boundary formula (\ref{formula1}).
We conclude that if either $\Theta_{1,2}$ or $\Theta_{2,1}$ is nonzero then at least one of the $N_\mu$'s must be even.

\begin{figure*}
\centering
\includegraphics{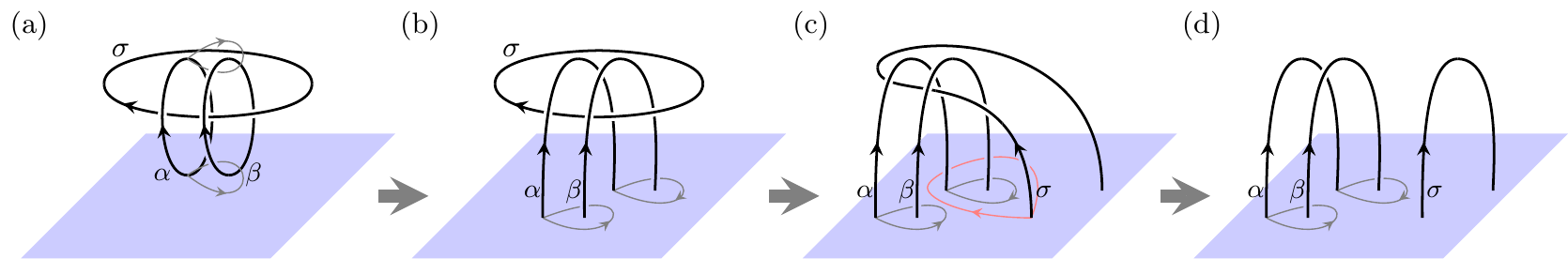}
\caption{Deforming a three-loop braiding process [panel (a)] in the bulk into a braiding process [panel (d)] of 
vortex arches near the surface. The deformation splits into three steps, (a)$\rightarrow$(b), (b)$\rightarrow$(c), and (c)$\rightarrow$(d).
The gray curves in (a), (b), (c) and (d) are trajectories of some points on $\alpha$; the red curve in (c) is the trajectory of one of the endpoints of $\sigma$. }\label{fig_deformation}
\end{figure*}

\section{Derivation of the bulk-boundary correspondence}
\label{sec:derivation}

In this section, we derive the bulk-boundary formula (\ref{formula2}) for $\Theta_{ij,l}$. The derivations of the other two formulas (\ref{formula1},\ref{formula3}) are similar and are given in Appendix \ref{sec:derive_formula3} and \ref{sec:derive_formula1}.

\subsection{Step 1: deforming the braiding process}
\label{sec:derive_formula2_deformation}

Our derivation proceeds in four steps. In the first step, we derive an equivalence between the three-loop braiding process associated with $\Theta_{ij,l}$ and another process which involves braiding vortex ``arches'' on the surface. The key to deriving this equivalence is the general principle that statistical Berry phases are invariant under ``smooth'' deformations of braiding processes: that is, if two braiding processes can be ``smoothly'' deformed into one another then the associated statistical Berry phases must be equal. Here, a ``smooth'' deformation is a sequence of local changes to the excitations involved in the braiding process. The local changes can be arbitrary except that the moving excitation must stay far apart from the other excitations at every step of the deformation. (Here, when we say ``local changes'' to the excitations we mean any changes that can be implemented by unitary operators supported in the neighborhood of the excitations).

To begin, let us imagine performing the three-loop braiding process in the bulk [Fig.~\ref{fig_deformation}(a)]. That is, we braid a loop $\alpha$ around another loop $\beta$ for $N^{ij}$ times while both are linked to a third loop $\sigma$. Here $\alpha, \beta, \sigma$ carry unit type-$i$, type-$j$ and type-$l$ flux, i.e. $\phi_{\alpha} = \frac{2\pi}{N_i}\mathbf e_i$, $\phi_\beta = \frac{2\pi}{N_j}\mathbf e_j$ and $\phi_\sigma = \frac{2\pi}{N_l}\mathbf e_l$. The Berry phase associated with this process is $\Theta_{ij,l}$.

Next, we stretch $\alpha$ and $\beta$ and absorb the bottoms of these loops into the surface.\footnote{The fundamental reason that this 
absorption/annihilation is possible is that the surface is \emph{symmetric} by assumption. This symmetry guarantees that the bottoms of the vortex loops can be annihilated at the surface using local unitary operators.} This step changes $\alpha,\beta$ into vortex \emph{arches} that terminate on the surface [Fig.~\ref{fig_deformation}(b)]. After this step, the deformed braiding process involves braiding vortex arches $\alpha, \beta$ while they are both linked to $\sigma$. By the general principle described above, this deformed process must yield the same Berry phase, $\Theta_{ij,l}$.

To proceed further, we now stretch $\sigma$ and absorb its bottom into the surface. This step changes $\sigma$ into another vortex arch [Fig.~\ref{fig_deformation}(c)]. Finally, we disentangle the three arches by unwinding one of the two ends of the arch $\sigma$ [Fig.~\ref{fig_deformation}(d)]. The braiding process now involves braiding two unlinked arches
$\alpha, \beta$ around one another. Again the Berry phase must be the same as in the original process. Putting this all together, we conclude that $\Theta_{ij,l}$ is equal to the Berry phase associated with braiding the arch $\alpha$ around the arch $\beta$ for $N^{ij}$ times, as shown in Fig.~\ref{fig_deformation}(d).

\subsection{Step 2: splitting the excitations}
\label{sec:derive_formula2_splitting}

Our task is now to analyze the vortex arch braiding process in Fig.~\ref{fig_deformation}(d). Before doing this, it is useful to first consider a thought experiment in which we shrink down the arches $\alpha, \beta$ so that all that is left are two localized surface excitations. From general considerations we know that the resulting surface excitations can be written as a linear superposition of states, each of which has a definite anyon type. We ask: what types of surface anyons appear in this linear superposition?

The answer is simple: if we shrink $\alpha$, we get a superposition of different surface anyons all of which have anyonic flux $x_{il}$; likewise, if we shrink $\beta$, we get a superposition of surface anyons all of which have anyonic flux $x_{jl}$. To see this, notice that shrinking down $\alpha, \beta, \sigma$ in Fig.~\ref{fig_deformation}(d) is very similar to shrinking $\alpha, \sigma$ in Fig.~\ref{fig_xdef}. Furthermore, in the latter case, we know that shrinking $\alpha$ gives a superposition of different surface anyons all of which have anyonic flux $x_{il}$ (see Eq. \ref{xileqs}). Therefore the same must be true for Fig.\ref{fig_deformation}(d).

The most important point from this discussion is that if we shrink $\alpha$ or $\beta$ to the surface, we will generally get \emph{non-trivial} 
surface anyons. This property is inconvenient for our subsequent analysis and motivates us to \emph{split} $\alpha, \beta$ into 
more ``elementary'' excitations. 

\begin{figure}
\centering
\includegraphics{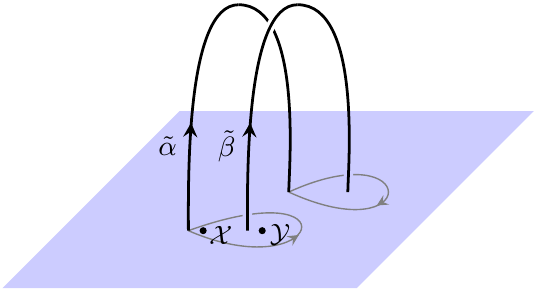}
\caption{Splitting $\alpha$ into $\tilde\alpha$ and $\mathcal X$, and splitting $\beta$ into $\tilde\beta$ and $\mathcal Y$.} \label{fig_splitting}
\end{figure}

For this reason, the next step in our analysis is to split the arch $\alpha$ into a surface anyon $\mathcal X$ and a new arch $\tilde\alpha$. Similarly we split $\beta$ into a surface anyon $\mathcal Y$ and a new arch $\tilde \beta$ (Fig.~\ref{fig_splitting}).
We choose the surface anyons $\mathcal X, \mathcal Y$ to be any anyons with anyonic flux
\begin{equation}
\xi_\mathcal X = x_{il}, \quad \xi_\mathcal Y = x_{jl}\label{eq_dev1}
\end{equation}
while we choose the arches $\tilde \alpha, \tilde \beta$ to be any arches with the property that $\alpha$ can be written as a fusion product of $\mathcal X$ and $\tilde \alpha$, and that $\beta$ can be written as a fusion product of $\mathcal Y$ and $\tilde \beta$.

The motivation for performing this splitting is that the new arches $\tilde \alpha, \tilde \beta$ have a nice property, by construction: if we shrink $\tilde \alpha, \tilde \beta$ to the surface, we get a superposition of \emph{charge} excitations instead of more complicated surface anyons. This property will play an important role below when we compute the Berry phase associated with braiding $\tilde \alpha$ around $\tilde \beta$ (see Eq. \ref{AB-formula} below).

\subsection{Step 3: decomposing the braiding process}
\label{sec:derive_formula2_decomposing}

After splitting $\alpha,\beta$ as described above, our braiding process reduces to one in which we braid $\tilde \alpha$ and $\mathcal X$ around $\tilde \beta$ and $\mathcal Y$ for $N^{ij}$ times (Fig.~\ref{fig_splitting}). The next step is to decompose this process into a series of simpler processes.

First we introduce some notation. Let $|\psi\>$ be the initial state at the beginning of the braiding, i.e. the state obtained from the deformation process shown in Fig. \ref{fig_deformation} followed by the splitting process in Fig. \ref{fig_splitting}. Let $\mathbb V$ be the subspace consisting of all states that are degenerate in energy with $|\psi\>$. Here, $\mathbb V$ could have dimension $1$ or higher. Let $W: \mathbb V \rightarrow \mathbb V$ be the unitary braiding matrix that describes the (possibly) non-Abelian Berry phase associated with braiding $\tilde \alpha$ and $\mathcal X$ around $\tilde \beta$ and $\mathcal Y$ \emph{once}.

\begin{figure}
\centering
\includegraphics{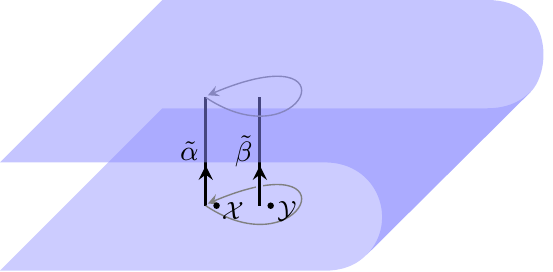}
\caption{Folding the surface and straightening the vortex lines in Fig.~\ref{fig_splitting}. In between the top and bottom parts 
of the surface is the gauged SPT model, and outside is the vacuum.} \label{fig_straighten}
\end{figure}

Then, with this notation, the fact that the Berry phase for our braiding process is $\Theta_{ij,l}$ translates to the equation
\begin{equation}
W^{N^{ij}} |\psi\> = e^{i \Theta_{ij,l}} |\psi\>
\label{Weq}
\end{equation}
To proceed further, it is helpful to represent this braiding process using a 2D picture. One way to do this is to imagine folding the surface and straightening the vortex arches,
as shown in Fig.~\ref{fig_straighten}. After doing this, the braiding process now involves vertical vortex lines. If we now take a top view of Fig.~\ref{fig_straighten}, then $W$ can be visualized as
\begin{equation}
\begin{tikzpicture}[baseline={([yshift=-.5ex]current  bounding  box.center)}]
\clip (-1.9,-0.8)rectangle+(5.5,1.55);
\draw [-stealth](0,0)..controls (0.5, 0.15) .. (0.7,0.15)..controls (1.2,0.15)  and (3,-1.4).. (3,0)..controls (3,1.2) and (0.9,0.5) .. (0.05,0.1);
\draw [white, line width=2.2 pt](0.7,0)..controls (1.2,-0.5)  and (3.3,-1.3).. (3.3,0)..controls (3.3,1.6) and (1.2,-0.1).. (0.8,0.0);
\draw [-stealth](0.7,0)..controls (1.2,-0.5)  and (3.3,-1.3).. (3.3,0)..controls (3.3,1.6) and (1.2,-0.1).. (0.8,0.0);
\fill (0,0) circle (0.05);
\fill (0.7,0) circle (0.05);
\fill (2,0) circle (0.05);
\fill (2.7,0) circle (0.05);
\node at (-1,0){$W =$};
\node at (0,-0.28){$\tilde\alpha$};
\node at (2,-0.28){$\tilde\beta$};
\node at (0.7,-0.28){$\mathcal X$};
\node at (2.7,-0.28){$\mathcal Y$};
\end{tikzpicture}
\label{eq_diag2}
\end{equation}

With this picture in mind, it is easy to see how to decompose the above braiding process into simpler processes. In particular, let $W_{\tilde\alpha \tilde\beta}$, $W_{\tilde \alpha \mathcal Y}$, $W_{\mathcal X\tilde \beta}$, and $W_{\mathcal X\mathcal Y}$ be the braid matrices corresponding to the following processes:
\begin{align}
\begin{tikzpicture}[baseline={([yshift=-.5ex]current  bounding  box.center)}]
\clip (-1.9,-0.55)rectangle+(5.5,1.1);
\fill (0.7,0) circle (0.05);
\fill (2,0) circle (0.05);
\fill (2.7,0) circle (0.05);
\draw [-stealth](0,0)..controls (0.3,0) and (0.9,0.2).. (1.3,0.2)..controls(1.6,0.2) and (1.8,-0.5).. (2,-0.5)..controls(2.6,-0.5) and (2.6,0.5)..(2,0.5)..controls(1.3,0.5) and (0.7,0.1).. (0.05,0.07);
\fill (0,0) circle (0.05);
\node at (-1,0.3)[anchor=north]{$W_{\tilde\alpha\tilde\beta} =$};
\node at (0,-0.28){$\tilde\alpha$};
\node at (2,-0.28){$\tilde\beta$};
\node at (0.7,-0.28){$\mathcal X$};
\node at (2.7,-0.28){$\mathcal Y$};
\end{tikzpicture}\label{w1}\\
\begin{tikzpicture}[baseline={([yshift=-.5ex]current  bounding  box.center)}]
\clip (-1.9,-0.55)rectangle+(5.5,1.1);
\fill (0.7,0) circle (0.05);
\fill (2,0) circle (0.05);
\fill (2.7,0) circle (0.05);
\draw [-stealth](0,0)..controls (0.3,0.2) and (1.6,0.2).. (2,0.2)..controls(2.3,0.2) and (2.5,-0.5).. (2.7,-0.5)..controls(3.3,-0.5) and (3.3,0.5)..(2.7,0.5)..controls(2,0.5) and (0.7,0.3).. (0.05,0.1);
\fill (0,0) circle (0.05);
\node at (-1,0.3)[anchor=north]{$W_{\tilde\alpha\mathcal Y}=$};
\node at (0,-0.28){$\tilde\alpha$};
\node at (2,-0.28){$\tilde\beta$};
\node at (0.7,-0.28){$\mathcal X$};
\node at (2.7,-0.28){$\mathcal Y$};
\end{tikzpicture}\label{w2}\\
\begin{tikzpicture}[baseline={([yshift=-.5ex]current  bounding  box.center)}]
\clip (-1.9,-0.55)rectangle+(5.5,1.1);
\fill (0.7,0) circle (0.05);
\fill (2,0) circle (0.05);
\fill (2.7,0) circle (0.05);
\draw [-stealth](0.7,0)..controls (1.2,0.2) and (1.5, -0.5)..(2,-0.5)..controls(2.2,-0.5) and (2.3,-0.3).. (2.3,0)..controls (2.3,0.6) and (1.2,0.2).. (0.75,0.08);
\fill (0,0) circle (0.05);
\node at (-1,0.3)[anchor=north]{$W_{\mathcal X \tilde \beta} =$};
\node at (0,-0.28){$\tilde\alpha$};
\node at (2,-0.28){$\tilde\beta$};
\node at (0.7,-0.28){$\mathcal X$};
\node at (2.7,-0.28){$\mathcal Y$};
\end{tikzpicture}\label{w3}\\
\begin{tikzpicture}[baseline={([yshift=-.5ex]current  bounding  box.center)}]
\clip (-1.9,-0.55)rectangle+(5.5,1.1);
\fill (0.7,0) circle (0.05);
\fill (2,0) circle (0.05);
\fill (2.7,0) circle (0.05);
\draw [-stealth](0.7,0)..controls (1,0) and (1.6,0.2).. (2,0.2)..controls(2.3,0.2) and (2.5,-0.5).. (2.7,-0.5)..controls(3.3,-0.5) and (3.3,0.5)..(2.7,0.5)..controls(2,0.5) and (1.4,0.2).. (0.75,0.07);
\fill (0,0) circle (0.05);
\node at (-1,0.3)[anchor=north]{$W_{\mathcal X\mathcal Y} =$};
\node at (0,-0.28){$\tilde\alpha$};
\node at (2,-0.28){$\tilde\beta$};
\node at (0.7,-0.28){$\mathcal X$};
\node at (2.7,-0.28){$\mathcal Y$};
\end{tikzpicture}\label{w4}
\end{align}
Then, it is easy to see that if we perform the above four processes sequentially, the result can be smoothly deformed into the process corresponding to $W$. Translating this into algebra, we derive:
\begin{equation}
W  = W_{\mathcal X\mathcal Y}W_{\mathcal X\tilde \beta}W_{\tilde \alpha \mathcal Y}W_{\tilde\alpha \tilde\beta}
\end{equation}
Substituting this into (\ref{Weq}), we obtain:
\begin{equation}
(W_{\mathcal X\mathcal Y}W_{\mathcal X\tilde \beta}W_{\tilde \alpha \mathcal Y}W_{\tilde\alpha \tilde\beta})^{N_{ij}} |\psi\> = e^{i \Theta_{ij,l}} |\psi\> \label{eq_dev5}
\end{equation}
Equation (\ref{eq_dev5}) is the main result of this step.

\subsection{Step 4: evaluating the expression (\ref{eq_dev5})}
We now evaluate the expression on the left-hand side of Eq. (\ref{eq_dev5}). We accomplish this using several algebraic properties of the braid matrices $W_{\mathcal X\mathcal Y}, W_{\mathcal X\tilde \beta}, W_{\tilde \alpha \mathcal Y} $, and $ W_{\tilde\alpha \tilde\beta}$ that we will derive below. The first property is that these braid matrices all commute with each other, except for the pairs $(W_{\tilde \alpha\tilde \beta}, W_{ \mathcal X\tilde\beta})$ and $(W_{\tilde \alpha\tilde \beta}, W_{\tilde\alpha \mathcal Y})$ which obey the commutation relations
\begin{align}
W_{\tilde \alpha\tilde \beta} W_{ \mathcal X\tilde\beta} & =e^{i \zeta_1} \cdot W_{ \mathcal X\tilde\beta}W_{\tilde \alpha\tilde \beta}  \label{eq_dev8}\\
W_{\tilde \alpha\tilde \beta} W_{\tilde\alpha \mathcal Y}& =e^{i \zeta_2} \cdot W_{\tilde\alpha\mathcal Y}W_{\tilde \alpha\tilde \beta}  \label{eq_dev9}
\end{align}
where $\zeta_1,\zeta_2$ are given by
\begin{align*}
\zeta_1 = \sum_\mu x_{il}^\mu \Omega_{ij\mu}, \quad
\zeta_2 = \sum_\mu x_{jl}^\mu \Omega_{ji\mu}
\end{align*}
Another important property of these matrices is that when we raise them to the power $N^{ij}$, they simplify considerably:
\begin{align}
W_{ \mathcal X\tilde\beta}^{N^{ij}} &  = e^{i \zeta_3} \hat{\rm I} \label{eq_dev11}\\
W_{\tilde\alpha \mathcal Y}^{N^{ij}} & = e^{i \zeta_4} \hat{\rm I} \label{eq_dev12} \\
W_{\mathcal X\mathcal Y}^{N^{ij}} & =\hat{\rm I} \label{eq_dev13}\\
W_{\tilde \alpha\tilde \beta}^{N^{ij} }|\psi\> & = |\psi\> \label{eq_dev14}
\end{align}
where $\hat {\rm I}$ is the identity matrix in $\mathbb V$, and $\zeta_3,\zeta_4$ are given by
\begin{align*}
\zeta_3 = \sum_\mu \frac{N^{ij}}{N^{j\mu}}x_{il}^\mu \Omega_{j\mu}, \quad
\zeta_4 = \sum_\mu \frac{N^{ij}}{N^{i\mu}}x_{jl}^\mu \Omega_{i\mu}
\end{align*}
Notice that the last relation (\ref{eq_dev14}) is different from the others because it only tells us about the action of a braid matrix on the 
specific state $|\psi\>$ defined above. In contrast,
the other relations are matrix equations that hold throughout the degenerate subspace $\mathbb V$.

We now use these relations to evaluate the left-hand side of Eq. (\ref{eq_dev5}) and thereby derive the bulk-boundary formula (\ref{formula2}). First, from the commutation relations (\ref{eq_dev8})-(\ref{eq_dev9}), we reorder operators at the cost of a phase factor:
\begin{align}
(W_{\mathcal X\mathcal Y} W_{\mathcal X\tilde \beta} W_{\tilde \alpha \mathcal Y} W_{\tilde\alpha \tilde\beta})^{N^{ij}} |\psi\> &= W_{\mathcal X\mathcal Y}^{N^{ij}} W_{\mathcal X\tilde \beta}^{N^{ij}} W_{\tilde \alpha \mathcal Y}^{N^{ij}} W_{\tilde\alpha \tilde\beta}^{N^{ij}}  |\psi\>  \nonumber \\
&\times e^{\frac{i N^{ij}(N^{ij}-1)}{2}\left(\zeta_1+ \zeta_2\right)}
\end{align}
Here the phase factor comes from the fact that the commutation relations (\ref{eq_dev8}) and (\ref{eq_dev9}) are used $\frac{N^{ij}}{2}(N^{ij}-1)$ times during reordering.

Next, we apply the identities (\ref{eq_dev11})-(\ref{eq_dev14}) to obtain
\begin{align}
W_{\mathcal X\mathcal Y}^{N^{ij}}W_{\mathcal X\tilde \beta}^{N^{ij}}W_{\tilde \alpha \mathcal Y}^{N^{ij}}W_{\tilde\alpha \tilde\beta}^{N^{ij}} |\psi\> =
e^{i (\zeta_3 + \zeta_4)} |\psi\>
\end{align}
Putting this together, we derive
\begin{align*}
(W_{\mathcal X\mathcal Y} W_{\mathcal X\tilde \beta} W_{\tilde \alpha \mathcal Y} W_{\tilde\alpha \tilde\beta})^{N^{ij}} |\psi\> &=
e^{i(\zeta_3 + \zeta_4)}|\psi\> \\
& \times e^{i\frac{N^{ij}(N^{ij}-1)}{2}\left(\zeta_1 + \zeta_2\right)}
\end{align*}
Comparing with Eq. (\ref{eq_dev5}), we conclude that
\begin{align}
\Theta_{ij,l} = \zeta_3 + \zeta_4 +\frac{N^{ij}(N^{ij}-1)}{2}\left(\zeta_1 + \zeta_2\right) \label{eq_dev19}
\end{align}
This is nothing other than the bulk-boundary formula (\ref{formula2}), as one can see from the definitions of $\zeta_1,\zeta_2,\zeta_3,\zeta_4$.

To complete the argument, we now derive the algebraic relations (\ref{eq_dev8})-(\ref{eq_dev14}). We begin by proving the statement that proceeds Eq. (\ref{eq_dev8}), namely that the braid matrices all commute with each other except for the pairs $(W_{\tilde \alpha\tilde \beta}, W_{ \mathcal X\tilde\beta})$ and $(W_{\tilde \alpha\tilde \beta}, W_{\tilde\alpha \mathcal Y})$. To establish this statement, it suffices to show that (1) $W_{\mathcal X\mathcal Y}$ commutes with the other three operators, and that (2) $W_{\tilde\alpha\mathcal Y}$ and $W_{\mathcal X\tilde\beta}$ commute with one another. The first result follows from the fact that the mutual statistics between surface anyons is \emph{Abelian}, so that $W_{\mathcal X\mathcal Y}$ is proportional to the identity $\hat{\rm I}$. The second result follows from the observation that the braiding paths associated with $W_{\tilde\alpha\mathcal Y}$ and $W_{\mathcal X\tilde\beta}$ do not overlap.

Next, we need to establish the commutation relations (\ref{eq_dev8}), (\ref{eq_dev9}). Proving these relations is more technical and hence we 
postpone their derivation to Appendix \ref{sec:app_formula}. Here, we only give the intuitive picture behind these relations. Consider for example (\ref{eq_dev8}). This relation can be equivalently written as
\begin{equation}
e^{i\zeta_1} \hat{\mathrm I} = W_{\tilde \alpha\tilde \beta}^{-1}W_{ \mathcal X\tilde\beta}^{-1}W_{\tilde \alpha\tilde \beta} W_{ \mathcal X\tilde\beta} \label{eq_dev20}
\end{equation}
To understand the meaning of the right hand side, remember that braiding processes are symmetrical in the sense that braiding $X$ around $Y$ is topologically equivalent to braiding $Y$ around $X$ (for properly chosen paths). Therefore, the product on the right hand side can be interpreted as a process in which $\tilde\beta$ is first braided around $\mathcal X$, then around $\tilde\alpha$, then around $\mathcal X$ in the opposite direction, and finally around $\tilde\alpha$ in the opposite direction. This is very similar to the braiding process that defines $\Omega_{ij\mu}$, except for two differences: (1) the moving excitation is the vortex $\tilde\beta$ rather than the anyon $\mathcal X$, and (2) the anyon $\mathcal X$ carries anyonic flux $\xi_\mathcal X = x_{il}$ instead of having unit type-$\mu$ flux. It turns out that the first difference is irrelevant and can be safely ignored. On the other hand, the second difference \emph{is} important and changes the product on the  left-hand side from $e^{i \Omega_{ij \mu}}\hat{\mathrm I}$ to $e^{i\sum_\mu x_{il}^\mu \Omega_{ij\mu}} \hat{\mathrm I}$ (see Appendix \ref{sec:app_formula}).

We now move on to prove the identities (\ref{eq_dev11})-(\ref{eq_dev14}). We begin with Eq.~(\ref{eq_dev13}) since it is the simplest to derive. To prove this identity, recall that the surface anyons $\mathcal X, \mathcal Y$ have Abelian statistics so the braid matrix $W_{\mathcal X\mathcal Y}$ takes the form
\begin{equation}
W_{\mathcal X\mathcal Y} = e^{i\theta_{\mathcal X \mathcal Y}} \hat{\rm I}
\end{equation}
Next, from Eqs. (\ref{eq_AB_stat}) and (\ref{mutual}) we see that
\begin{equation}
\theta_{\mathcal X \mathcal Y} = \theta_{\xi_\mathcal X \xi_\mathcal Y} = \sum_{\mu\nu} x_{il}^\mu x_{jl}^\nu \Phi_{\mu\nu}
\end{equation}
where we have used $\xi_\mathcal X = x_{il}$ and $ \xi_\mathcal Y = x_{jl}$. Combining these two results, we obtain the identity
\begin{equation}
W_{\mathcal X\mathcal Y} = e^{i \sum_{\mu\nu} x_{il}^\mu x_{jl}^\nu \Phi_{\mu\nu}} \hat{\rm I}
\end{equation}
If we now raise both sides to the $N^{ij}$ power, we can see that the right hand side reduces to the identity operator $\hat{\rm I}$ since $N^{ij} x_{il}^\mu$ is a multiple of $N_\mu$ according to Eq. (\ref{x_constraint1}), and $N_{\mu} \Phi_{\mu\nu}$ is a multiple of $2\pi$ according to Eq.~(\ref{phi_constraint1}). We conclude that $W_{\mathcal X\mathcal Y}^{N^{ij}} =  \hat{\rm I}$ as claimed.

Next we consider equation (\ref{eq_dev14}). To derive this result, we use a property of $|\psi\>$ that we discussed in section \ref{sec:derive_formula2_splitting}: if we shrink $\tilde\alpha$ or $\tilde\beta$ down to the surface, the result is a superposition of charge excitations. To see why this property is useful, let us consider the special case where shrinking $\tilde \alpha, \tilde \beta$ gives \emph{definite} charge excitations $q_{\tilde \alpha}, q_{\tilde \beta}$ instead of a superposition of different charges. In this case, it is easy to see that the statistical phase $\theta$ associated with braiding $\tilde \alpha$ around $\tilde \beta$ is given by the Aharonov-Bohm law: $\theta = q_{\tilde\alpha}\cdot\phi_{\tilde\beta} + q_{\tilde\beta}\cdot\phi_{\tilde\alpha}$. Here, the first term comes from braiding the charge $q_{\tilde \alpha}$ around the flux $\phi_{\tilde \beta}$, while the second term
comes from braiding the flux $\phi_{\tilde \alpha}$ around the charge $q_{\tilde \beta}$. (See Ref. \onlinecite{threeloop} for a derivation of this formula in a closely related context). Translating this equation into our algebraic notation gives
\begin{equation}
W_{\tilde \alpha \tilde\beta} |\psi\> = e^{i(q_{\tilde\alpha}\cdot\phi_{\tilde\beta} + q_{\tilde\beta}\cdot\phi_{\tilde\alpha})} |\psi\>
\label{AB-formula}
\end{equation}
Now recall that $\tilde \alpha$ is a unit type-$i$ flux while $\tilde \beta$ is a unit type-$j$ flux, so that $\phi_{\tilde \alpha} = \frac{2\pi}{N_i} \mathbf e_i$ and $\phi_{\tilde \beta} = \frac{2\pi}{N_j} \mathbf e_j$. Substituting this into the above equation, and raising both sides to the $N^{ij}$ power we see that $W_{\tilde \alpha \tilde\beta}^{N^{ij}} |\psi\> = |\psi\>$. This establishes equation (\ref{eq_dev14}) for the special case where shrinking $\tilde \alpha, \tilde \beta$ gives definite charge excitations. In fact, since the equation $W_{\tilde \alpha \tilde\beta}^{N^{ij}} |\psi\> = |\psi\>$ holds independent of the values of $q_{\tilde \alpha}, q_{\tilde \beta}$, the superposition principle implies that it must also hold in the case where shrinking $\tilde \alpha, \tilde \beta$ gives a superposition of charge excitations. This proves equation (\ref{eq_dev14}) in the general case.

Finally, we need to discuss Eqs. (\ref{eq_dev11}),(\ref{eq_dev12}). The proof of these relations is technical so we postpone it to Appendix \ref{sec:app_formula}. However, the physical picture for these relations is simple. For example, consider Eq. (\ref{eq_dev11}). We can see that $\zeta_3$ is the Berry phase associated with braiding $\mathcal X$ around $\tilde\beta$ for $N^{ij}$ times. If we compare this braiding process to the one that defines $\Omega_{j\mu}$, we see that they are very similar, except for two differences: (1) $\mathcal X$ carries anyonic flux $\xi_\mathcal X = x_{il}$ instead of carrying unit type-$\mu$ flux, and (2) $\mathcal X$ is braided $N^{ij}$ times instead of $N^{j\mu}$ times. Given these two differences, it is perhaps not surprising that the Berry phase for this process is $\sum_\mu \frac{N^{ij}}{N^{j\mu}}x_{il}^\mu \Omega_{j\mu}$ instead of $\Omega_{j \mu}$.

\begin{figure*}
\centering
\includegraphics{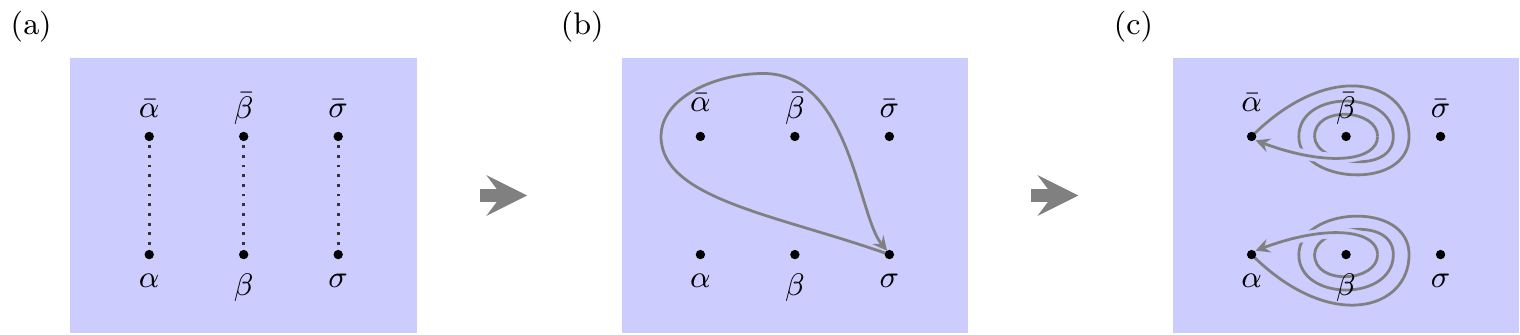}
\caption{A 2D braiding process that shows that $\Theta_{ij,l}$ vanishes for strictly 2D systems.} \label{fig_pure2d}
\end{figure*}

\section{Implications for purely 2D systems}
\label{sec:pure2d}

So far we have used the data $\{\Phi_\mu, \Phi_{\mu\nu}, \Omega_{i\mu}, \Omega_{ij\mu},x_{il}^\mu\}$ to describe surfaces of 3D systems. However, 
the same data can also be used to describe \emph{purely 2D} systems. More specifically, suppose we are given a 2D gapped lattice
boson model with Abelian symmetry group $G = \prod_{i=1}^K \mathbb{Z}_{N_i}$ and Abelian anyon excitations described by a group $A = \prod_{\mu=1}^M \mathbb{Z}_{N_\mu}$. Suppose in addition that the symmetry action does not permute the anyons. Then we can define the quantities $\{\Phi_\mu, \Phi_{\mu\nu}, \Omega_{i\mu}, \Omega_{ij\mu},x_{il}^\mu\}$ for this 2D system in the same way as we did for surfaces --- with $\Phi_\mu, \Phi_{\mu \nu}$ describing the exchange and mutual statistics of anyons, $\Omega_{i \mu}, \Omega_{ij \mu}$ describing the braiding statistics between anyons and vortices, and $x_{il}^\mu$ given by Eq. \ref{eq_x7}.

Now that we have defined $\{\Phi_\mu, \Phi_{\mu\nu}, \Omega_{i\mu}, \Omega_{ij\mu},x_{il}^\mu\}$ for 2D systems, we can ask what happens if we insert this data into the right hand side of Eqs. (\ref{formula1})-(\ref{formula3}) and compute the corresponding ``3D bulk'' quantities $\Theta_{i,l}, \Theta_{ij,l}, \Theta_{ijk,l}$. In this section we argue that if we do this, then these bulk quantities will always vanish. That is, for any 2D system we have
\begin{eqnarray}
\Theta_{i,l} = \Theta_{ij,l} = \Theta_{ijk,l} = 0
\label{2Dconstraints}
\end{eqnarray}
where $\Theta_{i,l}, \Theta_{ij,l}, \Theta_{ijk,l}$ are defined by Eqs. (\ref{formula1})-(\ref{formula3}).
We can think of the above equations as \emph{constraints} on which data $\{\Phi_\mu, \Phi_{\mu\nu}, \Omega_{i\mu}, \Omega_{ij\mu},x_{il}^\mu\}$ can be realized by 2D systems.

The simplest way to derive the constraints (\ref{2Dconstraints}) is to think of our 2D system as living on the surface of the 3D vacuum. 
It is clear that $\Theta_{i,l},\Theta_{ij,l}, \Theta_{ijk,l}$ vanish for the 3D vacuum, so by using the bulk-boundary formulas (\ref{formula1})-(\ref{formula3}) we obtain three constraints on $\{\Phi_\mu, \Phi_{\mu\nu}, \Omega_{i\mu}, \Omega_{ij\mu},x_{il}^\mu\}$ which are precisely Eqs. (\ref{2Dconstraints}).

While the above argument is perfectly solid, it is instructive to rederive the constraints using purely
two-dimensional arguments. In what follows, we will present such an argument for one of the three
constraints, namely $\Theta_{ij,l} = 0$. (Similar arguments can be used to establish the other constraints).
At the heart of our derivation is a particular braiding process that we will describe below. Our strategy
will be to compute the statistical phase associated with this process in two ways: in one approach we will see that the statistical phase is given by the right-hand side of (\ref{formula2}), while in the another approach we will see that the phase vanishes. Combining the two calculations, we then derive $\Theta_{ij,l}=0$.

Before describing the braiding process, we first need to describe the initial state $|\Psi\>$ at the beginning of the process. Imagine that we start in the ground state. We then create three vortex-antivortex pairs, $\{\alpha, \bar\alpha\}$, $\{\beta, \bar\beta\}$, and $\{\sigma, \bar\sigma\}$ where $\alpha, \beta, \sigma$ carry unit type-$i$, type-$j$ and type-$l$ gauge flux [Fig.~\ref{fig_pure2d}(a)]. We then braid $\sigma$ around $\bar\alpha$ and $\bar\beta$ [Fig.~\ref{fig_pure2d}(b)]. The state obtained in this way is the initial state $|\Psi\>$ for our braiding. The braiding process itself is rather simple: starting in the state $|\Psi\>$, we braid $\alpha$ around $\beta$ for $N^{ij}$ times in the counterclockwise direction, while simultaneously braiding $\bar\alpha$ around $\bar\beta$ for $N^{ij}$ times in the clockwise direction [Fig.~\ref{fig_pure2d}(c)].

Let us try to compute the statistical phase associated with this process. To this end, notice that there is a close analogy between the 2D braiding process shown in Fig.~\ref{fig_pure2d}(c) and the 3D arch braiding process shown in Fig.~\ref{fig_deformation}(d): the 2D braiding process looks like a horizontal cross-section of the 3D process. There is also a close connection between the initial state $|\Psi\>$ for the 2D process and the initial state $|\psi\>$ for the 3D process since the two states are obtained from similar manipulations
of vortices (compare Fig.~\ref{fig_pure2d}(a-b) with Fig.~\ref{fig_deformation}(a-c)). Because of these similarities, we can compute the statistical phase for the 2D process using essentially the same calculation as in the 3D case discussed in Sec.~\ref{sec:derivation}. In the first step, we split the vortex $\alpha$ into another vortex $\tilde \alpha$ together with an anyon $\mathcal X$ carrying anyonic flux $\xi_{\mathcal X} = x_{il}$. Also, we split $\beta$ into $\tilde \beta$ and $\mathcal Y$ where $\xi_{\mathcal Y} = x_{jl}$.
Then, after performing the splitting, we decompose the braiding process shown in Fig.~\ref{fig_pure2d}(c) into simpler
processes involving $\mathcal X, \mathcal Y, \tilde \alpha, \tilde \beta$, etc. Using the same arguments as in Sec.~\ref{sec:derivation},
we can express the statistical phases for these simpler processes in terms of
$\{\Phi_\mu, \Phi_{\mu\nu}, \Omega_{i\mu}, \Omega_{ij\mu},x_{il}^\mu\}$ and then put everything together to obtain the
statistical phase for the whole process. Since the computation is almost identical to the 3D case, the result is also the same: that is, one
finds that the statistical phase is given by the right-hand side of (\ref{formula2}).

\begin{figure}[b]
\centering
\includegraphics{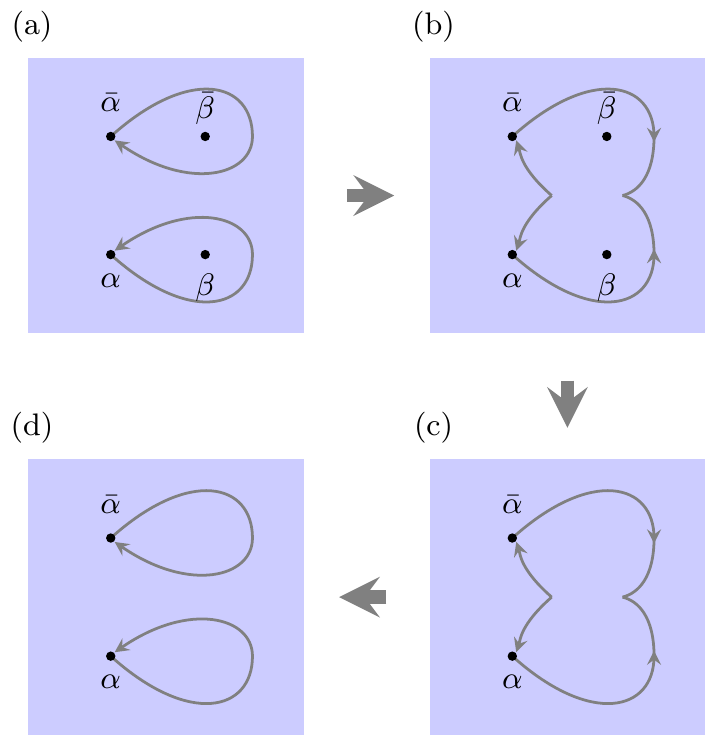}
\caption{Deforming a process in which $\alpha$ is braided around $\beta$ and simultaneously $\bar\alpha$ is braided around 
$\bar\beta$ in the opposite direction. In the initial state $|\Psi'\>$ before the braiding starts, the two pairs $\alpha, \bar\alpha$ and $\beta, \bar\beta$ are both in the vacuum fusion channel.} \label{fig_deform2d}
\end{figure}

Now we compute the statistical phase using a different approach and show that it vanishes. This alternate approach
is based on the observation that the two braiding processes shown in Fig.~\ref{fig_pure2d}(b) and Fig.~\ref{fig_pure2d}(c)
commute with one another, since they don't overlap. This commutativity means that instead of starting our braiding
process [Fig.~\ref{fig_pure2d}(c)] in the state $|\Psi\>$ which is obtained \emph{after} we do the braiding in
Fig.~\ref{fig_pure2d}(b), we can equally well start our braiding process in the state $|\Psi'\>$ which is obtained
\emph{before} we do the braiding in Fig.~\ref{fig_pure2d}(b). But if we start in the state $|\Psi'\>$ then it is easy
to see that the statistical phase for our braiding process must vanish. In fact, even a \emph{single} braiding of
$\alpha$ around $\beta$ in the counterclockwise direction together with a simultaneous braiding of
$\bar\alpha$ around $\bar\beta$ in the clockwise direction already gives a vanishing statistical phase
[Fig.~\ref{fig_deform2d}(a)]. To see this,
note that in the state $|\Psi'\>$, the two pairs $\alpha, \bar \alpha$ and $\beta, \bar \beta$ are both in the vacuum
fusion channel. This means that we can annihilate $\alpha$ and $\bar \alpha$ with local operators, if we bring them close together. 
Similarly, we can annihilate $\beta$ and $\bar \beta$.
Using this fact, we can deform the braiding process shown in Fig.~\ref{fig_deform2d}(a) so that we annihilate $\alpha$ and
$\bar\alpha$ at some stage
of the braiding and recreate them at a later stage [Fig.~\ref{fig_deform2d}(b)]. After this, we can further
annihilate the pair $\beta, \bar\beta$ [Fig.~\ref{fig_deform2d}(c)]. Finally, we can deform the process so that
$\alpha$ and $\bar\alpha$ are braided around the vacuum [Fig.~\ref{fig_deform2d}(d)]. Clearly the statistical
phase associated with Fig.~\ref{fig_deform2d}(d) is zero, so since the deformation cannot change the statistical
phase, we conclude that the statistical phase for the original braiding process shown in Fig.~\ref{fig_deform2d}(a)
must also vanish. Comparing this calculation with the previous one, we conclude that $\Theta_{ij,l} = 0$.


\section{Connection with group cohomology}
\label{sec:connection_set}

Recall that the \emph{group cohomology models} are exactly soluble lattice models that can realize SPT phases in arbitrary spatial 
dimension.\cite{chen13} One needs to specify two pieces of information to construct a $d$-dimensional group cohomology model: (1) a symmetry group $G$ and (2) a $d+1$-cocycle $\nu$, i.e. a function $\nu: G^{d+1} \rightarrow U(1)$ satisfying certain algebraic properties. It can be shown that if two cocycles $\nu_1$ and $\nu_2$ differ by a $d+1$-coboundary $\chi$, that is $\nu_1 = \nu_2 \cdot \chi$, then the corresponding models are identical to one another. Thus, the distinct group cohomology models are parameterized by elements of the cohomology group $H^{d+1}(G,U(1))$.

Focusing on the 3D case, the group cohomology models raise a basic question:
\begin{description}
\item[{\bf Q}] {\it Which surfaces can exist on the boundary of a 3D group cohomology model with $4$-cocycle $\nu$?}
\end{description}
Chen, Burnell, Vishwanath, and Fidkowski\cite{chen14} proposed a possible answer to this question for surfaces that are (1) gapped and symmetric and 
(2) have the property that the symmetry does not permute the surface anyons. Specifically, Chen et al. 
conjectured that any surface of this kind must obey the relation
\begin{align}
& \nu(a,b,c,d)   \sim R_{\omega(c,d) ,\omega(a,b)} \label{nu} \\
& \  \times F_{\omega(b,c),\omega(a,b+c),\omega(a+b+c,d)}
F^{-1}_{\omega(b,c),\omega(b+c,d),\omega(a,b+c+d)} \nonumber \\
& \ \times F_{\omega(a,b),\omega(c,d),\omega(a+b,c+d)}
F^{-1}_{\omega(a,b),\omega(a+b,c),\omega(a+b+c,d) } \nonumber\\
& \ \times F_{\omega(c,d),\omega(b,c+d),\omega(a,b+c+d)}
F^{-1}_{\omega(c,d),\omega(a,b),\omega(a+b,c+d)} \nonumber
\end{align}
where $a,b,c,d \in G$ and $R_{xy}, F_{xyz}$ and $\omega: G^2 \rightarrow A$ are various pieces of data that describe the properties of the surface. 
This conjecture was motivated by the authors' analysis of anomalies in 2D anyon systems.

Two comments about the notation in Eq. \ref{nu}: First, the `$\sim$' sign means that the left and right hand sides are equal up to multiplication by a $4$-coboundary $\chi(a,b,c,d)$. Second, we use
the `$+$' symbol for the group law because we will assume that $G$ is \emph{Abelian} in what follows.

It is interesting to compare Eq. \ref{nu} to the predictions of the bulk-boundary formulas (\ref{formula1}-\ref{formula3}). 
Indeed, for each group cohomology model we can compute the corresponding bulk data $\Theta_{i,l}, \Theta_{ij,l}, \Theta_{ijk,l}$. 
If we substitute this bulk data into the bulk-boundary formulas (\ref{formula1}-\ref{formula3}), we can obtain constraints on the 
surface data $\{\Phi_\mu, \Phi_{\mu\nu}, \Omega_{i\mu}, \Omega_{ij\mu},x_{il}^\mu\}$, as illustrated by the example in 
Sec.~\ref{sec:example2}. Since both the bulk-boundary formulas (\ref{formula1}-\ref{formula3}) and Eq. \ref{nu} give constraints 
on the set of allowed surfaces, we can ask how these constraints are related to one another. 
In this section, we will show that these constraints are exactly equivalent.

\begin{figure}[ptb]
	\centering
\includegraphics[height=1in,
width=3.0in]{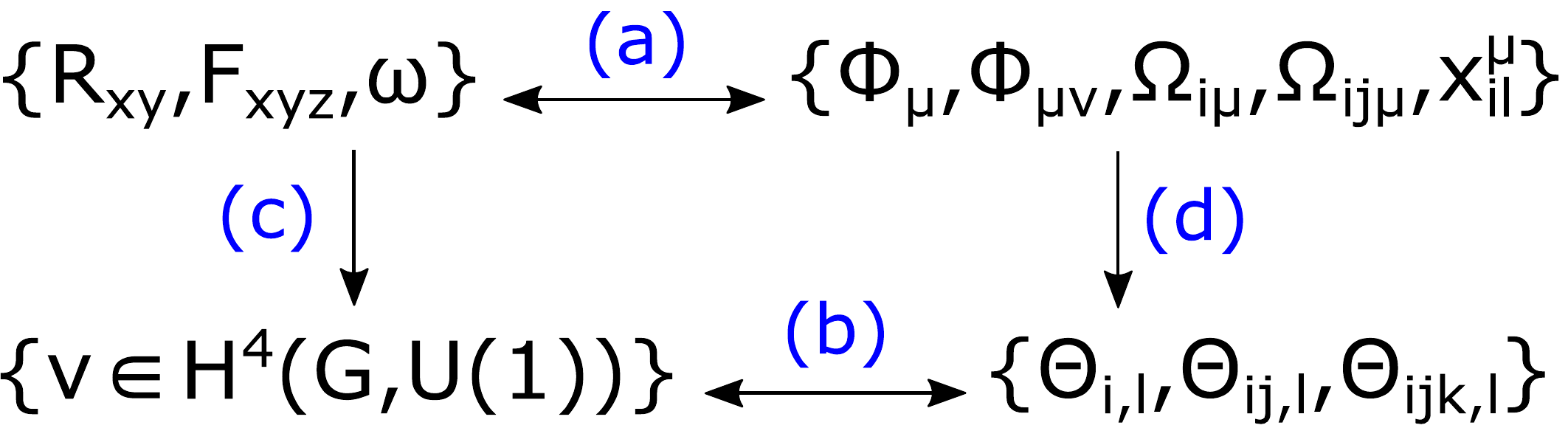}
\caption{The equivalence between Eq. \ref{nu} and the bulk-boundary formulas (\ref{formula1}-\ref{formula3}). The arrow (a) represents the one-to-one mapping between
the surface data used in Eq. \ref{nu} and in (\ref{formula1}-\ref{formula3}). The arrow (b) represents the one-to-one mapping between the bulk data used 
in Eq. \ref{nu} and in (\ref{formula1}-\ref{formula3}). Arrows (c) and (d) represent Eq \ref{nu} and Eqs. (\ref{formula1}-\ref{formula3}), respectively. 
We show that the whole diagram commutes in Sec.~\ref{sec:cohom_equi}.}
\label{fig_flow}
\end{figure}

We establish this equivalence in several steps. First, in section \ref{sec:cohom_review} we review the definition of the surface data
$\{R_{xy}, F_{xyz}, \omega\}$. Next, in section \ref{sec:cohom_surface} we show how to translate between the two types of surface data,
that is $\{R_{xy}, F_{xyz}, \omega\}$ and $\{\Phi_\mu, \Phi_{\mu\nu}, \Omega_{i\mu}, \Omega_{ij\mu},x_{il}^\mu\}$.
Similarly, in section \ref{sec:cohom_bulk}, we review how to compute the bulk data $\{\Theta_{i,l}, \Theta_{ij,l}, \Theta_{ijk,l}\}$
corresponding to a $4$-cocycle $\nu$. Finally, in section \ref{sec:cohom_equi}, we put everything together and derive the equivalence 
between Eq. \ref{nu} and the constraints coming from the bulk-boundary formulas (\ref{formula1}-\ref{formula3}). 
See Fig. \ref{fig_flow} for a summary of these results.

\subsection{Review of surface data \texorpdfstring{$\{R_{xy}, F_{xyz}, \omega\}$}{}}
\label{sec:cohom_review}

We begin by reviewing the definition of $\{R_{xy}, F_{xyz}, \omega\}$. Unlike 
$\{\Phi_\mu, \Phi_{\mu\nu}, \Omega_{i\mu}, \Omega_{ij\mu},x_{il}^\mu\}$, these
quantities are all defined using the \emph{ungauged} lattice boson models. The first two quantities, $R_{xy}$ and $F_{xyz}$, are relatively easy to
explain. 
These are the ``$R$-symbols'' and ``$F$-symbols''\cite{kitaev06} that describe the braiding and associativity relations of the Abelian anyons that live on the surface. 
These quantities take values in $U(1)$, while their indices $x,y,z$ run over the group $A$ of Abelian surface anyons. (Here we suppress extra indices 
that are often included in these symbols, e.g. $R_{xy}^z$, since these indices are redundant in the Abelian case).

The other piece of data, $\omega$, is an element of the cohomology group $H^2(G,A)$. More concretely, $\omega: G \times G \rightarrow A$
is a function that obeys the relation
\begin{equation}
\omega(a,b) + \omega(a+b,c) = \omega(b,c) + \omega(a, b+c) \label{eq_set1}
\end{equation}
and is defined up to the gauge transformation
\begin{equation}
\omega(a,b) \rightarrow \omega(a,b) + \chi(a+b) - \chi(a) - \chi(b) \label{eq_set2}
\end{equation}
where $\chi: G \rightarrow A$ is some arbitrary function. Here $\omega(a,b)+\omega(a+b,c)$ denotes the fusion product of the Abelian anyons 
$\omega(a,b)$ and $\omega(a+b,c)$, while $a+b$ denotes the group composition of $a,b \in G$.

The physical meaning of $\omega$ is that it describes how the symmetry acts on the surface anyons. (See Refs.~\onlinecite{kitaev06,essin13,fidkowski15,barkeshli14} for general discussions about symmetry actions on anyons.) To explain the precise definition, let us first 
consider the simpler case of a purely 2D anyon system (as opposed to a surface). In the purely 2D case, $\omega$ is defined as follows. 
For each group element $a \in G$, we can construct a corresponding ``defect line'' by twisting the Hamiltonian along some line 
running from some point $r_0$ to infinity. Here, by ``twisting'' the Hamiltonian we mean that we conjugate the Hamiltonian by 
a global symmetry transformation $S_a$ acting on one side of the defect line: $H_{\text{tw}} = S_a^{-1} H S_a$. Importantly, there is some ambiguity in 
defining this twisting procedure near the end of the defect line, $r_0$. Because of this ambiguity, we can construct many different defect line 
Hamiltonians $H_{\text{tw}},H_{\text{tw}}',...$ for the same group element $a \in G$. These Hamiltonians are all on an equal footing so their 
ground states are equally good definitions of defect lines. At the same time, if we compare the ground states of different Hamiltonians
$H_{\text{tw}},H_{\text{tw}}'$, they can differ in general by the
attachment of an anyon $x \in A$ at the end of the defect line.

\begin{figure}[ptb]
	\centering
\includegraphics[height=1in,
width=3in]{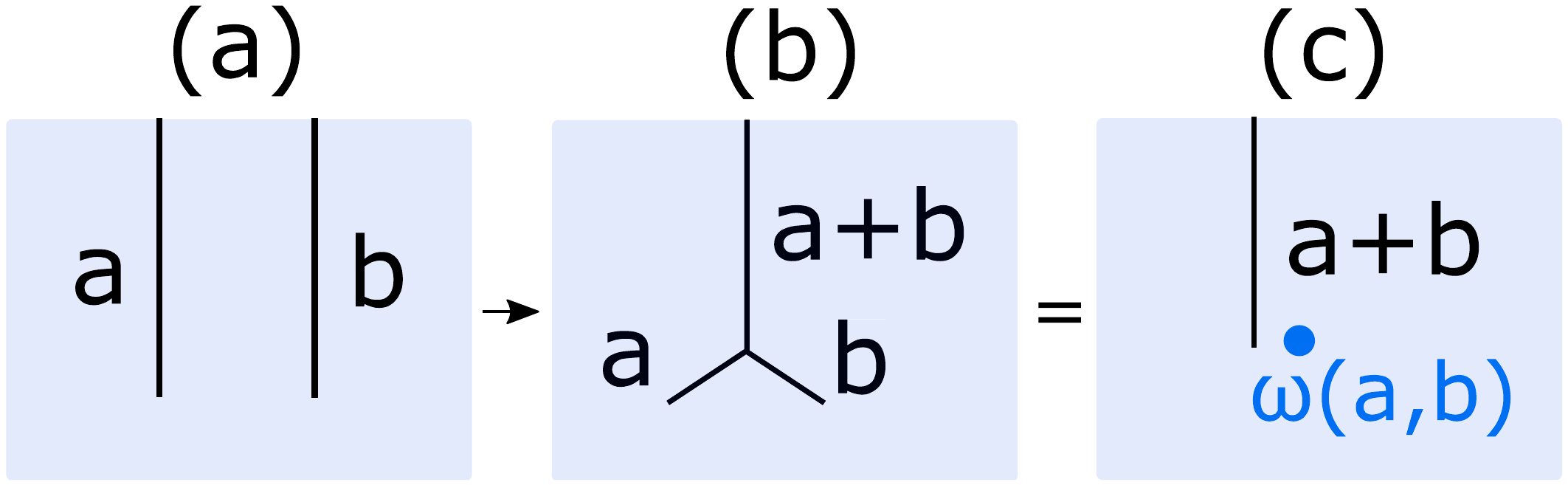}
\caption{Definition of $\omega(a,b)$. Panel (a) shows two defect lines labeled by $a$ and $b$. In (b), we fuse the two defect 
lines and get a defect line corresponding $a+b$. This defect line differs from the representative $a+b$ defect line shown in (c) by an anyon 
$\omega(a,b)$.}
\label{fig_omega}
\end{figure}

Now choose a \emph{representative} defect line for each $a \in G$. Consider two defect lines corresponding to $a,b \in G$. If we ``fuse'' the 
two lines together, we will get a defect line corresponding to 
$a+b$ (Fig. \ref{fig_omega}). In general, this defect line will differ from our representative $a+b$ 
defect line by an anyon $x \in A$, as discussed above. We then define a function $\omega: G \times G \rightarrow A$ by $\omega(a,b) = x$. 
Note that $\omega(a,b)\neq \omega(b,a)$ in general because, unlike point particles, the 
fusion of defect lines need not be commutative; i.e.
there is a well-defined distinction between the defect line on the left and the defect line on the right.

It is not hard to see that this function $\omega$ obeys the constraint (\ref{eq_set1}) and is well-defined up to the transformation (\ref{eq_set2}). 
Indeed, to derive the constraint (\ref{eq_set1}), consider three defect lines $a,b,c$, and imagine fusing them together in two different ways. Using 
the fact that the two different fusion outcomes must be consistent with one another, one can see that the function $\omega$ obeys the constraint 
(\ref{eq_set1}). As for the gauge transformation (\ref{eq_set2}), this follows from the fact that we can freely change our choice of representatives for each defect line $a$.

So far we have defined $\omega$ for purely 2D systems. We still need to explain how to define $\omega$ for a surface. In this case, we can use 
essentially the same definition as before but with one extra dimension; in particular, instead of twisting the Hamiltonian along a defect \emph{line} 
that ends at a point, we need to twist it along a defect \emph{plane} that ends at a line that is perpendicular to the surface. The rest of the 
definition follows the 2D case in the obvious way.\footnote{The reader may worry that the ambiguity in defining defect planes is 
fundamentally different from that for defect lines due to their larger dimensionality. This is not the case: if we compare two defect planes that correspond to the 
same group element {$a$}, they can differ at most by the attachment of a point-like surface anyon that lies at the intersection of the
end of the defect planes and the surface, since the bulk is short-range entangled.} 

\subsection{Translating between the two types of surface data}
\label{sec:cohom_surface}

Now we explain how to translate between the two types of surface data:
\begin{align*}
\{R_{xy}, F_{xyz}, \omega\} \leftrightarrow \{\Phi_\mu, \Phi_{\mu\nu}, \Omega_{i\mu}, \Omega_{ij\mu},x_{il}^\mu\}.
\end{align*}
This translation problem can be divided into two pieces, one of which involves the braiding statistics data, namely $\{R_{xy}, F_{xyz}\}$ and $\{\Phi_\mu, \Phi_{\mu\nu}\}$, and the other of which involves the symmetry data, namely $\omega$ and $\{\Omega_{i\mu}, \Omega_{ij\mu},x_{il}^\mu\}$.

First, we fix our notation. As in the previous sections, we will assume that the symmetry group is $G = \prod_{i=1}^K \mathbb{Z}_{N_i}$ while the anyon group is $A = \prod_{\mu = 1}^M \mathbb{Z}_{N_\mu}$. We parameterize group elements $a \in G$ by $K$ component integer vectors $a = (a_1,...,a_K)$, while we parameterize anyons $x \in A$ by $M$ component integer vectors $x = (x^1,...,x^M)$. We let the components $a_i$ take values in the range $0, 1,\dots, (N_i-1)$, and let the components $x^\mu$ take values in the range $0,1,\dots, (N_\mu-1)$.

We begin by describing the dictionary between the two types of braiding statistics data, $\{R_{xy}, F_{xyz}\}$ and $\{\Phi_\mu, \Phi_{\mu\nu}\}$. One direction is easy: it is clear from the definitions of $\Phi_\mu$ and $\Phi_{\mu\nu}$ that
\begin{equation}
e^{-i\Phi_\mu} = R_{\boldsymbol \epsilon_\mu \boldsymbol \epsilon_\mu}, \quad e^{-i\Phi_{\mu \nu}} = R_{\boldsymbol \epsilon_\mu \boldsymbol \epsilon_\nu} R_{\boldsymbol \epsilon_\nu \boldsymbol \epsilon_\mu}
\label{phiR}
\end{equation}
where $\boldsymbol \epsilon_\mu$ denotes the unit type-$\mu$ anyon, $(0,...,1,...,0)$ with a $1$ in the $\mu$th entry.
(Here the `$-$' sign in the exponent is not particularly significant and depends on conventions: the sign can be either `$+$' or `$-$' depending on 
whether $R$ is defined in terms of counterclockwise or clockwise braiding. The reason we choose `$-$' rather than `$+$' is that, with this 
convention, the bulk boundary formulas (\ref{formula1}-\ref{formula3}) are consistent with the cohomology formula (\ref{nu}). If we choose the `$+$' sign 
instead, then (\ref{formula1}-\ref{formula3}) are consistent with a modified version of (\ref{nu}) which is obtained by replacing 
$\nu \rightarrow \nu^{-1}$).

The opposite direction, in which we express $\{R_{xy}, F_{xyz}\}$ in terms of $\{\Phi_\mu, \Phi_{\mu\nu}\}$, is harder. One problem is 
that there are some gauge choices in the definition of $\{R_{xy}, F_{xyz}\}$, so the inverse map is not uniquely defined. Another problem is that 
there are complicated constraints on $\{R_{xy}, F_{xyz}\}$ coming from the pentagon and hexagon equations\cite{kitaev06}. Thus we not only have to 
invert Eq. \ref{phiR} but we also have to solve these constraints. Fortunately both of these problems can be overcome by doing things in the right 
order. In particular, we proceed by \emph{first} finding the most general solution $\{R_{xy}, F_{xyz}\}$ to the pentagon and hexagon equations\cite{kitaev06} that is 
consistent with the fusion rules specified by $A = \prod_{\mu = 1}^M \mathbb{Z}_{N_\mu}$, and \emph{then} matching 
the general expression for $R_{xy}$ to Eq. \ref{phiR}. Skipping over the intermediate steps, the final result is that we obtain the following expressions for 
$\{R_{xy}, F_{xyz}\}$:
\begin{align}
R_{xy} &  =\exp\left(-i\sum_{\mu} x^{\mu}y^{\mu} \Phi_{\mu} -i\sum_{\mu<\nu} x^{\mu}y^{\nu} \Phi_{\mu\nu}\right), \nonumber \\
F_{xyz} &  =\exp\left(-i\sum_{\mu} x^{\mu}\left(y^{\mu}+z^{\mu}-[y^{\mu}+z^{\mu}]\right)  \Phi_{\mu}  \right).
\label{Rphi}
\end{align}
Here the square bracket $[y^\mu + z^\mu]$ is defined to be $y^\mu + z^\mu \pmod{N_\mu}$ with values taken in the range $0,...,N_\mu-1$. 
The two mappings (\ref{phiR}) and (\ref{Rphi}) define a 
one-to-one correspondence between the data $\{\Phi_\mu, \Phi_{\mu\nu}\}$ and $\{R_{xy}, F_{xyz}\}$ (up to gauge equivalence).

Next we turn to the connection between $\omega$ and $\{\Omega_{i\mu}, \Omega_{ij\mu},x_{il}^\mu\}$. In order to make this connection, it
is helpful to introduce another quantity, namely a unitary matrix $U_a(x)$ that is associated to each surface anyon $x \in A$ and each group
element $a \in G$. This matrix $U_a(x)$ is defined as follows\cite{kitaev06, essin13,fidkowski15,barkeshli14}. Recall that the surface anyons are all Abelian (by assumption) so they do not have any 
topologically protected degeneracies. However they can have \emph{symmetry} protected degeneracies; that is, the surface anyons can have multiple internal
states that are degenerate with one another and that cannot be split without breaking one of the symmetries. As a consequence of this 
degeneracy, if we braid an anyon $x$ around a defect line $a$, then the resulting Berry phase can be non-Abelian since the internal states of 
$x$ can mix with one another. We define the matrix $U_a(x)$ to be the (possibly non-Abelian) Berry phase associated with this braiding process.
 
An important property of the $U_a(x)$ matrices is that they obey the following relation:\cite{essin13,barkeshli14}
\begin{equation}
        U_a(x) U_b(x) =\exp \left(i \sum_{\mu \nu} x^\mu  \omega^\nu(a,b) \Phi_{\mu\nu} \right) U_{a+b}(x).
        \label{omega1}
\end{equation}
Here $\omega^\nu(a,b)$ denotes the $\nu$th component of $\omega(a,b)$, which we think of as a $M$-component integer vector.
To derive this relation, imagine we braid an anyon $x$ around a type $b$ defect line and then around a type $a$ defect line. We can compute
the Berry phase associated with this process in two different ways. In the first approach, we simply compose the two braiding processes, giving a Berry
phase $U_a(x) \cdot U_b(x)$. In the second approach, we fuse the two defect lines together to form a type $a+b$ defect line together with an
additional surface anyon $\omega(a,b)$. The Berry phase is therefore equal to $e^{i \sum_{\mu \nu} x^\mu \omega^\nu(a,b)}\Phi_{\mu\nu} U_{a+b}(x)$ where 
the first factor comes from the statistical phase associated with braiding $x$ around the anyon $\omega(a,b)$. Demanding consistency between these
two calculations gives Eq. \ref{omega1}.
 
With the help of Eq. \ref{omega1}, we can now derive the connection between $\omega$ and $\{\Omega_{i\mu}, \Omega_{ij\mu},x_{il}^\mu\}$. 
Let's start by expressing $\Omega_{i\mu}$ in terms of $\omega$. Recall that $\Omega_{i\mu}$ is defined as the Berry phase associated with braiding a 
type $\mathbf \epsilon_\mu$ anyon around an $\mathbf e_i$ gauge flux for $N^{i\mu}$ times. Equivalently, given the connection between 
gauge fluxes and defect lines, $\Omega_{i \mu}$ is equal to the Berry phase associated with braiding a type $\mathbf \epsilon_\mu$ anyon around a type $\mathbf e_i$ \emph{defect line} 
for $N^{i\mu}$ times. Expressing the latter Berry phase in terms of the $U_a(x)$ matrices, we derive
\begin{equation}
	 e^{i \Omega_{i\mu}} = (U_{\mathbf e_i}(\mathbf \epsilon_\mu))^{N^{i\mu}} 
\label{omega2}
\end{equation} 
Next we express the right side of (\ref{omega2}) in terms of $\omega$ by using (\ref{omega1}) repeatedly:
\begin{equation}
(U_{\mathbf e_i}(\mathbf \epsilon_\mu))^{N^{i\mu}} = 
\exp \left(i\sum_{\nu} \sum_{n=1}^{N^{i\mu}} \omega^\nu(\mathbf e_{i},n \mathbf e_{i}) \Phi_{\mu \nu}\right) 
\label{omega2b}
\end{equation}
Combining (\ref{omega2}),(\ref{omega2b}), we conclude that
\begin{align}
\Omega_{i \mu} = \sum_{\nu} \sum_{n=1}^{N^{i\mu}} \omega^\nu(\mathbf e_{i},n \mathbf e_{i}) \Phi_{\mu \nu} \label{set_surdata1}
\end{align}

In a similar fashion we can express $\Omega_{ij\mu}$ in terms of $\omega$.
First we note that $\Omega_{ij\mu}$ is equal to the Berry phase associated with braiding a type $\mathbf \epsilon_\mu$ anyon
around a type $\mathbf e_{i}$ defect and then around a type $\mathbf e_{j}$ defect and then around the type $\mathbf e_{i}$ defect in the opposite direction
and finally around the type $\mathbf e_{j}$ defect in the opposite direction. Expressing this Berry phase in terms of the $U_a(x)$ matrices gives
\begin{equation}
	e^{i \Omega_{ij \mu}} =  
U_{\mathbf e_j}(\mathbf \epsilon_\mu)^{-1}U_{\mathbf e_i}(\mathbf \epsilon_\mu)^{-1}U_{\mathbf e_j}(\mathbf \epsilon_\mu)
U_{\mathbf e_i}(\mathbf \epsilon_\mu)
	\label{omega3}
\end{equation}
As before, we can now rewrite the right side in terms of $\omega$ to obtain
\begin{equation}
	\Omega_{ij\mu}  =\sum_\nu (\omega^\nu(\mathbf e_j, \mathbf e_i)-\omega^\nu(\mathbf e_i, \mathbf e_j)) \Phi_{\mu\nu}.
	\label{set_surdata2}
\end{equation}

Finally, we need to express $x_{il}^\mu$ in terms of $\omega$. Comparing the expression for $\Omega_{ij \mu}$ with the definition of 
$x_{il}^\mu$, (\ref{eq_x7}), we see that
\begin{equation}
x_{il}^\mu = \omega^{\mu}(\mathbf e_l, \mathbf e_i) -\omega^{\mu}(\mathbf e_i, \mathbf e_l). \label{set_surdata3}
\end{equation}
One can check the expressions (\ref{set_surdata1},\ref{set_surdata2},\ref{set_surdata3}) are all invariant under the gauge transformation (\ref{eq_set2}).

It is also possible to translate in the opposite direction: that is, we can express $\omega$ in terms of 
$\{\Omega_{i\mu}, \Omega_{ij\mu},x_{il}^\mu\}$ for a particular gauge choice. However, we will not need this expression in what follows, so we will not 
write it down explicitly. Instead, we will only need the fact that (\ref{set_surdata1}) and (\ref{set_surdata2}), (\ref{set_surdata3}) define a 
one-to-one correspondence between $\{\Omega_{i\mu}, \Omega_{ij\mu},x_{il}^\mu\}$ and $\{\omega \in H^2(G,A)\}$. We establish this
result in Appendix \ref{app:counting}.

\subsection{Translating between the two types of bulk data}
\label{sec:cohom_bulk}

Next we review how to compute the bulk data $\{\Theta_{i,l},\Theta_{ij,l},\Theta_{ijk,l}\}$ for a group cohomology model with $4$-cocycle $\nu$. This calculation was worked out in Ref. \onlinecite{wangcj15}. The result is
\begin{align}
e^{i\Theta_{i,l}}  &=\prod_{n=1}^{N_{i}}\chi_{\mathbf e_{l}, \mathbf e_{i}}(\mathbf e_{i}, n\mathbf e_{i}),\label{thetab1}\\
e^{i\Theta_{ij,l}} &=\prod_{n=1}^{N^{ij}}\chi_{\mathbf e_{l}, \mathbf e_{i}}(\mathbf e_{j},n \mathbf e_{j}) \chi_{\mathbf e_{l},\mathbf e_{j}}(\mathbf e_{i},n\mathbf e_{i}),\label{thetab2}\\
e^{i\Theta_{ijk,l}} &=\frac{\chi_{\mathbf e_{l},\mathbf e_{i}}(\mathbf e_{k},\mathbf e_{j})}{\chi_{\mathbf e_{l}, \mathbf e_{i}}(\mathbf e_{j}, \mathbf e_{k})}\label{thetab3}
\end{align}
where
\begin{equation}
\chi_{\mathbf e_l,\mathbf e_i}(b,c)  =\frac{\chi_{\mathbf e_l}(\mathbf e_i,b,c) \chi_{\mathbf e_l}(b,c,\mathbf e_i)}{\chi_{\mathbf e_l}(b,\mathbf e_i,c)}\label{ki1}
\end{equation}
and
\begin{equation}
\chi_{a}(b,c,d)  =\frac{\nu(b,a,c,d)  \nu(b,c,d,a)}{\nu(a,b,c,d) \nu(b,c,a,d)}\label{ki}
\end{equation}
Importantly, Ref. \onlinecite{wangcj15} showed that the bulk data $\{\Theta_{i,l},\Theta_{ij,l},\Theta_{ijk,l}\}$ uniquely distinguish every group cohomology model; in other words, the mapping between $\{\Theta_{i,l},\Theta_{ij,l},\Theta_{ijk,l}\}$ and $\nu$ is a one-to-one correspondence (up to gauge equivalence).

\subsection{Establishing the equivalence}
\label{sec:cohom_equi}

With the help of the dictionaries described above, we are now ready to compare the cohomology formula (\ref{nu}) to the bulk-boundary formulas 
(\ref{formula1}-\ref{formula3}). Our strategy will be to translate Eq. \ref{nu} into three equations relating 
$\{\Theta_{i,l},\Theta_{ij,l},\Theta_{ijk,l}\}$ and $\{\Phi_\mu, \Phi_{\mu\nu}, \Omega_{i\mu}, \Omega_{ij\mu},x_{il}^\mu\}$. We will then compare 
these equations directly to (\ref{formula1}-\ref{formula3}).

We begin with the equation for $\Theta_{i,l}$. To derive this equation, we first express $\Theta_{i,l}$ in terms of $\nu$ using Eqs. (\ref{thetab1}), (\ref{ki1}), (\ref{ki}):
\begin{equation}
e^{i\Theta_{i,l}} = \prod_{n=1}^{N_i}\frac{\nu(\mathbf e_i, n \mathbf e_i, \mathbf e_i, \mathbf e_l)\nu(\mathbf e_i, \mathbf e_l, n\mathbf e_i, \mathbf e_i)}{\nu(\mathbf e_i, n\mathbf e_i, \mathbf e_l, \mathbf e_i)\nu(\mathbf e_l,\mathbf  e_i, n\mathbf e_i,\mathbf  e_i)} \label{equi1}
\end{equation}
Next we substitute (\ref{nu}) into the right hand side. Since the resulting expression is complicated, it is helpful to separately evaluate the contributions from the $R$ and $F$ symbols. We start with the contribution from the $R$ symbol:
\begin{equation}
e^{i \Theta_{i,l}}[R] = \prod_{n=1}^{N_i}\frac{ R_{\omega(\mathbf e_i, \mathbf e_l) ,\omega(\mathbf e_i, n \mathbf e_i)} R_{\omega(n\mathbf e_i, \mathbf e_i) ,\omega(\mathbf e_i, \mathbf e_l)}}{R_{\omega(\mathbf e_l, \mathbf e_i) ,\omega(\mathbf e_i, n\mathbf e_i)} R_{\omega(n\mathbf e_i,\mathbf  e_i) ,\omega(\mathbf e_l,\mathbf  e_i)}} \label{Rcontr}
\end{equation}
Ultimately, we want to express the quantities on the right hand side in terms of the surface data $\{\Phi_\mu, \Phi_{\mu\nu}, \Omega_{i\mu}, \Omega_{ij\mu},x_{il}^\mu\}$. However, rather than doing this immediately, it is convenient to first use the identity\footnote{This identity follows easily from the cocycle condition (\ref{eq_set1}).}
\begin{equation}
\omega(n\mathbf e_i, \mathbf e_i) =\omega(\mathbf e_i, n\mathbf e_i)
\end{equation}
to rewrite the above expression as
\begin{equation}
e^{i \Theta_{i,l}}[R] = \prod_{n=1}^{N_i}\frac{ R_{\omega(\mathbf e_i, \mathbf e_l) ,\omega(\mathbf e_i, n \mathbf e_i)} R_{\omega(\mathbf e_i, n\mathbf e_i) ,\omega(\mathbf e_i, \mathbf e_l)}}{R_{\omega(\mathbf e_l, \mathbf e_i) ,\omega(\mathbf e_i, n\mathbf e_i)} R_{\omega(\mathbf e_i,n\mathbf  e_i) ,\omega(\mathbf e_l,\mathbf  e_i)}}
\end{equation}
We can then express the product of $R$ symbols in terms of $\Phi_{\mu \nu}$ using
\begin{equation}
R_{xy} R_{yx} = \exp(-i\theta_{xy}) = \exp \left(-i\sum_{\mu\nu}  x^\mu y^\nu \Phi_{\mu \nu} \right)
\end{equation}
thus giving
\begin{equation}
e^{i\Theta_{i,l}}[R] = \exp\Bigg\{  i\sum_{n=1}^{N_i} \sum_{\mu\nu}
X^\mu_{\mathbf e_i, \mathbf e_l}\omega^\nu(\mathbf e_i, n\mathbf e_i) \Phi_{\mu\nu} \Bigg\}
\end{equation}
where $X^\mu_{ab} \equiv \omega^\mu(b,a)-\omega^\mu(a,b)$. Identifying $X^\mu_{\mathbf e_i, \mathbf e_l}$ with $x_{il}^\mu$
(see Eq. \ref{set_surdata3}), and comparing with the expression for $\Omega_{i \mu}$ (\ref{set_surdata1}), we derive
\begin{equation}
e^{i\Theta_{i,l}}\left[R\right]  =\exp\left(i\sum_{\mu}\frac{N_{i}}{N^{i\mu}}x_{il}^{\mu}\Omega_{i\mu} \right)
\end{equation}

Now consider the contribution from the $F$ symbols, $e^{i\Theta_{i,l}}[F]$. Since each of the $\nu$'s in Eq. (\ref{equi1}) contributes $6$ $F$'s, the expression for $e^{i\Theta_{i,l}}[F]$ contains $24$ $F$'s. We need to relate this product of $24$ $F$'s to the surface data $\{\Phi_\mu, \Phi_{\mu\nu}, \Omega_{i\mu}, \Omega_{ij\mu},x_{il}^\mu\}$. To do this, we express each $F$ in terms of $\Phi_\mu, \Phi_{\mu \nu}$ using (\ref{Rphi}), and then simplify the resulting expression. After some straightforward but tedious algebra, we obtain
\begin{align}
e^{i\Theta_{i,l}}[F] = \exp\bigg\{i\sum_\mu N_i \Phi_\mu (X^\mu_{\mathbf e_i, \mathbf e_l})^2\bigg\}
\end{align}
Combining the $R$ and $F$ contributions and using the fact that $X^\mu_{\mathbf e_i, \mathbf e_l} = x_{il}^\mu$, we finally derive an expression for $\Theta_{i,l}$:
\begin{equation}
e^{i\Theta_{i,l}}= \exp\bigg\{i\sum_{\mu}\frac{N_{i}}{N^{i\mu}}x_{il}^{\mu}\Omega_{i\mu}+N_{i}(x_{il}^{\mu})^{2}\Phi_{\mu}\bigg\}
\label{thetailnu}
\end{equation}

At this point we can see that the above expression for $\Theta_{i,l}$ is identical to the bulk-boundary formula (\ref{formula1}). In a similar fashion, we can show that the formulas for
$\Theta_{ij,l}$ and $\Theta_{ijk,l}$ also match the corresponding bulk-boundary formulas (\ref{formula2}) and (\ref{formula3}). Since the expressions for $\{\Theta_{i,l},\Theta_{ij,l},\Theta_{ijk,l}\}$ all match, we conclude that the cohomology formula (\ref{nu}) and the bulk-boundary formulas (\ref{formula1}-\ref{formula3}) give \emph{equivalent} constraints for the surfaces of the group cohomology models.

\section{Conclusion}

In summary, we have derived a bulk-boundary correspondence for 3D bosonic SPT phases with finite unitary Abelian symmetry group. Our correspondence relates the bulk properties of these phases to the properties of their gapped symmetric surfaces. This relationship is described by three equations (\ref{formula1}-\ref{formula3}) which express the bulk data $\{\Theta_{i,l}, \Theta_{ij,l}, \Theta_{ijk,l}\}$ in terms of the surface data $\{\Phi_\mu, \Phi_{\mu\nu}, \Omega_{i\mu}, \Omega_{ij\mu}, x_{il}^\mu\}$.

It should be possible to generalize our bulk-boundary correspondence beyond the case considered in this paper. For example, we made the simplifying assumption that the surface supports only Abelian anyons.
However, there is reason to think that this assumption is actually unnecessary --- that is, the bulk-boundary formulas (\ref{formula1}-\ref{formula3}) continue to hold even if the surface supports \emph{non-Abelian} anyons.\footnote{Here, we assume that the definitions of the surface data can be modified appropriately so that these quantities remain well-defined whether or not the surface
supports non-Abelian anyons.} One piece of evidence for this is the close connection between (\ref{formula1}-\ref{formula3}) and the conjecture (\ref{nu}) of Ref.~\onlinecite{chen14}. Indeed, the conjecture (\ref{nu}) is believed to hold whether or not the surface supports non-Abelian anyons, suggesting that our correspondence should hold more generally as well. 

Another natural extension of this work would be to \emph{fermionic} SPT phases. It is not hard to see that two of the bulk-boundary formulas (\ref{formula2}-\ref{formula3}) generalize trivially to the fermionic case since they do not involve exchange statistics in any way. On the other hand, the formula (\ref{formula1}) likely needs to be modified since this relation makes use of exchange statistics and topological
spin.

An interesting direction for future work would be to study bulk-boundary relations for 3D SPT phases with \emph{gapless} surfaces. A recent paper\cite{chenx15} proposed a bulk-boundary correspondence in this context by considering modular transformations on a three dimensional torus; it would be interesting to understand the relationship between this correspondence and the bulk-boundary relations derived in this paper.

\begin{acknowledgments}
We thank M. Cheng and Z.-C. Gu for helpful discussions. This work is supported in part by the NSF under grant No. DMR-1254741. Research at Perimeter Institute is supported by the Government of Canada through Industry Canada and by the Province of Ontario through the Ministry of Economic Development \&  Innovation.
\end{acknowledgments}

\appendix

\section{Showing \texorpdfstring{$\Omega_{i\mu}$}{omegaimu} and \texorpdfstring{$\Omega_{ij\mu}$}{omegaijmu} are well defined}
\label{sec:app_def}


\begin{figure*}
\includegraphics[width=7in]{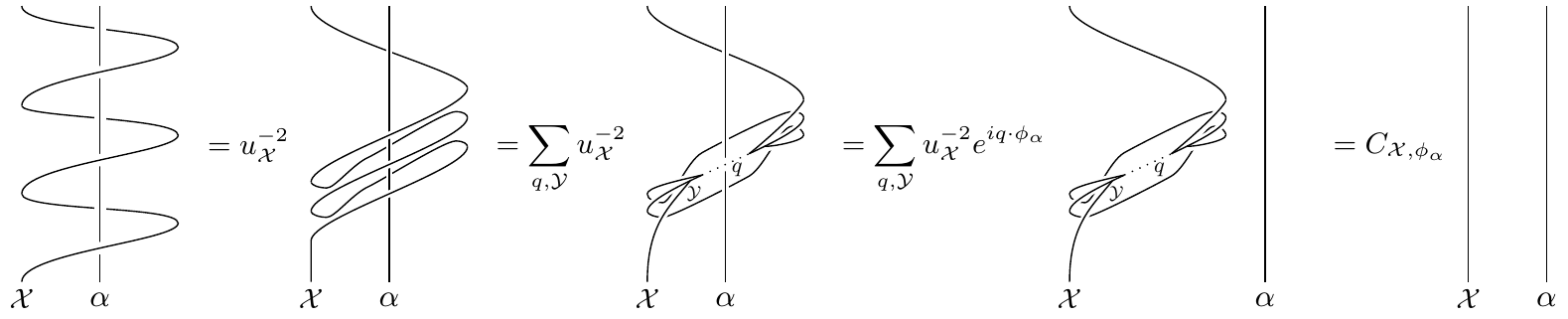}
\caption{Diagrammatic proof that $\Omega_{i\mu}$ is well defined. }\label{fig_diagram1}
\end{figure*}

\subsection{\texorpdfstring{$\Omega_{i\mu}$}{omegaimu} is well defined}

\label{sec:appd1}

Recall that $\Omega_{i\mu}$ is defined as the Berry phase associated with braiding a surface anyon $\mathcal X$ around a vortex line $\alpha$ for $N^{i\mu}$ times. Here $\mathcal X$ is any surface anyon carrying unit type-$\mu$ anyonic flux while $\alpha$ is any vortex line carrying unit type-$i$ gauge flux. The goal of this section is to show that this definition is sensible: that is, (1) the above Berry phase is Abelian and (2) the Berry phase does not depend on the choice of $\mathcal X$ or $\alpha$.

To establish these points, we analyze the above braiding process using a diagrammatic technique. This diagrammatic technique was originally developed for 2D anyon systems; in order to apply it to our system, we imagine folding the surface and straightening the vortex line $\alpha$ as in Fig.~\ref{fig_straighten}. After doing this, we can view our system as two dimensional, and we can view $\alpha$ and $\mathcal X$ as 2D anyons. We will assume this quasi-2D point of view in what follows.

While the diagrammatic method has sophisticated rules, we only need a few ingredients in our proof. (For more details about the diagrammatic method, see Ref.~\onlinecite{kitaev06}). The diagrams that we will use are built out of lines and trivalent vertices and are drawn in a (2+1) dimensional space. Roughly speaking, the lines represent space-time trajectories of 2D anyons with the arrow of time being upward, while the vertices are where the anyons fuse or split. Each diagram defines a complex number or matrix, which can be interpreted
as the quantum mechanical amplitude for the process shown in the diagram. These amplitudes can be evaluated using certain relations. What we will need below are two relations:
\begin{equation}
\begin{tikzpicture}[baseline={([yshift=-.5ex]current  bounding  box.center)}]
\draw (0,-0.25)..controls(0, 0.7) and(-0.3, 1)..(-0.5,1)..controls(-0.7,1) and(-0.8,0.5) ..(-1,0.5)..controls(-1.2,0.5) and(-1.5,0.8)..(-1.5,1.75);
\draw (1.2,-0.25)--(1.2,1.75);
\node at (0.6,0.75)[scale=1]{$\displaystyle{=u_a}$};
\node at (0,-0.4)[scale=0.8]{$a$};
\node at (-1.5,1.9)[scale=0.8]{$a$};
\node at (-0.85,0.85)[scale=0.8]{$\bar a$};
\node at (1.2,-0.4)[scale=0.8]{$a$};
\node at (1.2,1.9)[scale=0.8]{$a$};
\end{tikzpicture}\label{eq_app_a1}
\end{equation}
where $\bar a$ is the anti-particle of an anyon $a$ and $u_a$ is a complex number associated with $a$, and
\begin{equation}
\begin{tikzpicture}
\draw (0,0)--(0,2);
\draw (0.7,0)--(0.7,2);
\node at (1.6,0.9)[scale=1]{$=\displaystyle{\sum_{c,n}}$};
\node at (0,-0.15)[scale=0.8]{$a$};
\node at (0.7,-0.15)[scale=0.8]{$b$};
\node at (0,2.15)[scale=0.8]{$a$};
\node at (0.7,2.2)[scale=0.8]{$b$};
\begin{scope}[xshift=0.5cm]
\draw (2,0)--(2.35,0.7)--(2.35,1.3)--(2,2);
\draw (2.7,0)--(2.35,0.7)--(2.35,1.3)--(2.7,2);
\node at (2,-0.15)[scale=0.8]{$a$};
\node at (2.7,-0.15)[scale=0.8]{$b$};
\node at (2,2.15)[scale=0.8]{$a$};
\node at (2.7,2.2)[scale=0.8]{$b$};
\node at (2.2,1)[scale=0.8]{$c$};
\node at (2.5,0.7)[scale=0.8]{$n$};
\node at (2.5,1.3)[scale=0.8]{$n$};
\end{scope}
\end{tikzpicture}\label{eq_app_a2}
\end{equation}
where $n$ labels the states in the fusion space $\mathbb V_{ab}^c$, i.e., the different ways to split $c$ into $a$ and $b$.

\begin{figure*}
\includegraphics[width=7in]{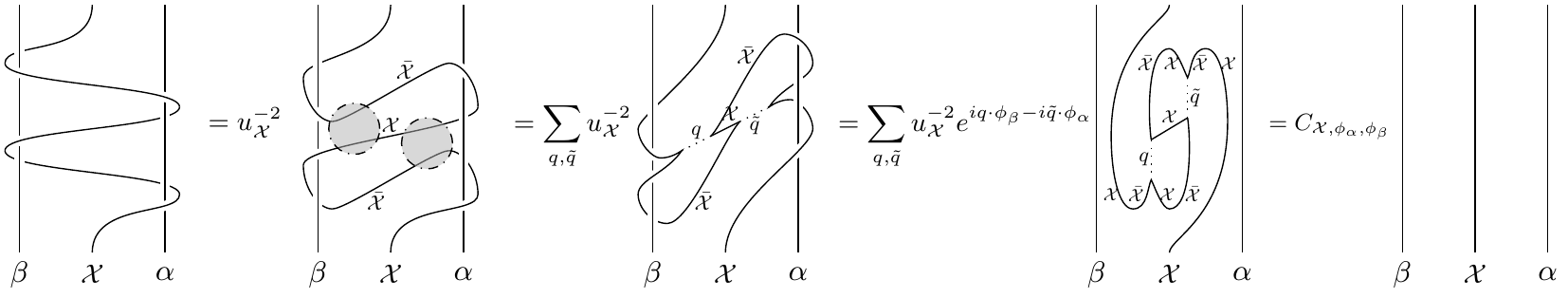}
\caption{Diagrammatic proof that \texorpdfstring{$\Omega_{ij\mu}$}{omegaijmu} is well defined. }\label{fig_diagram2}
\end{figure*}

With this background, we are now ready to establish points (1) and (2) listed above. The main calculation is shown in Fig. \ref{fig_diagram1}. We start with the diagram on the far left, which shows a braiding process in which $\mathcal X$ is braided around $\alpha$ for $N^{i\mu}$ times. (The case $N^{i\mu}=3$ is shown for illustration). The first equation follows from the relation (\ref{eq_app_a1}), while the second equation follows from (\ref{eq_app_a2}). Here, the index $\mathcal Y$ runs over the different fusion products of $\mathcal X \times \mathcal X$, while the index $q$ runs over the fusion products of $\mathcal Y \times \mathcal X$. Importantly, we know that $q$ must be a charge excitation, since if we fuse $\mathcal X$ with itself $N^{i \mu}$ times, the resulting fusion products are all charge excitations according to Eqs.~(\ref{eq_fusion},\ref{eq_fusion2}). To derive the third equation, we note that the mutual statistics between the charge $q$ and the vortex $\alpha$ is Abelian and is given by the Aharonov-Bohm law, i.e. $\exp(i q\cdot\phi_\alpha)$; therefore we can pass the charge $q$ worldline through the $\alpha$ worldline at the cost of introducing a factor of $\exp(i q\cdot\phi_\alpha)$. Moving on to the last equation, we observe that in the fourth diagram of
Fig. \ref{fig_diagram1}, the $\alpha$ worldline is completely decoupled from $\mathcal X$, but $\mathcal X$ still has some ``self-interaction''. This self-interaction may contribute a complicated numerical factor to the amplitude, but no matter how complicated it is, we can see that it depends only on $\mathcal X$, $\mathcal Y$ and $q$, and not on the fusion channel of $\mathcal X$ and $\alpha$. Therefore, after we perform the sum over $\mathcal Y$ and $q$, we can group all the numerical factors into a single complex number $C_{\mathcal X, \phi_\alpha}$ and thereby derive the last equation in Fig. \ref{fig_diagram1}. We note that since the braiding process is unitary, the constant $C_{\mathcal X, \phi_\alpha}$ must be a phase factor.

How does this calculation help us establish the two claims (1) and (2)? The key point is that $C_{\mathcal X, \phi_\alpha}$ only depends on $\mathcal X$ and $\alpha$ and doesn't depend on their fusion channel. This means that the braid matrix associated with this process is proportional to the identity matrix, i.e. it is an Abelian phase. This establishes property (1). To establish (2) we need to show that the phase factor $C_{\mathcal X, \phi_\alpha}$ is independent of the choice of $\mathcal X$ or $\alpha$ as long as $\mathcal X$ carries unit type-$\mu$ anyonic flux and $\alpha$ carries unit type-$i$ gauge flux. It is obvious that $C_{\mathcal X, \phi_\alpha}$ is independent of the choice of $\alpha$; to see that it is independent of the choice of $\mathcal X$, let $\mathcal X'$ be another anyon that carries unit type-$\mu$ anyonic flux. We need to show that $C_{\mathcal X', \phi_\alpha} = C_{\mathcal X, \phi_\alpha}$. To prove this, we use the fact that $\mathcal X'$ can be obtained by fusing some charge $q$ to $\mathcal X$. Then, by the Aharonov-Bohm law we have
\begin{equation}
C_{\mathcal X', \phi_\alpha} = C_{\mathcal X, \phi_\alpha}e^{i q\cdot\phi_\alpha N^{i\mu}} = C_{\mathcal X, \phi_\alpha} \label{eq_app_a3}
\end{equation}
where the last equality follows from the fact that $N^{i \mu} \phi_\alpha$ is a multiple of $2\pi$.

\subsection{\texorpdfstring{$\Omega_{ij\mu}$}{omegaijmu} is well defined}

Recall that $\Omega_{ij\mu}$ is defined as the Berry phase associated with the following process: a surface anyon $\mathcal X$ is first braided around a vortex $\alpha$, then around a vortex $\beta$, then around $\alpha$ in the opposite direction, and finally around $\beta$ in the opposite direction. Here, $\mathcal X$ is any surface anyon carrying unit type-$\mu$ anyonic flux, and $\alpha, \beta$ are any vortex lines carrying unit type-$i$ and type-$j$ gauge flux. Our goal is to show that this definition is sensible: that is (1) the above Berry phase is Abelian and (2) the Berry phase doesn't depend on the choice of $\mathcal X$, $\alpha$ or $\beta$. As in the previous section, we will establish these claims using the 2D diagrammatic technique; just as before, this 2D technique is applicable even though our system is three dimensional since we can fold the surface and straighten the vortex lines as in Fig.~\ref{fig_straighten} so that $\mathcal X$, $\alpha$, and $\beta$ can be viewed as 2D anyons.

The main calculation is shown in Fig.~\ref{fig_diagram2}. The diagram on the far left shows the space-time trajectories of $\mathcal X, \alpha,\beta$ during the above braiding process. The first equation follows from the relation (\ref{eq_app_a1}) while the second equation follows from the relation (\ref{eq_app_a2}) which we apply to the two shaded regions shown in the second diagram. Here, the indices $q, \tilde{q}$ run over the fusion products of $\mathcal X \times \bar{\mathcal X}$. Importantly $q$ and $\tilde{q}$ can only be charge excitations since all the fusion products of $\mathcal X \times \bar{\mathcal X}$ are charges according to Eqs.~(\ref{eq_fusion},\ref{eq_fusion2}). To derive the third equation, we note that
the Aharonov-Bohm law tells us that the mutual statistics between $q$ and $\beta$ is $\exp(i q\cdot\phi_\beta)$, while the statistics between $\tilde{q}$ and $\alpha$ is $\exp(i \tilde{q}\cdot\phi_\alpha)$; therefore we can pass the $q, \tilde{q}$ worldlines through the $\beta, \alpha$ worldlines at the cost of introducing factors of $\exp(i q\cdot\phi_\beta)$ and $\exp(-i \tilde{q}\cdot\phi_\alpha)$ respectively. Moving on to the last equation, we note that in the fourth diagram of Fig.~\ref{fig_diagram2}, the $\alpha, \beta$ worldlines are completely decoupled from $\mathcal X$, but $\mathcal X$ still has some ``self-interaction.'' By the same arguments as in the previous section, the contribution from this self-interaction can be grouped together with the other numerical factors into an overall constant $C_{\mathcal X, \phi_\alpha,\phi_\beta}$. In this way, we derive the last equation in Fig.~\ref{fig_diagram2}. We note that since the braiding process is unitary, the constant $C_{\mathcal X, \phi_\alpha,\phi_\beta}$ must be a phase factor.

We are now ready to prove the two claims (1), (2). To prove (1), we observe that $C_{\mathcal X, \phi_\alpha,\phi_\beta}$ does not depend on the fusion channel between $\mathcal X, \alpha, \beta$. This implies that the braid matrix associated with this process is proportional to the identity matrix, i.e. it is an Abelian phase. To prove (2), we need to show that the phase factor $C_{\mathcal X, \phi_\alpha,\phi_\beta}$ is independent of the choice of $\mathcal X$, $\alpha$ or $\beta$ as long as $\mathcal X$ carries unit type-$\mu$ anyonic flux and $\alpha, \beta$ carry unit type-$i$ and type-$j$ gauge flux.  It is obvious that $C_{\mathcal X, \phi_\alpha,\phi_\beta}$ is independent of the choice of $\alpha, \beta$; to see that it is independent of the choice of $\mathcal X$, let $\mathcal X'$ be another anyon that carries unit type-$\mu$ anyonic flux. We need to show that $C_{\mathcal X', \phi_\alpha,\phi_\beta} = C_{\mathcal X, \phi_\alpha,\phi_\beta}$. To prove this, we note that $\mathcal X'$ can be obtained by fusing some charge $q$ to $\mathcal X$. Therefore, by the Aharonov-Bohm law we have
\begin{align}
C_{\mathcal X', \phi_\alpha, \phi_\beta} & = C_{\mathcal X, \phi_\alpha, \phi_\beta}e^{i q\cdot \phi_\alpha + i q\cdot \phi_\beta- i q\cdot \phi_\alpha- i q\cdot \phi_\beta} \nonumber\\
 & = C_{\mathcal X, \phi_\alpha, \phi_\beta}.
\end{align}
This establishes claim (2).

\section{The physical interpretation of \texorpdfstring{$x_{il}$}{xil}}
\label{sec:app_xil}

In Sec.~\ref{sec:defx}, we gave a physical interpretation of the quantity $x_{il}$ in terms of the thought experiment depicted in Fig.~\ref{fig_xdef}. More precisely, we made the claim that (see Eq.~(\ref{xileqs}) in the main text):
\begin{align}
\xi_\mathcal X & = \xi_{\mathcal X'} = \dots = x_{il}\nonumber\\
\xi_\mathcal S & = \xi_{\mathcal S'}  = \dots = -x_{il} \label{eq_app_xil_1}
\end{align}
where $\mathcal X,\mathcal X',\dots$ are the different surface anyons obtained by shrinking the arch $\alpha$ in Fig.~\ref{fig_xdef}(c) while $\mathcal S, \mathcal S',\dots$ are the
different surface anyons obtained by shrinking the arch $\sigma$. In this Appendix, we prove Eq.~(\ref{eq_app_xil_1}).

The key step in our derivation is to consider a process in which we braid a surface anyon $\mathcal F$ around
the two vortex arches shown in Fig.~\ref{fig_xdef}(b):
\begin{equation}
\begin{tikzpicture}[baseline={([yshift=-.0ex]current  bounding  box.center)},scale=0.9]
\clip (-1.8,-0.05)rectangle +(6.3,2.4);
\fill [blue!20](0,0)--(1.5,1.5)--(4.5,1.5)--(3,0)--cycle;
\draw [black!50, -stealth] (0.8,0.3)..controls (3.5,0.2) and (3,1.0)..(2.4,1.0)..controls(1.9,1.0) and (1.7,0.6)..(2,0.6)..controls (2.3,0.6) and (2.1,0.9)..(2.4,0.9)..controls (2.7,0.9) and (2.7,0.5)..(1.9, 0.5)..controls(1.1,0.5) and (1.6,0.9)..(1.3,0.9)..controls(1,0.9) and (0.9,0.7) ..(0.8,0.35);
\begin{scope}
\clip (1.1,0.7)rectangle+(2.2,0.35);
\draw [blue!20, line width=2.7pt](1.6,2)..controls (1.7,2.5) and (3.2,2.5)..(3.2,0.75);
\draw [blue!20, line width=2.7pt](1.2,0.75)..controls(1.2,2.8)and (2,2.8)..(2,0.75);
\draw [blue!20, line width=2.7pt](2.4,0.75)..controls(2.4,2)and (1.5,1.5)..(1.6,2);
\end{scope}
\draw [thick](1.6,2)..controls (1.7,2.5) and (3.2,2.5)..(3.2,0.75);
\begin{scope}
\clip (1.6,1.6)rectangle+(0.6,0.8);
\draw [white, line width=2.7pt](1.2,0.75)..controls(1.2,2.8)and (2,2.8)..(2,0.75);
\end{scope}
\begin{scope}
\clip (1.9,0.75)rectangle+(0.3,0.4);
\draw [blue!20, line width=2.7pt](1.2,0.75)..controls(1.2,2.8)and (2,2.8)..(2,0.75);
\end{scope}
\draw [thick](1.2,0.75)..controls(1.2,2.8)and (2,2.8)..(2,0.75);
\begin{scope}
\clip (1.7,1.6)rectangle + (0.6,0.6);
\draw [white, line width=2.7pt](2.4,0.75)..controls(2.4,2)and (1.5,1.5)..(1.6,2);
\end{scope}
\draw [thick](2.4,0.75)..controls(2.4,2)and (1.5,1.5)..(1.6,2);
\draw [thick, -stealth] (1.215,1.2)--(1.22,1.27);
\draw [thick, -stealth] (2.345,1.2)--(2.33,1.27);
\fill (0.8,0.3) circle (0.04);
\node at (0.55,0.3)[scale=0.9 ]{$\mathcal F$};
\node at (1.15,2.1){$\alpha$};
\node at (3,2.1){$\sigma$};
\node at (-0.7, 1.2) {$W =$};
\end{tikzpicture} \label{eq_app_xil_3}
\end{equation}
Assuming that $\mathcal F$ carries unit type-$\mu$ anyonic flux, what we will show is that the above process gives rise to an Abelian Berry phase which is equal to $\Omega_{il\mu}$. Equivalently, in more formal language, we will show that
\begin{equation}
W |\psi\> = e^{i\Omega_{il\mu}} |\psi\>
\label{Womegaapp}
\end{equation}
where $|\psi\>$ denotes the initial state at the beginning of the braiding process.

Before we establish Eq.~\ref{Womegaapp}, we now argue that it implies Eq.~(\ref{eq_app_xil_1}). To see this, notice that the above braiding process $W$ can be smoothly deformed into a process in which $\mathcal F$ is braided around both ends of $\alpha$ in Fig.~\ref{fig_xdef}(c). The latter process can then be deformed into a process in which $\mathcal F$ is braided around the surface excitation obtained by shrinking down $\alpha$ (see Fig.~\ref{fig_xdef}(d)). Since the Berry phase must remain unchanged during these deformations, it follows that
\begin{equation}
e^{i \theta_{\mathcal F \mathcal X}}|\psi\> = e^{i \theta_{\mathcal F \mathcal X'}}|\psi\> = ... = W |\psi\>
\label{Wthetaapp}
\end{equation}
where $\mathcal X,\mathcal X',\dots$ are the different surface anyons obtained by shrinking the arch $\alpha$. At the same time, we can straightforwardly compute the statistical phase $\theta_{\mathcal F \mathcal X}$ using Eqs. (\ref{eq_AB_stat}) and (\ref{mutual}):
\begin{equation}
\theta_{\mathcal F \mathcal X} = \theta_{\xi_{\mathcal F} \xi_{\mathcal X}} = \sum_\nu (\xi_{\mathcal X})^\nu \Phi_{\mu \nu}
\label{thetafa}
\end{equation}
Similarly we have
\begin{equation}
\theta_{\mathcal F \mathcal X'} = \sum_\nu  (\xi_{\mathcal X'})^\nu \Phi_{\mu \nu},
\label{thetafa2}
\end{equation}
etc. Putting together Eqs. \ref{Womegaapp}-\ref{thetafa2}, we derive:
\begin{equation*}
\sum_\nu (\xi_{\mathcal X})^\nu \Phi_{\mu \nu}  = \sum_\nu  (\xi_{\mathcal X'})^\nu \Phi_{\mu \nu} = ... = \Omega_{il\mu} \pmod{2\pi}
\end{equation*}
We conclude that
\begin{align*}
(\xi_{\mathcal X})^\nu = (\xi_{\mathcal X'})^\nu = ... = x_{il}^\nu
\end{align*}
where $x_{il}^\nu$ is the unique integer solution to
\begin{equation*}
\sum_\nu x_{il}^\nu \Phi_{\mu\nu} = \Omega_{il\mu} \pmod{2\pi}
\end{equation*}
This establishes the first relation in Eq.~(\ref{eq_app_xil_1}).

As for the second relation in Eq.~(\ref{eq_app_xil_1}), this follows from the fact that if we further fuse the two surface anyons in Fig.~\ref{fig_xdef}(d), the outcome must be a charge excitation or the vacuum; hence, $\mathcal S,\mathcal S',\dots$ etc. must carry opposite anyonic flux to $\mathcal X, \mathcal X',\dots$ etc.

All that remains is to derive Eq. \ref{Womegaapp}. To do this, we decompose $W$ into simpler processes. In particular, let
\begin{align}
\begin{tikzpicture}[baseline={([yshift=-1.5ex]current  bounding  box.center)}, scale=0.9]
\clip (-1.6,-0.05)rectangle +(6.3,2.4);
\fill [blue!20](0,0)--(1.5,1.5)--(4.5,1.5)--(3,0)--cycle;
\draw [black!50, -stealth] (0.8,0.3)..controls (3,0.5) and (2.7,1.0)..(2.55,1)..controls (2.2,1) and (2.5,0.7)..(0.85,0.35);
\begin{scope}
\clip (1.1,0.7)rectangle+(2.2,0.35);
\draw [blue!20, line width=2.7pt](1.6,2)..controls (1.7,2.5) and (3.2,2.5)..(3.2,0.75);
\draw [blue!20, line width=2.7pt](1.2,0.75)..controls(1.2,2.8)and (2,2.8)..(2,0.75);
\draw [blue!20, line width=2.7pt](2.4,0.75)..controls(2.4,2)and (1.5,1.5)..(1.6,2);
\end{scope}
\draw [thick](1.6,2)..controls (1.7,2.5) and (3.2,2.5)..(3.2,0.75);
\begin{scope}
\clip (1.6,1.6)rectangle+(0.6,0.8);
\draw [white, line width=2.7pt](1.2,0.75)..controls(1.2,2.8)and (2,2.8)..(2,0.75);
\end{scope}
\begin{scope}
\clip (1.9,0.75)rectangle+(0.3,0.4);
\draw [blue!20, line width=2.7pt](1.2,0.75)..controls(1.2,2.8)and (2,2.8)..(2,0.75);
\end{scope}
\draw [thick](1.2,0.75)..controls(1.2,2.8)and (2,2.8)..(2,0.75);
\begin{scope}
\clip (1.7,1.6)rectangle + (0.6,0.6);
\draw [white, line width=2.7pt](2.4,0.75)..controls(2.4,2)and (1.5,1.5)..(1.6,2);
\end{scope}
\draw [thick](2.4,0.75)..controls(2.4,2)and (1.5,1.5)..(1.6,2);
\draw [thick, -stealth] (1.215,1.2)--(1.22,1.27);
\draw [thick, -stealth] (2.345,1.2)--(2.33,1.27);
\fill (0.8,0.3) circle (0.04);
\node at (0.55,0.3)[scale=0.9 ]{$\mathcal F$};
\node at (1.15,2.1){$\alpha$};
\node at (3,2.1){$\sigma$};
\node at (-0.9, 1) {$W_1\ =$};
\end{tikzpicture}\\
\begin{tikzpicture}[baseline={([yshift=-1.5ex]current  bounding  box.center)}, scale=0.9]
\clip (-1.6,-0.05)rectangle +(6.3,2.4);
\fill [blue!20](0,0)--(1.5,1.5)--(4.5,1.5)--(3,0)--cycle;
\draw [black!50, -stealth] (0.8,0.3)..controls (2.4,0.5) and (2.3,1.0)..(2.15,1)..controls (1.9,1) and (1.7,0.7)..(0.85,0.35);
\begin{scope}
\clip (1.1,0.7)rectangle+(2.2,0.35);
\draw [blue!20, line width=2.7pt](1.6,2)..controls (1.7,2.5) and (3.2,2.5)..(3.2,0.75);
\draw [blue!20, line width=2.7pt](1.2,0.75)..controls(1.2,2.8)and (2,2.8)..(2,0.75);
\draw [blue!20, line width=2.7pt](2.4,0.75)..controls(2.4,2)and (1.5,1.5)..(1.6,2);
\end{scope}
\draw [thick](1.6,2)..controls (1.7,2.5) and (3.2,2.5)..(3.2,0.75);
\begin{scope}
\clip (1.6,1.6)rectangle+(0.6,0.8);
\draw [white, line width=2.7pt](1.2,0.75)..controls(1.2,2.8)and (2,2.8)..(2,0.75);
\end{scope}
\begin{scope}
\clip (1.9,0.75)rectangle+(0.3,0.4);
\draw [blue!20, line width=2.7pt](1.2,0.75)..controls(1.2,2.8)and (2,2.8)..(2,0.75);
\end{scope}
\draw [thick](1.2,0.75)..controls(1.2,2.8)and (2,2.8)..(2,0.75);
\begin{scope}
\clip (1.7,1.6)rectangle + (0.6,0.6);
\draw [white, line width=2.7pt](2.4,0.75)..controls(2.4,2)and (1.5,1.5)..(1.6,2);
\end{scope}
\draw [thick](2.4,0.75)..controls(2.4,2)and (1.5,1.5)..(1.6,2);
\draw [thick, -stealth] (1.215,1.2)--(1.22,1.27);
\draw [thick, -stealth] (2.345,1.2)--(2.33,1.27);
\fill (0.8,0.3) circle (0.04);
\node at (0.55,0.3)[scale=0.9 ]{$\mathcal F$};
\node at (1.15,2.1){$\alpha$};
\node at (3,2.1){$\sigma$};
\node at (-0.9, 1) {$W_2\ =$};
\end{tikzpicture}\\
\begin{tikzpicture}[baseline={([yshift=-1.5ex]current  bounding  box.center)}, scale=0.9]
\clip (-1.6,-0.05)rectangle +(6.3,2.4);
\fill [blue!20](0,0)--(1.5,1.5)--(4.5,1.5)--(3,0)--cycle;
\draw [black!50, -stealth] (0.8,0.3)..controls (2.6,1.) and (1,1.5)..(0.8,0.35);
\begin{scope}
\clip (1.1,0.7)rectangle+(2.2,0.35);
\draw [blue!20, line width=2.7pt](1.6,2)..controls (1.7,2.5) and (3.2,2.5)..(3.2,0.75);
\draw [blue!20, line width=2.7pt](1.2,0.75)..controls(1.2,2.8)and (2,2.8)..(2,0.75);
\draw [blue!20, line width=2.7pt](2.4,0.75)..controls(2.4,2)and (1.5,1.5)..(1.6,2);
\end{scope}
\draw [thick](1.6,2)..controls (1.7,2.5) and (3.2,2.5)..(3.2,0.75);
\begin{scope}
\clip (1.6,1.6)rectangle+(0.6,0.8);
\draw [white, line width=2.7pt](1.2,0.75)..controls(1.2,2.8)and (2,2.8)..(2,0.75);
\end{scope}
\begin{scope}
\clip (1.9,0.75)rectangle+(0.3,0.4);
\draw [blue!20, line width=2.7pt](1.2,0.75)..controls(1.2,2.8)and (2,2.8)..(2,0.75);
\end{scope}
\draw [thick](1.2,0.75)..controls(1.2,2.8)and (2,2.8)..(2,0.75);
\begin{scope}
\clip (1.7,1.6)rectangle + (0.6,0.6);
\draw [white, line width=2.7pt](2.4,0.75)..controls(2.4,2)and (1.5,1.5)..(1.6,2);
\end{scope}
\draw [thick](2.4,0.75)..controls(2.4,2)and (1.5,1.5)..(1.6,2);
\draw [thick, -stealth] (1.215,1.2)--(1.22,1.27);
\draw [thick, -stealth] (2.345,1.2)--(2.33,1.27);
\fill (0.8,0.3) circle (0.04);
\node at (0.55,0.3)[scale=0.9 ]{$\mathcal F$};
\node at (1.15,2.1){$\alpha$};
\node at (3,2.1){$\sigma$};
\node at (-0.9, 1) {$W_3\ =$};
\end{tikzpicture}
\end{align}
With these definitions, we can see that
\begin{align}
W & = W_3W_1^{-1}W_2W_1 \label{eq_app_xil_5}
\end{align}
Next we observe that $W_1$ and $W_2$ obey the commutation relation
\begin{equation}
W_2^{-1}W_1^{-1}W_2W_1 =e^{-i \Omega_{li\mu}} \hat{\rm I}  \label{eq_app_xil_6}
\end{equation}
This relation follows from the definition of $\Omega_{li\mu}$, where the minus sign comes from the fact that $\mathcal F$ is braided around the ``incoming-flux'' end of $\alpha$, instead of the ``outgoing-flux'' end (see Fig.~\ref{fig_trajectories}).

In addition, we have the relation
\begin{equation}
W_3W_2 |\psi\rangle = |\psi\rangle \label{eq_app_xil_7}
\end{equation}
To see this, note that $W_3W_2$ describes a process in which $\mathcal F$ is braided around both ends of $\alpha$ in the configuration shown in Fig.~\ref{fig_xdef}(b). This braiding must give
a trivial Berry phase since we can deform the vortex arches from Fig.~\ref{fig_xdef}(b) back to Fig.~\ref{fig_xdef}(a) by fusing together the ends of $\alpha$ and lifting $\alpha$ off the surface, and this lifting process commutes with the braiding of $\mathcal F$.

Combining equations \ref{eq_app_xil_5}-\ref{eq_app_xil_7}, we derive
\begin{align*}
W| \psi\> = (W_3 W_2)  (W_2^{-1}W_1^{-1}W_2W_1) |\psi\> = e^{-i \Omega_{li\mu}} |\psi\>
\end{align*}
Eq.~\ref{Womegaapp} now follows immediately from the fact that $\Omega_{li\mu} = -\Omega_{il\mu}$.

\section{Deriving the formula (\ref{formula3})}
\label{sec:derive_formula3}

In this appendix, we derive the bulk-boundary formula (\ref{formula3}). The derivation closely follows that of (\ref{formula2}).

To begin, let us recall the braiding process that defines $\Theta_{ijk,l}$: first we braid a loop $\alpha$ around a loop $\beta$, then we braid $\alpha$ around another loop $\gamma$, and then finally we braid $\alpha$ around $\beta$ and $\gamma$ in the opposite direction. Here, $\alpha, \beta, \gamma$ are linked with a base loop $\sigma$ and the four loops carry unit flux of type $i,j,k,l$ respectively, i.e. $\phi_\alpha = \frac{2\pi}{N_i}\mathbf e_i$, $\phi_\beta = \frac{2\pi}{N_j}\mathbf e_j$, $\phi_\gamma = \frac{2\pi}{N_k}\mathbf e_k $ and  $\phi_\sigma = \frac{2\pi}{N_l}\mathbf e_l$.

The first step of the derivation is to deform the above process onto the surface using the same procedure as in Fig. \ref{fig_deformation}. After this deformation, $\alpha, \beta, \gamma, \sigma$ become vortex arches, and the braiding process involves braiding the arch $\alpha$ around the arch $\beta$, and then braiding $\alpha$ around $\gamma$, and then finally braiding $\alpha$ around $\beta$ and $\gamma$ in the opposite direction. By the same arguments as in section \ref{sec:derive_formula2_deformation}, the Berry phase for this new process must be the same as for the old one, that is, it must be equal to $\Theta_{ijk,l}$.

In the next step, we split $\alpha$ into two excitations: a surface anyon $\mathcal X$ and a vortex arch $\tilde \alpha$. Similarly, we split $\beta$ into an anyon $\mathcal Y$ and an arch $\tilde \beta$, and we split $\gamma$ into an anyon $\mathcal Z$ and an arch $\tilde \gamma$. Similarly to section \ref{sec:derive_formula2_splitting}, we choose $\mathcal X$, $\mathcal Y$ and $\mathcal Z$ to be any anyons carrying anyonic flux
\begin{equation}
\xi_\mathcal X = x_{il}, \quad  \xi_\mathcal Y = x_{jl}, \quad  \xi_\mathcal Z = x_{kl}
\end{equation}
while we choose $\tilde \alpha$, $\tilde \beta$, and $\tilde \gamma$ to be any arches with the property that $\alpha$ can be written as a fusion product of $\mathcal X$ and $\tilde \alpha$,
and similarly for $\tilde \beta, \tilde \gamma$. This splitting is designed so that $\tilde \alpha, \tilde \beta, \tilde \gamma$ have a special property, namely if we shrink these arches to the surface, we get a superposition of charge excitations rather than more complicated surface anyons.

At this point our braiding process involves braiding the pair $\{\tilde \alpha, \mathcal X\}$ around $\{\tilde \beta, \mathcal Y\}$, and then
around $\{\tilde \gamma, \mathcal Z\}$, and then finally
around $\{\tilde \beta, \mathcal Y\}$ and $\{\tilde \gamma, \mathcal Z\}$ in the opposite direction. Following the same notation as in section \ref{sec:derive_formula2_decomposing}, let $|\psi\>$ denote the initial state at the beginning of the braiding, and let $U$ denote the braid matrix associated with this braiding. Then, in this notation, the fact that the Berry phase for this process is $\Theta_{ijk,l}$ translates to the equation
\begin{equation}
U |\psi\> = e^{i \Theta_{ijk,l}} |\psi\> \label{eq_devnew1}
\end{equation}
Next, we write $U$ as
\begin{equation}
U = U_{2}^{-1} U_{1}^{-1} U_{2} U_{1} \label{eq_devnew2}
\end{equation}
where $U_1, U_2$ are the braid matrices corresponding to the following processes:
\begin{align}
\begin{tikzpicture}[baseline={([yshift=-.5ex]current  bounding  box.center)}]
\clip (-1.9,-0.75)rectangle+(7.5,1.5);
\draw [-stealth](0,0)..controls (0.5, 0.15) .. (0.7,0.15)..controls (1.2,0.15)  and (3,-1.4).. (3,0)..controls (3,1.2) and (0.9,0.5) .. (0.05,0.1);
\draw [white, line width=2.2 pt](0.7,0)..controls (1.2,-0.5)  and (3.3,-1.3).. (3.3,0)..controls (3.3,1.6) and (1.2,-0.1).. (0.8,-0.05);
\draw [-stealth](0.7,0)..controls (1.2,-0.5)  and (3.3,-1.3).. (3.3,0)..controls (3.3,1.6) and (1.2,-0.1).. (0.8,0.0);
\fill (0,0) circle (0.05);
\fill (0.7,0) circle (0.05);
\fill (2,0) circle (0.05);
\fill (2.7,0) circle (0.05);
\fill (4,0) circle (0.05);
\fill (4.7,0) circle (0.05);
\node at (-1,0){$U_{1}=$};
\node at (0,-0.28){$\tilde\alpha$};
\node at (2,-0.28){$\tilde\beta$};
\node at (4,-0.28){$\tilde\gamma$};
\node at (0.7,-0.28){$\mathcal X$};
\node at (2.7,-0.28){$\mathcal Y$};
\node at (4.7,-0.28){$\mathcal Z$};
\end{tikzpicture}\label{eq_dev23}\\
\begin{tikzpicture}[baseline={([yshift=-.5ex]current  bounding  box.center)}]
\clip (-1.9,-1.2)rectangle+(7.5,1.95);
\draw [-stealth](0,0)..controls (1,0.3) and (1.8,0.5).. (2.5, 0.5)..controls(3.5,0.5) and (3.5,-0.3)..(4, -0.5)..controls (4.5,-0.7) and(5,-0.5).. (5,0)..controls (5,1.3) and (1.2,0.5) .. (0.07,0.1);
\draw [white, line width=2.2 pt](0.7,0)..controls (1.6,-0.1) and (1.7,0.2).. (2.5,0.2).. controls (3.5,0.2) and(3.5,-0.6) ..(4,-0.8).. controls(4.5,-1)  and (5.3,-0.7).. (5.3,0)..controls (5.3,1.6) and (2,0).. (0.8,0.1);
\draw [-stealth](0.7,0)..controls (1.6,-0.1) and (1.7,0.2).. (2.5,0.2).. controls (3.5,0.2) and(3.5,-0.6) ..(4,-0.8).. controls(4.5,-1)  and (5.3,-0.7).. (5.3,0)..controls (5.3,1.6) and (2,0).. (0.8,0.1);
\fill (0,0) circle (0.05);
\fill (0.7,0) circle (0.05);
\fill (2,0) circle (0.05);
\fill (2.7,0) circle (0.05);
\fill (4,0) circle (0.05);
\fill (4.7,0) circle (0.05);
\node at (-1,0){$U_{2}=$};
\node at (0,-0.28){$\tilde\alpha$};
\node at (2,-0.28){$\tilde\beta$};
\node at (4,-0.28){$\tilde\gamma$};
\node at (0.7,-0.28){$\mathcal X$};
\node at (2.7,-0.28){$\mathcal Y$};
\node at (4.7,-0.28){$\mathcal Z$};
\end{tikzpicture}\label{eq_dev24}
\end{align}
We then further decompose $U_1, U_2$ into simpler processes. In particular, define
\begin{align}
\begin{tikzpicture}[baseline={([yshift=-.5ex]current  bounding  box.center)}]
\clip (-1.9,-0.7)rectangle+(7.5,1.35);
\fill (0.7,0) circle (0.05);
\fill (2,0) circle (0.05);
\fill (2.7,0) circle (0.05);
\draw [-stealth](0,0)..controls (0.3,0) and (0.9,0.2).. (1.3,0.2)..controls(1.6,0.2) and (1.8,-0.5).. (2,-0.5)..controls(2.6,-0.5) and (2.6,0.5)..(2,0.5)..controls(1.3,0.5) and (0.7,0.1).. (0.05,0.07);
\fill (0,0) circle (0.05);
\node at (-1,0.3)[anchor=north]{$U_{\tilde\alpha\tilde\beta} =$};
\node at (0,-0.28){$\tilde\alpha$};
\node at (2,-0.28){$\tilde\beta$};
\node at (0.7,-0.28){$\mathcal X$};
\node at (2.7,-0.28){$\mathcal Y$};
\fill (4,0) circle (0.05);
\fill (4.7,0) circle (0.05);
\node at (4,-0.28){$\tilde\gamma$};
\node at (4.7,-0.28){$\mathcal Z$};
\end{tikzpicture}\\
\begin{tikzpicture}[baseline={([yshift=-.5ex]current  bounding  box.center)}]
\clip (-1.9,-0.7)rectangle+(7.5,1.35);
\fill (0.7,0) circle (0.05);
\fill (2,0) circle (0.05);
\fill (2.7,0) circle (0.05);
\draw [-stealth](0,0)..controls (0.3,0.2) and (1.6,0.2).. (2,0.2)..controls(2.3,0.2) and (2.5,-0.5).. (2.7,-0.5)..controls(3.3,-0.5) and (3.3,0.5)..(2.7,0.5)..controls(2,0.5) and (0.7,0.3).. (0.05,0.1);
\fill (0,0) circle (0.05);
\node at (-1,0.3)[anchor=north]{$U_{\tilde\alpha\mathcal Y}=$};
\node at (0,-0.28){$\tilde\alpha$};
\node at (2,-0.28){$\tilde\beta$};
\node at (0.7,-0.28){$\mathcal X$};
\node at (2.7,-0.28){$\mathcal Y$};
\fill (4,0) circle (0.05);
\fill (4.7,0) circle (0.05);
\node at (4,-0.28){$\tilde\gamma$};
\node at (4.7,-0.28){$\mathcal Z$};
\end{tikzpicture}\\
\begin{tikzpicture}[baseline={([yshift=-.5ex]current  bounding  box.center)}]
\clip (-1.9,-0.7)rectangle+(7.5,1.35);
\fill (0.7,0) circle (0.05);
\fill (2,0) circle (0.05);
\fill (2.7,0) circle (0.05);
\draw [-stealth](0.7,0)..controls (1.2,0.2) and (1.5, -0.5)..(2,-0.5)..controls(2.2,-0.5) and (2.3,-0.3).. (2.3,0)..controls (2.3,0.6) and (1.2,0.2).. (0.75,0.08);
\fill (0,0) circle (0.05);
\node at (-1,0.3)[anchor=north]{$U_{\mathcal X \tilde \beta} =$};
\node at (0,-0.28){$\tilde\alpha$};
\node at (2,-0.28){$\tilde\beta$};
\node at (0.7,-0.28){$\mathcal X$};
\node at (2.7,-0.28){$\mathcal Y$};
\fill (4,0) circle (0.05);
\fill (4.7,0) circle (0.05);
\node at (4,-0.28){$\tilde\gamma$};
\node at (4.7,-0.28){$\mathcal Z$};
\end{tikzpicture}\\
\begin{tikzpicture}[baseline={([yshift=-.5ex]current  bounding  box.center)}]
\clip (-1.9,-0.7)rectangle+(7.5,1.35);
\fill (0.7,0) circle (0.05);
\fill (2,0) circle (0.05);
\fill (2.7,0) circle (0.05);
\draw [-stealth](0.7,0)..controls (1,0) and (1.6,0.2).. (2,0.2)..controls(2.3,0.2) and (2.5,-0.5).. (2.7,-0.5)..controls(3.3,-0.5) and (3.3,0.5)..(2.7,0.5)..controls(2,0.5) and (1.4,0.1).. (0.75,0.07);
\fill (0,0) circle (0.05);
\node at (-1,0.3)[anchor=north]{$U_{\mathcal X\mathcal Y} =$};
\node at (0,-0.28){$\tilde\alpha$};
\node at (2,-0.28){$\tilde\beta$};
\node at (0.7,-0.28){$\mathcal X$};
\node at (2.7,-0.28){$\mathcal Y$};
\fill (4,0) circle (0.05);
\fill (4.7,0) circle (0.05);
\node at (4,-0.28){$\tilde\gamma$};
\node at (4.7,-0.28){$\mathcal Z$};
\end{tikzpicture}\\
\begin{tikzpicture}[baseline={([yshift=-.5ex]current  bounding  box.center)}]
\clip (-1.9,-0.8)rectangle+(7.5,1.35);
\fill (0.7,0) circle (0.05);
\fill (2,0) circle (0.05);
\fill (2.7,0) circle (0.05);
\draw [-stealth](0,0)..controls (1,0.2) and (2.5,0.3).. (3.1,0.3)..controls(3.8,0.3) and (3.8,-0.5).. (4,-0.5)..controls(4.6,-0.5) and (4.6,0.5)..(4,0.5)..controls(3.3,0.5) and (2,0.4).. (0.05,0.07);
\fill (0,0) circle (0.05);
\node at (-1,0.3)[anchor=north]{$U_{\tilde\alpha\tilde\gamma} =$};
\node at (0,-0.28){$\tilde\alpha$};
\node at (2,-0.28){$\tilde\beta$};
\node at (0.7,-0.28){$\mathcal X$};
\node at (2.7,-0.28){$\mathcal Y$};
\fill (4,0) circle (0.05);
\fill (4.7,0) circle (0.05);
\node at (4,-0.28){$\tilde\gamma$};
\node at (4.7,-0.28){$\mathcal Z$};
\end{tikzpicture}\\
\begin{tikzpicture}[baseline={([yshift=-.5ex]current  bounding  box.center)}]
\clip (-1.9,-0.8)rectangle+(7.5,1.35);
\fill (0.7,0) circle (0.05);
\fill (2,0) circle (0.05);
\fill (2.7,0) circle (0.05);
\draw [-stealth](0,0)..controls (1,0.2) and (2.5,0.3).. (3.8,0.3)..controls(4.5,0.3) and (4.5,-0.5).. (4.7,-0.5)..controls(5.3,-0.5) and (5.3,0.5)..(4.7,0.5)..controls(3.3,0.5) and (2,0.4).. (0.05,0.07);
\fill (0,0) circle (0.05);
\node at (-1,0.3)[anchor=north]{$U_{\tilde\alpha\mathcal Z} =$};
\node at (0,-0.28){$\tilde\alpha$};
\node at (2,-0.28){$\tilde\beta$};
\node at (0.7,-0.28){$\mathcal X$};
\node at (2.7,-0.28){$\mathcal Y$};
\fill (4,0) circle (0.05);
\fill (4.7,0) circle (0.05);
\node at (4,-0.28){$\tilde\gamma$};
\node at (4.7,-0.28){$\mathcal Z$};
\end{tikzpicture}\\
\begin{tikzpicture}[baseline={([yshift=-.5ex]current  bounding  box.center)}]
\clip (-1.9,-0.8)rectangle+(7.5,1.35);
\fill (0.7,0) circle (0.05);
\fill (2,0) circle (0.05);
\fill (2.7,0) circle (0.05);
\draw [-stealth](0.7,0)..controls (1.5,0.2) and (2.5,0.3).. (3.1,0.3)..controls(3.8,0.3) and (3.8,-0.5).. (4,-0.5)..controls(4.6,-0.5) and (4.6,0.5)..(4,0.5)..controls(3.3,0.5) and (2,0.4).. (0.75,0.07);
\fill (0,0) circle (0.05);
\node at (-1,0.3)[anchor=north]{$U_{\mathcal X\tilde\gamma} =$};
\node at (0,-0.28){$\tilde\alpha$};
\node at (2,-0.28){$\tilde\beta$};
\node at (0.7,-0.28){$\mathcal X$};
\node at (2.7,-0.28){$\mathcal Y$};
\fill (4,0) circle (0.05);
\fill (4.7,0) circle (0.05);
\node at (4,-0.28){$\tilde\gamma$};
\node at (4.7,-0.28){$\mathcal Z$};
\end{tikzpicture}\\
\begin{tikzpicture}[baseline={([yshift=-.5ex]current  bounding  box.center)}]
\clip (-1.9,-0.8)rectangle+(7.5,1.35);
\fill (0.7,0) circle (0.05);
\fill (2,0) circle (0.05);
\fill (2.7,0) circle (0.05);
\draw [-stealth](0.7,0)..controls (1.5,0.2) and (2.5,0.3).. (3.8,0.3)..controls(4.5,0.3) and (4.5,-0.5).. (4.7,-0.5)..controls(5.3,-0.5) and (5.3,0.5)..(4.7,0.5)..controls(3.3,0.5) and (2,0.4).. (0.75,0.07);
\fill (0,0) circle (0.05);
\node at (-1,0.3)[anchor=north]{$U_{\mathcal X\mathcal Z} =$};
\node at (0,-0.28){$\tilde\alpha$};
\node at (2,-0.28){$\tilde\beta$};
\node at (0.7,-0.28){$\mathcal X$};
\node at (2.7,-0.28){$\mathcal Y$};
\fill (4,0) circle (0.05);
\fill (4.7,0) circle (0.05);
\node at (4,-0.28){$\tilde\gamma$};
\node at (4.7,-0.28){$\mathcal Z$};
\end{tikzpicture}
\end{align}
With these definitions, it is easy to see that
\begin{align}
U_{1} &=U_{\mathcal X\mathcal Y}U_{\mathcal X\tilde\beta}U_{\tilde\alpha\mathcal Y} U_{\tilde\alpha\tilde\beta} \label{eq_dev25}\\
U_{2} &=U_{\mathcal X\mathcal Z}U_{\mathcal X\tilde\gamma}U_{\tilde\alpha\mathcal Z} U_{\tilde\alpha\tilde\gamma} \label{eq_dev26}
\end{align}
Combining Eqs. (\ref{eq_devnew1}) and (\ref{eq_devnew2}), we conclude that
\begin{align}
e^{i\Theta_{ijk,l}}|\psi\> = &(U_{\tilde\alpha\tilde\gamma}^{-1}U_{\tilde\alpha\mathcal Z}^{-1}U_{\mathcal X\tilde\gamma} ^{-1}U_{\mathcal X\mathcal Z}^{-1})(U_{\tilde\alpha\tilde\beta}^{-1}U_{\tilde\alpha\mathcal Y}^{-1} U_{\mathcal X\tilde\beta}^{-1}U_{\mathcal X\mathcal Y}^{-1})\nonumber \\
& \times (U_{\mathcal X\mathcal Z}U_{\mathcal X\tilde\gamma}U_{\tilde\alpha\mathcal Z} U_{\tilde\alpha\tilde\gamma} ) (U_{\mathcal X\mathcal Y}U_{\mathcal X\tilde\beta}U_{\tilde\alpha\mathcal Y} U_{\tilde\alpha\tilde\beta})|\psi\>  \label{eq_dev28}
\end{align}

The last step is to evaluate the expression on the right hand side of (\ref{eq_dev28}). Our strategy is straightforward: since every operator $U$ appears along with its inverse $U^{-1}$, we use the commutation relations satisfied by these braid matrices to bring each $U$ and $U^{-1}$ next to one another and then cancel each pair. These commutation relations are as follows:
\begin{align}
U_{\mathcal X\tilde\gamma} U_{\mathcal X\tilde\beta} & = e^{i \eta_1} U_{\mathcal X\tilde\beta} U_{\mathcal X\tilde\gamma}, \quad
U_{\tilde\alpha\tilde\gamma} U_{\tilde\alpha\mathcal Y} = e^{i \eta_2} U_{\tilde\alpha\mathcal Y}  U_{\tilde\alpha\tilde\gamma}\nonumber \\
U_{\tilde\alpha\mathcal Z}   U_{\tilde\alpha\tilde\beta}   & = e^{i \eta_3} U_{\tilde\alpha\tilde\beta} U_{\tilde\alpha\mathcal Z}, \quad
U_{\mathcal X \tilde \beta} U_{\tilde\alpha\tilde\beta}     = e^{i \eta_4} U_{\tilde\alpha\tilde\beta} U_{\mathcal X \tilde \beta} \nonumber \\
U_{\tilde \alpha \tilde\beta} U_{\tilde \alpha \mathcal Y} & = e^{i \eta_5} U_{\tilde \alpha \mathcal Y} U_{\tilde \alpha \tilde\beta}, \quad
U_{\tilde\alpha\tilde \gamma}   U_{\tilde\alpha \mathcal Z}    = e^{i \eta_6} U_{\tilde\alpha \mathcal Z} U_{\tilde\alpha\tilde \gamma} \label{eq_dev33}
\end{align}
where
\begin{align}
\eta_1 &= \sum_\mu  x_{il}^\mu \Omega_{jk\mu}, \quad
\eta_2 = \sum_\mu x_{jl}^\mu \Omega_{ki\mu}  \nonumber \\
\eta_3 &= \sum_\mu x_{kl}^\mu \Omega_{ij\mu}, \quad
\eta_4 = \sum_\mu x_{il}^\mu \Omega_{ji\mu}, \nonumber \\
\eta_5 &= \sum_\mu x_{jl}^\mu \Omega_{ji\mu}, \quad
\eta_6 = \sum_\mu x_{kl}^\mu \Omega_{ki\mu}
\label{eq_dev36}
\end{align}
All other pairs of the $U$'s commute with one another, with the only exception being $(U_{\tilde\alpha\tilde\gamma}, U_{\tilde\alpha\tilde\beta})$: this pair satisfies the weaker commutation relation
\begin{align}
U_{\tilde\alpha\tilde\gamma} U_{\tilde\alpha\tilde\beta} |\psi\>   = U_{\tilde\alpha\tilde\beta} U_{\tilde\alpha\tilde\gamma} |\psi\>  \label{eq_dev30}
\end{align}
The derivations of Eqs. \ref{eq_dev33} are similar to those of Eqs.~(\ref{eq_dev8})-(\ref{eq_dev9}), which are given in 
Appendix \ref{sec:app_formula}. Likewise, the derivation of Eq. \ref{eq_dev30} is similar to that of Eq. \ref{eq_dev14}.

Let us now use these commutation relations to derive the bulk-boundary formula. Inserting these relations into (\ref{eq_dev28}), and commuting through the different $U$ matrices, we obtain
\begin{equation}
\Theta_{ijk,l} = \eta_1+\eta_2+\eta_3
\end{equation}
If we now examine the definitions of $\eta_1,\eta_2, \eta_3$, we see that this is precisely the bulk-boundary formula (\ref{formula3}).

Before concluding, it is worth pointing out that the relations in Eq. (\ref{eq_dev33}) are not as complicated as they appear. In particular, these relations all share a common structure: they all involve pairs $(U_{XY}, U_{X'Y'})$ in which (1) three of the four indices $\{X,Y,X',Y'\}$ are distinct, (2) two of the three distinct indices are vortices $\tilde \alpha, \tilde \beta, \tilde \gamma$, and (3) the remaining distinct index is a surface anyon $\mathcal X, \mathcal Y, \mathcal Z$.

\section{Deriving the formula (\ref{formula1})}
\label{sec:derive_formula1}

In this appendix, we derive the bulk-boundary formula (\ref{formula1}) for $\Theta_{i,l}$. For the purposes of our derivation, it is helpful to separate the cases where $N_i$ is odd and $N_i$ is even.

The case where $N_i$ is odd is very simple: in this case, we can express $\Theta_{i,l}$ in terms of $\Theta_{ii,l}$ using the relation (\ref{thetailodd}). Making use of the bulk-boundary formula for $\Theta_{ii,l}$ (\ref{formula2}) together with the constraints (\ref{phi_constraint}) and (\ref{x_constraint1}), the required formula (\ref{formula1}) follows immediately.

Now consider the case where $N_i$ is even. To derive (\ref{formula1}) in this case, we recall the alternative definition of $\Theta_{i,l}$ discussed at the end of section \ref{sec:bulkinv_def}: $\Theta_{i,l}$ is equal to $(-1)$ times the Berry phase associated with braiding a vortex loop $\alpha$ around its anti-vortex $\bar\alpha$ for $\frac{N_i}{2}$ times while both $\alpha$ and $\bar\alpha$ are linked to another vortex loop $\sigma$. Here $\alpha$ and $\sigma$ carry unit type-$i$ and type-$l$ flux respectively, i.e.
$\phi_\alpha = \frac{2\pi}{N_i}\mathbf e_i$, $\phi_\sigma = \frac{2\pi}{N_l}\mathbf e_l$. (Note that this alternative definition is only applicable in the case where $N_i$ is even).

The first step of the derivation is to deform the above braiding process onto the surface following the same procedure as in Fig. \ref{fig_deformation}. After the deformation, the vortex loops $\alpha, \bar \alpha$ become vortex arches, and the deformed braiding process involves braiding the arch $\alpha$ around its ``anti-arch'' $\bar{\alpha}$ for $\frac{N_i}{2}$ times. \footnote{One subtlety is that, depending on how the deformation is performed, the arch obtained from the loop {$\bar \alpha$} may or may not be the ``anti-arch'' of that obtained from {$\alpha$}. However, we can always design our deformation so that this property holds.} By the same arguments as in section \ref{sec:derive_formula2_deformation}, the Berry phase for this process must be the same as the original process, i.e. it must be equal to $-\Theta_{i,l}$.

Let us translate this statement into more formal language. Following the notation of section \ref{sec:derive_formula2_decomposing}, let $|\psi\>$ be the initial state at the beginning of the braiding, and let $V$ be the braid matrix associated with braiding $\alpha$ around $\bar \alpha$ \emph{once}. Then, the fact that the Berry phase for our process is $-\Theta_{i,l}$ translates into the equation
\begin{equation}
V^{N_i/2}|\psi\> = e^{-i \Theta_{i,l}} |\psi\>
\label{Veq}
\end{equation}

To proceed further, we imagine folding the surface and straightening the vortex lines as shown in Fig.~\ref{fig_straighten}. After doing this, we then view the folded surface as a 2D system --- a 2D slab. This point of view is convenient because it allows us to think about the braiding process as one involving \emph{2D anyons} so we can make use of the powerful tools that have been developed for 2D systems. In particular, using these tools, one can show that (see Appendix \ref{sec:app_topospin})
\begin{equation}
V |\psi\> = e^{-4\pi i s_\alpha} |\psi\> \label{eq_dev59}
\end{equation}
where $s_\alpha$ is the topological spin of $\alpha$ when viewed as a 2D anyon.
Putting together Eqs. (\ref{Veq}) and (\ref{eq_dev59}), we derive
\begin{align}
e^{i\Theta_{i,l}} = e^{2\pi i N_i s_\alpha} \label{eq_dev60}
\end{align}
Thus, our problem reduces to computing the topological spin $s_{\alpha}$. To do this, we split $\alpha$ into two excitations: a surface anyon $\mathcal X$ and a vortex arch $\tilde \alpha$. We choose $\mathcal X$ to be any surface anyon that carries anyonic flux $\xi_\mathcal X = x_{il}$, and we choose $\tilde \alpha$ to be any arch such that $\alpha$ can be written as a fusion product of $\tilde \alpha$ and $\mathcal X$. As in the previous sections, this splitting is designed so that $\tilde \alpha$ has a special property: if we shrink $\tilde \alpha$ to the surface, we get a superposition of different charge excitations (instead of more complicated surface anyons).

After this splitting, we apply the following result from 2D anyon theory:
\begin{equation}
R_{\mathcal X\tilde\alpha}^{\alpha}R_{\tilde\alpha\mathcal X}^{\alpha} = e^{2\pi i(s_{\alpha} - s_{\tilde\alpha}-s_{\mathcal X})} \hat{\rm I} \label{eq_dev61}
\end{equation}
Here $R_{\mathcal X\tilde\alpha}^{\alpha}$ is the $R$ symbol, so $R_{\mathcal X\tilde\alpha}^{\alpha}R_{\tilde\alpha\mathcal X}^{\alpha}$ describes the effect of a full braiding of $\mathcal X$ around $\tilde\alpha$ in the fusion channel $\alpha$. Comparing this relation with (\ref{eq_dev60}), we derive
\begin{equation}
e^{i\Theta_{i,l}} = e^{2\pi i N_i(s_{\tilde\alpha}+s_{\mathcal X})}  (R_{\mathcal X\tilde\alpha}^{\alpha}R_{\tilde\alpha\mathcal X}^{\alpha})^{N_i} \label{eq_dev62}
\end{equation}

Now we evaluate the terms on the right hand side of Eq. (\ref{eq_dev62}). The quantity $s_\mathcal X$ can be computed as
\begin{align}
e^{2\pi i s_{\mathcal X}} &= e^{i\theta_{\xi_\mathcal X}} \nonumber \\
&= e^{i\sum_\mu (x_{il}^\mu)^2 \Phi_{\mu}}  \cdot e^{i\sum_{\mu<\nu} x_{il}^\mu x_{il}^\nu \Phi_{\mu\nu}}  \label{eq_dev63}
\end{align}
where the first equality follows from Eq. (\ref{eq_spin}), and the second equality follows from $\xi_\mathcal X = x_{il}$ together with Eq. (\ref{exchange}). If we now raise this equation to the $N_i$th power, the second term on the right hand side drops out since $N_ix_{il}^{\mu}$ is a multiple of $N_\mu$ and $N_\mu\Phi_{\mu\nu}$ is a multiple
of $2\pi$ according to the constraints (\ref{x_constraint1}) and (\ref{phi_constraint1}). Thus we obtain
\begin{equation}
e^{2\pi i N_i s_{\mathcal X}} = e^{i\sum_\mu N_i (x_{il}^\mu)^2 \Phi_{\mu}} \label{eq_dev64}
\end{equation}
Next consider the mutual braiding statistics term
$(R_{\mathcal X\tilde\alpha}^{\alpha}R_{\tilde\alpha\mathcal X}^{\alpha})^{N_i}$. This term is given by
\begin{align}
(R_{\mathcal X\tilde\alpha}^{\alpha}R_{\tilde\alpha\mathcal X}^{\alpha})^{N_i}  & = \exp \left(i \sum_\mu \frac{N_i}{N^{i\mu}} x_{il}^\mu \Omega_{i\mu} \right)  \label{eq_dev65}
\end{align}
according to Eqs. (\ref{eq_dev11}) and (\ref{eq_dev12}) with $i=j$.

The final piece to compute is $e^{2\pi i N_i s_{\tilde\alpha}}$. We will now argue that this term is trivial:
\begin{equation}
e^{2\pi i N_i s_{\tilde\alpha}} = 1 \label{eq_dev66}
\end{equation}
To prove this, recall that $\tilde \alpha$ has a special property: if we shrink $\tilde \alpha$ down to the surface, we get a superposition of charges rather than more complicated surface anyons. Let $-q$ be one of the charges that appear in this superposition, and let $\tilde \alpha'$ be another vortex arch obtained by fusing $\tilde \alpha$ with $q$. Clearly $\tilde \alpha'$, like $\tilde \alpha$, has the property that if we shrink it down to the surface, we get a superposition of charges. Furthermore, by construction, one of the charges that appear in this superposition is $q=0$, i.e. the vacuum. This means that if we shrink $\tilde \alpha'$ down to the surface and then apply a projection operator that projects onto the $q=0$ state, then we can annihilate $\tilde \alpha'$ \emph{locally}.

Let us think about this property of $\tilde \alpha'$ in a geometry where we fold the surface and straighten $\tilde \alpha'$ as in Fig.~\ref{fig_straighten}. Then, if we view our system as two dimensional, the above property is equivalent to the statement that $\tilde \alpha'$ can be annihilated locally at the \emph{edge} of our system (where the edge runs along the folding axis).

At this point, we use a general result about 2D systems: any anyon excitation $a$ that can be annihilated locally at a gapped edge of a 2D system must have vanishing topological spin, $s_a = 0$. This result was discussed in Ref. \onlinecite{levin13} in the case where the anyon $a$ has \emph{Abelian} quasiparticle statistics, but we expect that this result holds in the non-Abelian case as well, and we will assume this in what follows.

Assuming the above result, we deduce that $s_{\tilde \alpha'} = 0$. With this identity, it is now simple to derive Eq. (\ref{eq_dev66}). First, we express $s_{\tilde \alpha}$ in terms of $s_{\tilde\alpha'}$:
\begin{equation}
e^{2\pi i s_{\tilde \alpha}} \hat{\rm I} = e^{2\pi i (s_{\tilde \alpha'} + s_{q})}
R_{\tilde \alpha' q}^{\tilde \alpha} R_{q \tilde \alpha'}^{\tilde \alpha}
\end{equation}
Next, we note that the product $R_{\tilde \alpha' q}^{\tilde \alpha} R_{q \tilde \alpha'}^{\tilde \alpha}$ describes a full braiding of $q$ around $\tilde \alpha'$ and therefore is an Aharonov-Bohm phase of the form $e^{2\pi i \text{(integer)}/N_i}$. At the same time, we have $s_q = s_{\tilde \alpha'} =0$. Making these substitutions, we conclude that $s_{\tilde \alpha}$ must be a multiple of $1/N_i$, which implies Eq. (\ref{eq_dev66}).

The derivation is now complete: substituting Eqs.~(\ref{eq_dev64}), (\ref{eq_dev65}) and (\ref{eq_dev66}) into Eq.~(\ref{eq_dev62}), we derive the formula (\ref{formula1}).


\section{Some equations from Sec.~\ref{sec:derivation}}
\label{sec:app_formula}

In this Appendix, we establish several equations that were used in the derivation of bulk-boundary correspondence.
To be specific, we prove the commutation relations (\ref{eq_dev8}), (\ref{eq_dev9}), as well as
the relations (\ref{eq_dev11}) and (\ref{eq_dev12}) which involve taking the $N^{ij}$th power of various braid matrices.


\subsection{Commutation relations}
\label{sec:app_formula1}

To avoid repetition, we will only prove the commutation relation (\ref{eq_dev8}); the derivation of the other commutation relation (\ref{eq_dev9}) follows similar arguments. We begin by rewriting Eq.~(\ref{eq_dev8}) as
\begin{equation}
W_{\tilde\alpha\tilde\beta}^{-1}W_{\mathcal X\tilde\beta}^{-1} W_{\tilde\alpha\tilde\beta}W_{\mathcal X\tilde\beta} \label{eq_app_c1} = e^{i \sum_\mu x_{il}^\mu \Omega_{ij\mu}} \hat{\rm I}
\end{equation}
Here, the braid matrices $W_{\tilde\alpha\tilde\beta}$ and $W_{\mathcal X\tilde\beta}$ are defined as in Eqs.~(\ref{w1}, \ref{w2}), while $\tilde\alpha$, $\tilde\beta$ are vortex lines carrying unit type-$i$ and type-$j$ gauge flux, and $\mathcal X$ is a surface anyon carrying anyonic flux $x_{il}$.

Next, we recall that braiding statistics is symmetric in the sense that braiding $X$ around $Y$ is topologically equivalent to braiding $Y$ around $X$. This means that $W_{\mathcal X\tilde\beta} = W_{\tilde\beta\mathcal X}$ and $W_{\tilde\alpha\tilde\beta} = W_{\tilde\beta\tilde\alpha}$. Therefore, instead of proving (\ref{eq_app_c1}), it suffices to prove
\begin{equation}
W_{\tilde\beta\tilde\alpha}^{-1}W_{\tilde\beta\mathcal X}^{-1} W_{\tilde\beta\tilde\alpha}W_{\tilde\beta\mathcal X}  = e^{i \sum_\mu x_{il}^\mu \Omega_{ij\mu}} \hat{\rm I} \label{eq_app_c3}
\end{equation}

To proceed further, we imagine splitting $\mathcal X$ into a collection of more ``elementary'' surface anyons $\{\mathcal X_{\mu}^n \}$. More specifically, we split $\mathcal X$ into $\{\mathcal X_{\mu}^n \}$ where each $\mathcal X_\mu^{n}$ carries unit type-$\mu$ anyonic flux. Here the index $\mu$ runs from $1,...,M$, while $n$ runs from $1,..., x_{il}^\mu$.

After this splitting, we decompose the process $W_{\tilde\beta\mathcal X}$ into a sequence of sub-processes in which $\tilde\beta$ is braided around each elementary anyon $\mathcal X_\mu^{n}$. Translating this into algebra, we obtain:
\begin{equation}
W_{\tilde\beta\mathcal X} = W_{\tilde\beta\mathcal X_1^{1}} \dots W_{\tilde\beta\mathcal X_M^{x}} \label{eq_app_c4}
\end{equation}
where we have used the abbreviation $x \equiv x_{il}^M$.

Given Eq.~(\ref{eq_app_c4}), our task reduces to proving a commutation relation involving the elementary anyons $\mathcal X_\mu^n$:
\begin{equation}
W_{\tilde\beta\tilde\alpha}^{-1} W_{\tilde\beta\mathcal X_\mu^{n} }^{-1} W_{\tilde\beta\tilde\alpha} W_{\tilde\beta\mathcal X_\mu^{n} } =e^{i\Omega_{ij\mu}} \hat{\rm I} \label{eq_app_c5}
\end{equation}
Indeed, it is easy to see that Eq.(\ref{eq_app_c3}) follows from Eqs. (\ref{eq_app_c4}) and (\ref{eq_app_c5}) since our decomposition $\mathcal X \rightarrow \{\mathcal X_\mu^n \}$ contains $x_{il}^\mu$ anyons carrying unit type-$\mu$ anyonic flux.

To complete the proof, we now derive (\ref{eq_app_c5}). Consider the following braiding processes:
\begin{align}
\begin{tikzpicture}[baseline={([yshift=-.5ex]current  bounding  box.center)}]
\clip(-1.9,-0.5)rectangle+(4.5,1);
\fill (0,0) circle (0.05);
\fill (1,0) circle (0.05);
\fill (2,0) circle (0.05);
\draw [-stealth](0,0)..controls (0.5,-0.5) and (1.5,-0.7).. (1.5,0)..controls (1.5,0.7) and (0.5,0.5).. (0.05,0.05);
\node at (-0.3,0)[anchor=east]{$W_{\tilde\alpha\tilde\beta} =$};
\node at (0,-0.25){$\tilde\alpha$};
\node at (1,-0.25){$\tilde\beta$};
\node at (2,-0.27){$\mathcal X_\mu^{n}$};
\end{tikzpicture}\nonumber\\
\begin{tikzpicture}[baseline={([yshift=-.5ex]current  bounding  box.center)}]
\clip(-1.9,-0.65)rectangle+(4.5,1.2);
\fill (0,0) circle (0.05);
\fill (1,0) circle (0.05);
\fill (2,0) circle (0.05);
\draw [-stealth](0,0)..controls (1.4,0.65) and (1.2,-0.6)..(2,-0.5)..controls (2.5,-0.4) and (2.5,0.4) ..  (2,0.4)..controls (1.6,0.4) and (0.5,0.3) .. (0.05,0.09);
\node at (-0.3,0)[anchor=east]{$W_{\tilde\alpha\mathcal X_\mu^{n}} =$};
\node at (0,-0.25){$\tilde \alpha$};
\node at (1,-0.25){$\tilde\beta$};
\node at (2,-0.27){$\mathcal X_\mu^{n}$};
\end{tikzpicture}\nonumber\\
\begin{tikzpicture}[baseline={([yshift=-.5ex]current  bounding  box.center)}]
\clip(-1.9,-0.65)rectangle+(4.5,1.2);
\fill (0,0) circle (0.05);
\fill (1,0) circle (0.05);
\fill (2,0) circle (0.05);
\draw [-stealth](1,0)..controls (1.5,-0.5) and (2.5,-0.7).. (2.5,0)..controls (2.5,0.7) and (1.5,0.5).. (1.05,0.05);
\node at (-0.3,0)[anchor=east]{$W_{\tilde\beta\mathcal X_\mu^{n}} =$};
\node at (0,-0.25){$\tilde\alpha$};
\node at (1,-0.25){$\tilde\beta$};
\node at (2,-0.27){$\mathcal X_\mu^{n}$};
\end{tikzpicture}\nonumber
\end{align}
Now consider the product $W_{\tilde\alpha\mathcal X_{\mu}^{n}} W_{\tilde\alpha\tilde\beta}$. We can see that this product corresponds to a process in which $\tilde\alpha$ is braided around $\tilde\beta$ and $\mathcal X_\mu^{n}$ as a whole. It follows that this product commutes with $W_{\tilde\beta\mathcal X_\mu^{n}}$, since the two braiding processes do not overlap. Therefore, we derive:
\begin{equation}
(W_{\tilde\alpha\mathcal X_\mu^{n}}W_{\tilde\alpha\tilde\beta}) W_{\tilde\beta\mathcal X_\mu^{n}} =
W_{\tilde\beta\mathcal X_\mu^{n}} (W_{\tilde\alpha\mathcal X_\mu^{n}}W_{\tilde\alpha\tilde\beta}) \label{eq_app_c6}
\end{equation}
At the same time, we have the relation
\begin{equation}
W_{\tilde\beta\mathcal X_\mu^{n}}^{-1}W_{\tilde\alpha\mathcal X_\mu^{n}}^{-1}W_{\tilde\beta\mathcal X_\mu^{n}} W_{\tilde\alpha\mathcal X_\mu^{n}}  = e^{i\Omega_{ij\mu}} \hat{\rm I}\label{eq_app_c7}
\end{equation}
which follows from the definition of $\Omega_{ij\mu}$ and the property $W_{XY} = W_{YX}$
for any $X$ and $Y$. Combining (\ref{eq_app_c6}), (\ref{eq_app_c7}) and using $W_{\tilde\alpha\tilde\beta} = W_{\tilde\beta\tilde\alpha}$, straightforward algebra gives
equation (\ref{eq_app_c5}).


\subsection{Relations with \texorpdfstring{$N^{ij}$}{Nij}th power}
\label{sec:app_formula2}

We now derive the relation (\ref{eq_dev11}), which we reproduce below for convenience:
\begin{equation}
\left(W_{\mathcal X\tilde\beta} \right)^{N^{ij}} = \exp \left(i \sum_\mu \frac{N^{ij}}{N^{ju}} x_{il}^\mu \Omega_{ju} \right) \hat{\rm I} \label{eq_app_c20}
\end{equation}
The other relation (\ref{eq_dev12}) can be obtained in the same way, so we will not discuss it here.

To begin, we recall that braiding is symmetric so it suffices to show that
\begin{equation}
\left(W_{\tilde\beta\mathcal X} \right)^{N^{ij}} = \exp \left(i \sum_\mu \frac{N^{ij}}{N^{ju}} x_{il}^\mu \Omega_{ju} \right) \hat{\rm I} \label{eq_app_c22}
\end{equation}
Next, we split $\mathcal X$ into elementary anyons $\{\mathcal X_\mu^{n}\}$, as in the previous section. As before, each elementary anyon $\mathcal X_\mu^{n}$ carries unit type-$\mu$ anyonic flux and the index $\mu$ runs from $1,...,M$ while $n$ runs from $1,...,x_{il}^\mu$.

We then decompose the process of braiding $\tilde\beta$ around $\mathcal X$ into a sequence of sub-processes in which $\tilde\beta$ is braided around each $\mathcal X_\mu^{n}$. Translating this into algebra gives
\begin{equation}
W_{\tilde\beta\mathcal X} = W_{\tilde\beta\mathcal X_1^{1}} \dots W_{\tilde\beta\mathcal X_M^{x}} \label{eq_app_c221}
\end{equation}
where we have used the abbreviation $x \equiv x_{il}^M$.

Next we claim that $W_{\tilde\beta\mathcal X_\mu^{n}}$ commutes with $W_{\tilde\beta\mathcal X_\nu^{m}}$ for arbitrary $n,m,\mu,\nu$. To see this, note that the following identity holds by the same reasoning as Eq.~\ref{eq_app_c6}:
\begin{equation*}
(W_{\mathcal X_\nu^{m}\tilde\beta} W_{\mathcal X_\nu^{m}\mathcal X_\mu^{n}})   W_{\mathcal X_\mu^{n}\tilde\beta}  =
W_{\mathcal X_\mu^{n} \tilde\beta} (W_{\mathcal X_\nu^{m}\tilde\beta} W_{\mathcal X_\nu^{m}\mathcal X_\mu^{n}})
\end{equation*}
Furthermore, the mutual statistics of $\mathcal X_\mu^{n}$ and $\mathcal X_\nu^{m}$ is Abelian (see Eq. (\ref{eq_AB_stat})), so $W_{\mathcal X_\nu^{m}\mathcal X_\mu^{n} }$ is proportional to the identity operator. It follows that $W_{\mathcal X_\nu^{m}\tilde\beta} W_{\mathcal X_\mu^{n}\tilde\beta}  =  W_{\mathcal X_\mu^{n} \tilde\beta} W_{\mathcal X_\nu^{m}\tilde\beta}$ so $W_{\tilde\beta\mathcal X_\mu^{n}}$ and $W_{\tilde\beta\mathcal X_\nu^{m}}$ commute.

Using the above commutativity property together with Eq. \ref{eq_app_c221}, we derive the relation
\begin{equation}
W_{\tilde\beta\mathcal X}^{N_{ij}} =W_{\tilde\beta\mathcal X_1^{1}}^{N_{ij}} \dots W_{\tilde\beta\mathcal X_M^{x}}^{N_{ij}} \label{eq_app_c23}
\end{equation}
To proceed further, we make use of the following result:
\begin{equation}
W_{\tilde\beta\mathcal X_\mu^{n}}^{N_{j}} = e^{i \vartheta_{\tilde\beta \mu}} \hat{\rm I}
\label{WNj}
\end{equation}
where $\vartheta_{\tilde\beta \mu}$ is some phase that depends only on $\tilde \beta$ and $\mu$ and is \emph{independent} of $n$.

Before establishing Eq.~\ref{WNj}, we now show that it implies our claim, Eq.~\ref{eq_app_c22}. To see this, we insert Eq.~\ref{WNj} into Eq.~\ref{eq_app_c23}. Then using the fact that our decomposition $\mathcal X \rightarrow \{A_\mu^n\}$ contains $x_{il}^\mu$ anyons carrying unit type-$\mu$ anyonic flux, we derive
\begin{equation}
W_{\tilde\beta\mathcal X}^{N^{ij}} = \exp \left(i \sum_\mu \frac{N^{ij} }{N_j}x_{il}^\mu\vartheta_{\tilde\beta \mu} \right)  \hat{\rm I} \label{eq_app_c24}
\end{equation}
At the same time, we know that
\begin{equation}
W_{\tilde\beta\mathcal X_\mu^{n}}^{N^{j\mu}} = e^{i \Omega_{j\mu}} \hat{\rm I}
\label{WNjmu}
\end{equation}
by the definition of $\Omega_{j\mu}$. Comparing (\ref{WNj}) and (\ref{WNjmu}) we see that
\begin{equation}
\Omega_{j\mu}= \frac{N^{j\mu}}{N_j} \vartheta_{\tilde\beta \mu} \pmod{2\pi} \label{eq_app_c26}
\end{equation}
Therefore, we can rewrite (\ref{eq_app_c24}) as
\begin{align}
W_{\tilde\beta\mathcal X}^{N^{ij}} & = \exp \left( i \sum_\mu \frac{N^{ij}x_{il}^\mu }{N^{j\mu}}\frac{N^{j\mu}}{N_j}\vartheta_{\tilde\beta \mu} \right) \hat{\rm I} \nonumber\\
 & = \exp \left( i\sum_\mu \frac{N^{ij}x_{il}^\mu }{N^{j\mu}}\Omega_{j\mu} \right) \hat{\rm I} \label{eq_app_c25}
\end{align}
Here, in the second line, we use the fact that $\frac{N^{ij}x_{il}^\mu }{N^{j\mu}}$ is an \emph{integer} which follows from the general constraint (\ref{x_constraint1}) with a little
algebra. This proves the claim, Eq.~\ref{eq_app_c22}.

All that remains is to derive Eq.~\ref{WNj}. To this end, we observe that $W_{\tilde\beta\mathcal X_\mu^{n}}^{N_{j}}$ is the braid matrix for a process in which a vortex line $\tilde \beta$ is braided
around a surface anyon $\mathcal X_\mu^{n}$ for $N_j$ times. This process is similar to the one discussed in Appendix \ref{sec:appd1}, and we can use similar methods to analyze it. In fact, we can
essentially repeat the diagrammatic computation shown in Fig.~\ref{fig_diagram1} with the substitutions $\mathcal X\rightarrow \tilde\beta$, $\alpha\rightarrow \mathcal X_\mu^{n}$ and $N^{i\mu} \rightarrow N_j$. The first equation in Fig.~\ref{fig_diagram1} carries over without change. As for the second equation, we note that if we fuse $\tilde\beta$ with itself $N_j$ times, the only possible fusion products are surface anyons and charges, since $\tilde\beta$ carries gauge flux
$\phi_{\tilde{\beta}} = \frac{2\pi}{N_j}\mathbf e_j$. Therefore, in the second equation, the summation over charges $q$ is replaced by a summation over surface anyons and charges.
The third equation also carries over with the only modification being that the Aharonov-Bohm phases $\exp(i q\cdot\phi_\alpha)$ are replaced by the statistical phase between these
surface anyons/charges and $\mathcal X_\mu^n$. 
The latter only depends on the anyonic flux $\xi_{\mathcal X_{\mu}^n} = \boldsymbol \epsilon_\mu$, but not on the anyon $\mathcal X_{\mu}^n$ itself. The last equation also holds, since we can group all the numerical factors into a single complex constant $C_{\tilde \beta,\boldsymbol\epsilon_\mu}$.
This constant must be a phase factor, since the braiding process is unitary. Thus, we can write $C_{\tilde \beta,\boldsymbol\epsilon_\mu} = e^{i \vartheta_{\tilde\beta \mu}}$ for some $\vartheta_{\tilde \beta \mu}$. This
establishes Eq.~\ref{WNj}.

\section{Deriving Equation (\ref{eq_dev59})}

\label{sec:app_topospin}

\begin{figure}
\centering
\includegraphics{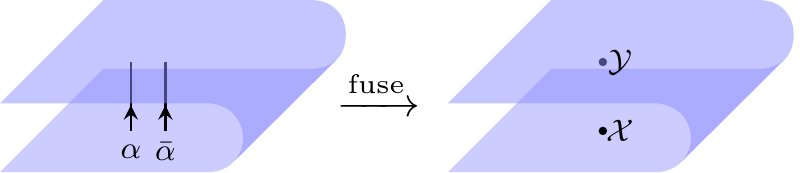}
\caption{Fusing vortex $\alpha$ and its anti-vortex $\bar\alpha$. The resulting doublet $(\mathcal X, \mathcal Y)$ is viewed as a single anyon from the 2D slab point of view.} \label{fig_fusion}
\end{figure}

In this appendix, we derive Eq.~(\ref{eq_dev59}), which we reproduce below for convenience:
\begin{equation}
V |\psi\> = e^{-4\pi i s_\alpha}  |\psi\> \label{eq_app_d1}
\end{equation}
Let us recall the meaning of the different symbols in this equation: $|\psi\>$ is a particular state containing a vortex arch $\alpha$ and its anti-arch $\bar \alpha$; $V$ is the braid matrix associated with braiding $\alpha$ around $\bar \alpha$; and $s_\alpha$ is the topological spin of $\alpha$. Here, the precise definition of $s_\alpha$ involves folding the surface and straightening the vortex lines as in Fig.~\ref{fig_straighten}. After this folding and straightening, we view $\alpha$ as a 2D anyon and we define its topological spin $s_\alpha$ in the usual way.

To prove (\ref{eq_app_d1}) we use a special property of the state $|\psi\>$. To explain this property, imagine fusing together the two arches $\alpha$ and $\bar \alpha$. Since $\alpha$ and $\bar\alpha$ carry opposite gauge flux, this fusion process annihilates the vortex arches in the bulk, leaving behind a pair of surface excitations --- one excitation at each end of the arches. Next imagine that we fuse together the pair of surface excitations. The special property of $|\psi\>$ is that this second fusion process yields a superposition of \emph{charges} rather than more complicated surface anyons. To prove this property, recall that the state $|\psi\>$ is constructed through a particular procedure: we take two bulk vortex loops $\alpha$ and $\bar \alpha$, which are linked to a third loop $\sigma$, and then we absorb them into the boundary and unwind $\sigma$ as in Fig. \ref{fig_deformation}. Now, notice that if we fuse the loops $\alpha$ and
$\bar{\alpha}$ in the bulk, i.e. \emph{before} absorbing them to the surface, then we will get a superposition of charge excitations. At the same time, is not hard to argue that the deformation to the surface cannot change this property. Putting these two facts together, the above property of $|\psi\>$ follows immediately.

To understand the implications of this property, imagine that we fold the surface and then view $\alpha$ and $\bar{\alpha}$ as 2D anyons as
shown in Fig.~\ref{fig_straighten}. We can then think of the $\alpha \times \bar \alpha$ fusion process from a 2D point of view. Clearly
the different fusion outcomes are of the general form $C = (\mathcal X, \mathcal Y)$ where $(\mathcal X, \mathcal Y)$ denotes an excitation
with a surface anyon $\mathcal X$ on the bottom surface and $\mathcal Y$ on the top surface (see Fig.~\ref{fig_fusion}). In this setup,
the above property of $|\psi\>$ is equivalent to the statement that
\begin{equation}
\xi_\mathcal X + \xi_{\mathcal Y} =0.
\label{xiAtilde}
\end{equation}
for any of the possible fusion outcomes $C$.

Equation (\ref{xiAtilde}) is useful because it implies that all the fusion outcomes $C = (\mathcal X, \mathcal Y)$ have vanishing topological spin, that is:
\begin{equation}
s_C = 0
\label{sx0}
\end{equation}
To see this, note that $C$ is the fusion product
of $\mathcal X \equiv (\mathcal X, \mathbf{1})$ and $\mathcal Y^{op} \equiv (\mathbf 1,\mathcal Y)$. Therefore, using a formula from the general algebraic theory\cite{kitaev06} of anyons, we have
\begin{equation}
e^{i 2\pi s_C} \hat{\rm I} = e^{i 2\pi (s_{\mathcal X} + s_{\mathcal Y^{op}})} \cdot R^C_{\mathcal X \mathcal Y^{op}}
R^C_{\mathcal Y^{op} \mathcal X}
\end{equation}
where $R^C_{\mathcal X \mathcal Y^{op}} R^C_{\mathcal Y^{op} \mathcal X}$ describes a full braiding of $\mathcal X$ and
$\mathcal Y^{op}$ in the fusion channel $C$. Let us compute the different terms on the right hand side. We have:
\begin{align}
s_{\mathcal X} = s_{\xi_\mathcal X}
\end{align}
by Eq. (\ref{eq_spin}). Similarly
\begin{align}
s_{\mathcal Y^{op}} = - s_{\mathcal Y} = -s_{\xi_{\mathcal Y}} = -s_{-\xi_{\mathcal Y}} =-s_{\xi_\mathcal X}
\end{align}
where the first equality follows from the fact that the anyons on the top and bottom surfaces have opposite statistics,
the third equality follows from the relation $s_x = s_{-x}$, and the last equality follows from (\ref{xiAtilde}).
At the same time, it is clear that $R^C_{\mathcal X \mathcal Y^{op}} R^C_{\mathcal Y^{op} \mathcal X} = \hat{\rm I}$ since $\mathcal X$ and ${\mathcal Y}^{op}$ live on different surfaces and therefore must have trivial braiding statistics with one another. Combining all these results, equation (\ref{sx0}) follows immediately.

Once we have Eq. \ref{sx0}, we can easily derive Eq. \ref{eq_app_d1}. Indeed, for each of the possible fusion outcomes $C$ we have
\begin{align}
R_{\bar\alpha\alpha}^C R_{\alpha\bar\alpha}^{C} &  = e^{i2\pi(s_C - s_\alpha-s_{\bar\alpha})} \hat{\rm I} \nonumber\\
& =  e^{-4\pi i s_\alpha} \hat{\rm I}
\end{align}
where in the second line we have used the fact that $s_C = 0$ and $s_\alpha = s_{\bar\alpha}$. Since this equation holds for all possible fusion
outcomes of $|\psi\>$, we immediately derive Eq. \ref{eq_app_d1}.

\section{Constraints on the surface and bulk data}
\label{sec:app_con}
In this Appendix, we derive the constraints (\ref{surf_constraint1}-\ref{surf_constraint4}) on $\Omega_{i\mu}$ and $\Omega_{ij\mu}$.
We also discuss the relationship between these surface constraints and previously known constraints
on the bulk data.

\subsection{Constraints on \texorpdfstring{$\Omega_{i\mu}$}{omegaimu} and \texorpdfstring{$\Omega_{ij\mu}$}{omegaijmu}}
\label{sec:app_surin_con}
In the main text, we claimed that the surface data satisfies the constraints (\ref{surf_constraint1} - \ref{surf_constraint4}), which
we reproduce below for convenience:
\begin{align}
N_{i\mu} \Omega_{i\mu} & = 0 \pmod{2\pi} \label{app_con1}\\
N_{ij\mu} \Omega_{ij\mu}  & = 0 \pmod{2\pi} \label{app_con2}\\
\Omega_{ij\mu} + \Omega_{ji\mu} & = 0 \pmod{2\pi} \label{app_con3}\\
\Omega_{ii\mu} &= 0 \pmod{2\pi} \label{app_con4}
\end{align}
We now prove these constraints.

To establish (\ref{app_con1}), it suffices to show that $N_\mu \Omega_{i\mu} =0$ and $N_i \Omega_{i\mu} =0$.
To prove $N_\mu \Omega_{i\mu} =0$, imagine braiding a vortex $\alpha$, carrying unit type-$i$ gauge flux,
around a composite of $N_\mu$ identical anyons $\mathcal X$, each carrying a unit type $\mu$ anyonic flux. Imagine further that we perform
this braiding $N^{i\mu}$ times. We can compute the resulting statistical phase in two different ways. In the first approach, we note
that the fusion product of $N_\mu$ anyons $\mathcal X$ yields a superposition of charge excitations, so the statistical phase must be $0$.
In the second approach, we decompose the braiding process into a sequence of elementary processes in which $\alpha$ is braided around each
$\mathcal X$ individually. From this alternate point of view, it is not hard to see that the total statistical phase is given by
$N_\mu \Omega_{i\mu}$, since each $\mathcal X$ contributes a phase of $\Omega_{i\mu}$. Comparing these two calculations, we conclude that
$N_\mu \Omega_{i\mu} =0$. Similar reasoning shows that $N_i \Omega_{i\mu} =0$, thus proving (\ref{app_con1}).

The proof of (\ref{app_con2}) is similar to (\ref{app_con1}), so we will not present it here. As for (\ref{app_con3}), this follows from
the fact that the braiding process associated
with $\Omega_{ji \mu}$ is identical to the process associated with $\Omega_{ij \mu}$, but performed in the reverse direction.

Finally, we need to show (\ref{app_con4}). To this end, consider a state containing an anyon $\mathcal X$, a vortex $\alpha$ and its anti-vortex
$\bar\alpha$. Suppose that $\mathcal X$ carries unit type-$\mu$ anyonic flux while $\alpha$ carries unit type-$i$ gauge flux. Let us fold the
surface and straighten the vortex lines as in Fig.~\ref{fig_straighten}, and imagine braiding $\mathcal \alpha$ around $\mathcal X$ and
then around $\bar\alpha$, and finally around $\mathcal X$ and $\bar{\alpha}$ in the opposite direction. We define the operators
$V_{\alpha\mathcal X}$ and
$V_{\alpha\bar\alpha}$ as braiding $\alpha$ around $\mathcal X$ and $\bar \alpha$ respectively. Then, the braiding process is algebraically given
by the commutator $V_{\alpha\bar\alpha}^{-1}V_{\alpha\mathcal X}^{-1} V_{\alpha\bar\alpha} V_{\alpha\mathcal X}$.

To derive (\ref{app_con4}), we evaluate this commutator in two different ways. In the first approach,  we let
the commutator act on an initial state $|\psi\rangle$ that satisfies the condition that if we shrink $\alpha$ and $\bar\alpha$ into the
surface, we obtain a superposition of charges. Therefore, Eq.~(\ref{eq_dev59}) is applicable, i.e.,
\begin{equation*}
V_{\alpha\bar\alpha}|\psi\rangle = e^{-4\pi i s_\alpha} |\psi\rangle 
\end{equation*}
In addition, it is clear that the state $V_{\alpha\mathcal X}|\psi\rangle$ satisfies the same condition as $|\psi\rangle$, and therefore
we also have
\begin{equation*}
V_{\alpha\bar\alpha}V_{\alpha\mathcal X}|\psi\rangle = e^{-4\pi i s_\alpha} V_{\alpha\mathcal X}|\psi\rangle 
\end{equation*}
Combining the above two relations, we immediately derive
\begin{equation*}
V_{\alpha\bar\alpha}^{-1}V_{\alpha\mathcal X}^{-1} V_{\alpha\bar\alpha} V_{\alpha\mathcal X}|\psi\rangle = |\psi\rangle 
\end{equation*}
In the second approach, we compute the commutator using the same arguments that we
used to derive Eq.~(\ref{eq_dev8}). Proceeding as in that derivation, a straightforward calculation shows that the commutator is equal to
$e^{-i\Omega_{ii\mu}}\hat{\rm I}$. Comparing the results from the two approaches, we deduce that $\Omega_{ii \mu} = 0$.

\subsection{Relationship with constraints on bulk data}
\label{sec:app_bulkin_con}

Ref.~\onlinecite{wangcj15} showed that, in the case of the group cohomology models\cite{chen13}, the
bulk data satisfies the following constraints (modulo $2\pi$):
\begin{align}
&\Theta_{ii, l}   = 2 \Theta_{i,l}   \label{eq_binv1}\\
&\Theta_{ij,l}  =\Theta_{ji, l}  \label{eq_binv2}\\
&N_{ijl} \Theta_{ij,l}   = 0  \label{eq_binv3}\\
& N_{il} \Theta_{i,l}  = 0  \label{eq_binv4}\\
&\frac{N^{ijl}}{N^{ij}} \Theta_{ij,l}  + \frac{N^{ijl}}{N^{jl}} \Theta_{jl,i} + \frac{N^{ijl}}{N^{li}} \Theta_{li,j} =0  \label{eq_binv5}\\
&\frac{N^{il}}{N_i} \Theta_{i,l}  + \Theta_{il,i} =0  \label{eq_binv6}\\
&\Theta_{i,i}    =0 \label{eq_binv7}  \\
&\Theta_{ijk,l}    = {\rm sgn}(\hat p) \Theta_{\hat p(i)\hat p(j)\hat p(k), \hat p(l)} \label{eq_binv8}\\
&\Theta_{iij,l}   = 0 \label{eq_binv9}  \\
&N_{ijkl} \Theta_{ijk,l}  = 0  \label{eq_binv10}
\end{align}
Here $\hat p$ is a permutation of indices $i,j,k,l$ and ${\rm sgn}(\hat p)$ is its signature. In addition, Ref.~\onlinecite{wangcj15} proved
the converse statement, that is, every solution to these constraints can be realized by a group cohomology model.

Interestingly, one can check that the bulk data defined by the bulk-boundary correspondence (\ref{formula1})-(\ref{formula3})
\emph{automatically} obeys the constraints (\ref{eq_binv1}-\ref{eq_binv10}) as long as the surface data obeys the constraints
(\ref{phi_constraint1})-(\ref{x_constraint3}). The proof is a straightforward mathematical exercise.

\section{Equivalence between \texorpdfstring{$\{\Omega_{i\mu}, \Omega_{ij\mu}\}$ and $\{\omega \in H^2(G, A)\}$}{two types of surface data}}
\label{app:counting}

In this appendix, we show that the mapping defined by Eqs.~(\ref{set_surdata1}) and (\ref{set_surdata2}) is both {\it injective} and {\it surjective}. 
That is, we show that this mapping gives a one-to-one correspondence between the elements $\{\omega\}$ of $H^2(G,A)$ and the
values of the surface data $\{\Omega_{i\mu},\Omega_{ij\mu}\}$ that obey the constraints 
(\ref{surf_constraint1}-\ref{surf_constraint4}).

To show the mapping is injective, we begin by counting the elements in $H^2(G,A)$. According to the Kunneth formula\cite{chen13}, 
\begin{equation}
H^2(G, A) = \prod_{i,\mu}\mathbb Z_{N_{i\mu}} \prod_{ i<j, \mu} \mathbb Z_{N_{ij\mu}},
\end{equation}
where $G = \prod_{i=1}^K \mathbb Z_{N_i}$ and $A =\prod_{\mu=1}^M \mathbb Z_{N_\mu}$. Therefore, the number $\mathcal N$ of elements 
in $H^2(G, A)$ is equal to
\begin{equation}
\mathcal N = \prod_{i,\mu} N_{i\mu} \prod_{ i<j, \mu} N_{ij\mu}
\end{equation}
Given this counting, we can establish injectivity if we can show that image of the map defined by (\ref{set_surdata1}) and (\ref{set_surdata2}) contains $\mathcal N$ distinct values of 
$\{\Omega_{i\mu}, \Omega_{ij\mu}\}$. To this end, consider the following class of 2-cocycles $\omega \in H^2(G,A)$, parameterized by integers $p_{i\nu}$ and $p_{ij\nu}$:
\begin{equation}
\omega^\nu(a, b) =  \sum_i\frac{p_{i\nu}}{N_i}(a_i+b_i-[a_i+b_i]) + \sum_{ij} \frac{N_\nu p_{ij\nu}}{N_{ij\nu}}a_i b_j \label{2cocycle}
\end{equation}
(Here $a,b\in G$).
It is not hard to check that (\ref{2cocycle}) satisfies the 2-cocycle condition (\ref{eq_set1}). 
Inserting (\ref{2cocycle}) into (\ref{set_surdata1}) and (\ref{set_surdata2}), we obtain
\begin{align}
\Omega_{i\mu} & = \sum_{\nu}\frac{p_{i\nu}N_\nu}{N_{i\nu}} \Phi_{\mu\nu} \nonumber \\
\Omega_{ij\mu} &= \sum_{\nu}\frac{(p_{ij\nu}-p_{ji\nu})N_\nu}{N_{ij\nu}}  \Phi_{\mu\nu}\end{align}
By varying $p_{i\mu}$ and $p_{ij\mu}$, it is easy to see that we can obtain $\mathcal N$ distinct values of $\{\Omega_{i\mu},\Omega_{ij\mu}\}$. 
This proves that the mapping given by (\ref{set_surdata1}) and (\ref{set_surdata2}) is injective.

To prove that the mapping is surjective, it suffices to show that there are only $\mathcal N$ values of  
$\{\Omega_{i\mu},\Omega_{ij\mu}\}$ obeying the constraints (\ref{surf_constraint1}-\ref{surf_constraint4}).
The latter result can be established straightforwardly by constructing the most general solution to these constraints.

\bibliography{spt}

\end{document}